\documentclass[sn-mathphys-ay]{sn-jnl}

\usepackage{graphicx}%
\usepackage{multirow}%
\usepackage{amsmath,amssymb,amsfonts}%
\usepackage{amsthm}%
\usepackage{mathrsfs}%
\usepackage[title]{appendix}%
\usepackage{xcolor}%
\usepackage{textcomp}%
\usepackage{manyfoot}%
\usepackage{booktabs}%
\usepackage{algorithm}%
\usepackage{algorithmicx}%
\usepackage{algpseudocode}%
\usepackage{listings}%
\usepackage{adjustbox}

\theoremstyle{thmstyleone}%
\theoremstyle{thmstyletwo}%
\usepackage{amsmath}
\usepackage{mathtools}
\usepackage{bm}
\usepackage{bbm}
\usepackage{caption}
\usepackage{subcaption}
\usepackage{indentfirst}

\newcommand{\nub}{\bm{\nu}}
\newcommand{\thetab}{\bm{\theta}}

\theoremstyle{definition}
\newtheorem{thm}{Teorema}[section]

\newtheorem{dfn}[thm]{Definition}

\everymath{\displaystyle}

\newcommand{\das}{\stackrel{d}{=}}
\newcommand{\distras}[1]{%
	\savebox{\mybox}{\hbox{\kern3pt$\scriptstyle#1$\kern3pt}}%
	\savebox{\mysim}{\hbox{$\sim$}}%
	\mathbin{\overset{#1}{\kern\z@\resizebox{\wd\mybox}{\ht\mysim}{$\sim$}}}%
}
\begin{document}

\title[Bayesian inference for SMSN linear models under the CP]{Bayesian inference for scale mixtures of skew-normal linear models under the centered parameterization}


\author*[1]{\fnm{João} \sur{Victor B. de Freitas}}\email{jvbfreitas@ime.unicamp.br}\equalcont{These authors contributed equally to this work.}

\author[1]{\fnm{Caio} \sur{L. N. Azevedo}}\email{cnaber@ime.unicamp.br}
\equalcont{These authors contributed equally to this work.}

\affil[1]{\orgdiv{Departamento de Estatística, Instituto de Matemática, Estatística e Computação Científica}, \orgname{Universidade Estadual de Campinas}, \orgaddress{\city{Campinas}, \state{São Paulo}, \country{Brazil}}}


\abstract{In many situations we are interested in modeling real data where the response distribution, even conditionally on the covariates, presents asymmetry and/or heavy/light tails. In these situations, it is more suitable to consider models based on the skewed and/or heavy/light tailed distributions, such as the class of scale mixtures of skew-normal distributions. The classical parameterization of this distributions may not be good due to the some inferential issues when the skewness parameter is in a neighborhood of 0, then, the centered parameterization becomes more appropriate. In this paper, we developed a class of scale mixtures of skew-normal distributions under the centered parameterization, also a linear regression model based on them was proposed. We explore a hierarchical representation and set up a MCMC scheme for parameter estimation. Furthermore, we developed residuals and influence analysis tools. A Monte Carlo experiment is conducted to evaluate the performance of the MCMC algorithm and the behavior of the residual distribution. The methodology is illustrated with the analysis of a real data set.}




\maketitle

\section{Introduction}
The normal distribution has been used for many years on diverse fields of knowledge. Despite its simplicity and popularity, it is well known that several phenomena cannot be properly modeled by this distribution, due the presence of asymmetry, heavy/light tails or multi-modality. For example, as noted by \cite{ArellanoValle2009}, the (empirical) distribution of data sets (response variable) often present skewness and heavier or lighter tails than the normal distribution. 

The skew-normal (SN) distribution, proposed by \cite{Azzalini_1985}, has been attracted a lot of attention, due to its mathematical tractability and for having the normal distribution as a special case \citep{Gupta2004}. For more details about the SN distribution (and some interesting extensions) see \cite{azzalini_capitanio_2013} and \cite{genton2004skew}. In the SN distribution the asymmetry is entirely modeled by a single parameter. For this reason, it is a useful alternative to normal distribution for non-symmetric random variables. On the other hand, a very useful class of models that can handle both asymmetry and heavy/light tails is the scale mixtures of skew-Normal distribution (SMSN) (see \cite{Clecio_VH_2011}). This class, as described in \cite{Clecio_VH_2011}, includes the normal and SN distributions as special cases as well as several asymmetric distributions, as the skew-t, skew slash, skew generalized t and skew contaminated normal and the scale mixtures of normal distributions~\citep{Andrews_Mallows}.  

Due to its flexibility the SMSN has been widely used in regression models, see for example \cite{cancho2011bayesian}, \cite{zeller2011local} and \cite{schumacher2021scale}. However, under the direct parameterization, the SN and SMSN distributions may have some problems due to the non-quadratic shape of its likelihood when the skewness is in a neighborhood of 0~\citep{ArellanoAzzalini2008}. To circumvent this problem, the centered parameterization becomes useful~\citep{Azzalini_1985}.

In this paper we introduce the SMSN distributions under the centered parameterization, or centered scale mixtures of SN (CSMSN) distributions, as an alternative to the parameterization given in \cite{Clecio_VH_2011}, which circumvents some problems related to the use of the SN and SMSN distributions under direct parameterization. Also, we present the regression model based on the CSMSN under bayesian paradigm. Residuals and influence diagnostics are discussed, and a application to a real dataset is given after.

\section{Scale Mixtures of Skew-Normal Distribution } \label{s1: smsn}

\subsection{SN distribution} \label{ss1: sn}

We say that Y has SN distribution with location parameter  $\alpha \in \mathbb{R}$, scale parameter  $\beta \in \mathbb{R}^{+}$ and skewness parameter $\lambda \in \mathbb{R}$, denoted by $Y \sim SN(\alpha,\beta ^2,\lambda)$, if its probability density function (p.d.f) is given by $f(y\rvert \alpha,\beta,\lambda)= 2\beta^{-1} \phi\left(\frac{y-\alpha}{\beta}\right)
\Phi\left(\lambda\left(\frac{y-\alpha}{\beta}\right)\right) I_{(-\infty,\infty)}(y)$, where $\phi(.)$ and $\Phi(.)$ denote the the standard normal density and the cumulative distribution function (CDF), respectively.
It is straightforward to see that when $\lambda =0$, the SN reduces to the normal distribution.
Using the results from \cite{Azzalini_1985}, \cite{Pewsey2000} and \cite{Azzalini2005Multi}, the mean, variance, kurtosis ($\gamma_2$) and Pearson's index of skewness ($\gamma_1$) are given, respectively, by: $E(Y) = \alpha + \beta b \delta$, $Var(Y) = \beta^2(1-b^2\delta^2)$, $\gamma_2 = 2(\pi - 3)\frac{(b\delta)^4}{(1 - b^2 \delta^2)^2}$ and $\gamma_1 = \frac{E((Y-E(Y))^3)}{Var(Y)^{3/2}} = \frac{4 - \pi}{2} \frac{(b \delta)^3}{(1 - b^2 \delta^2)^{3/2}}$, where $b= \sqrt{\frac{2}{\pi}}$ and $\delta = \frac{\lambda}{\sqrt{1+\lambda^2}}$.

As noted by \cite{ArellanoAzzalini2008}, the direct parameterization of the SN distribution has some problems in terms of parameter estimation, at least near $\lambda=0$, since the log-likelihood presents a non-quadratic shape. Even under the Bayesian paradigm, this fact can lead to some problems. 
\cite{Pewsey2000} addressed various issues related to the direct parameterization and explained why it should not be used for estimation procedures. \cite{Azzalini_1985} noticed that when $\lambda=0$ the Fisher Information is singular and \cite{Pewsey2000} linked this singularity to the parameter redundancy of the parameterization for the normal case.

\begin{figure}[h!]
     \centering
     \begin{subfigure}[b]{0.49\textwidth}
         \centering
         \includegraphics[width=\textwidth]{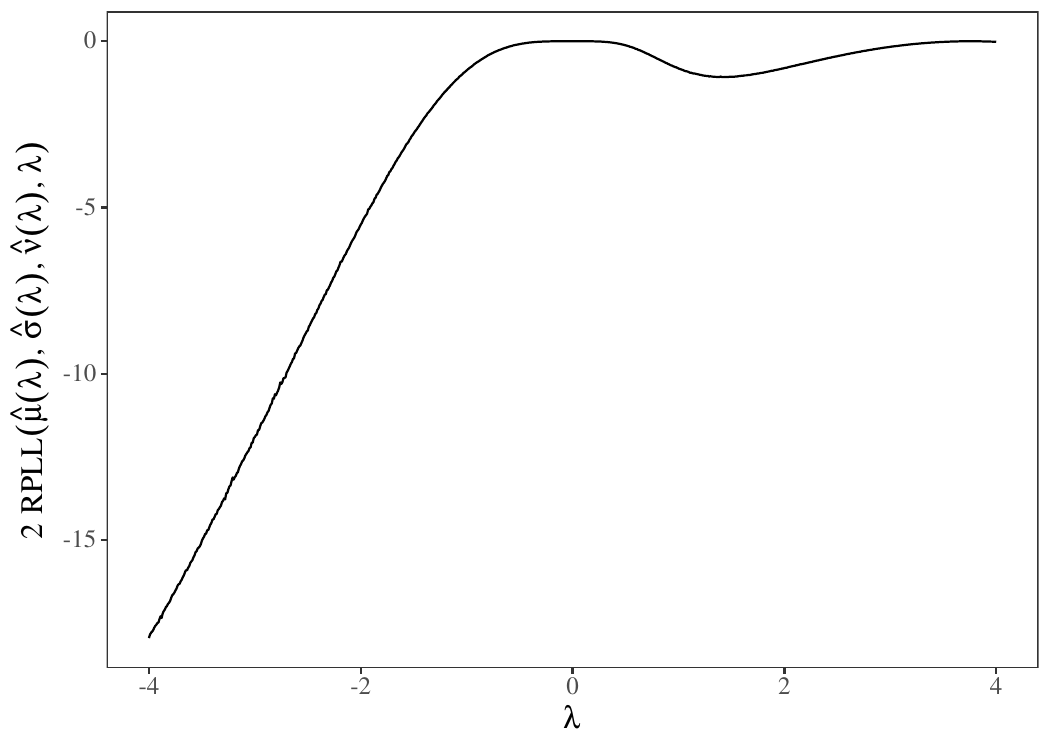}
     \end{subfigure}
     \hfill
     \begin{subfigure}[b]{0.49\textwidth}
         \centering
         \includegraphics[width=\textwidth]{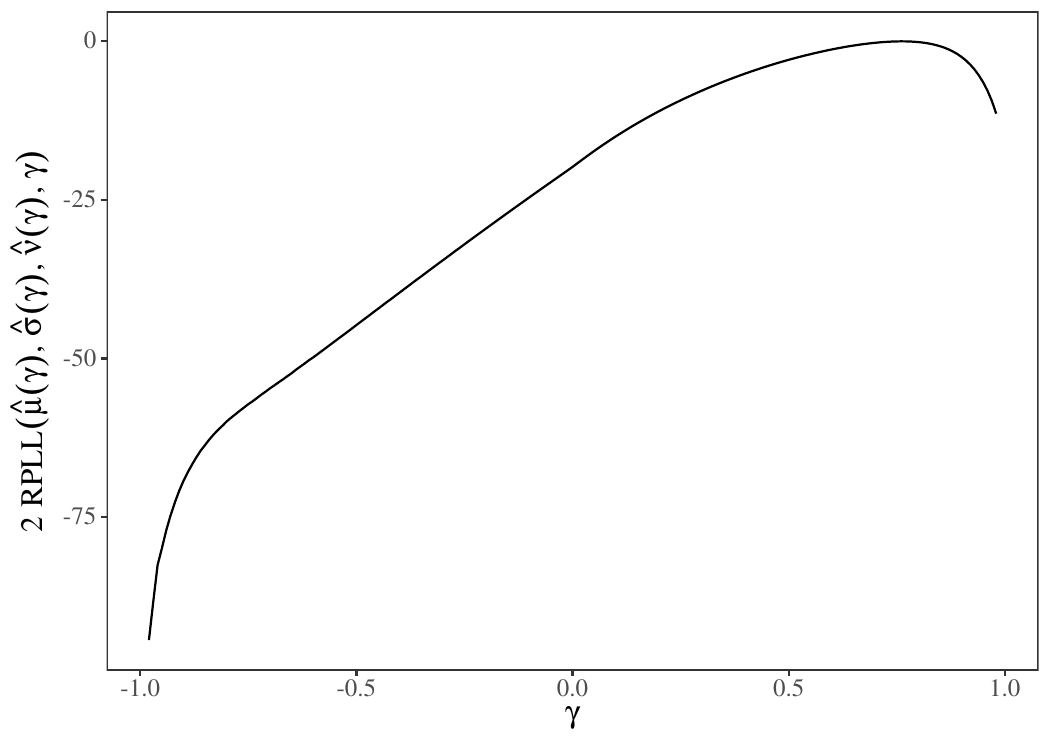}
     \end{subfigure}
     \hfill
     \caption{Twice profiled relative log-likelihood for $\lambda$ in the direct parameterization (left panel) and for $\gamma$ in the centered parameterization (right panel)}
	\label{fig:lvpr_sn}
\end{figure}

In order to show the behavior of the likelihood under the centered and direct parameterization, we calculated the profiled log-likelihood as described in \cite{ZeRoberto} and \cite{azzalini_capitanio_2013}. First, 200 samples were generated from the SN distribution with  $\alpha=0$, $\beta=1$ and $\lambda=4$, and twice the
profiled relative log-likelihood was calculated for both centered and direct parameterization. In Figure \ref{fig:lvpr_sn} it is possible to see that under the direct parameterization the log-likelihood exhibits a non-quadratic shape besides presenting plateaus causing that there is not a single estimate for $\lambda$. However, under the centered parameterization, the log-likelihood presents a concave shape. More details on how to obtain the profiled relative log-likelihood can be found in section 3.

As illustrated in Figure \ref{fig:lvpr_sn}, the problems cited earlier are solved by using the centered parameterization of the SN distribution. Indeed, \cite{Azzalini_1985} proposed an alternative parameterization for $Y \sim SN(\alpha,\beta^2,\lambda)$, which is defined by 
\begin{align}
Y = \mu + \sigma Z_0
\label{centred_sn}
\end{align}
where $Z_0 = \frac{Z - \mu_z}{\sigma_z}$ with $Z \sim SN(0,1,\lambda)$, $\mu_z = b\delta$ and $\sigma_z = \sqrt{1-b^2\delta^2}$.

The alternative parameterization is then formed by the centered parameters $\mu \in \mathbb{R}$,  $ \sigma \in \mathbb{R}^{+} $ and $\gamma \in (-0.99527, 0.99527)$, whose explicit expression are given by $\mu = E(Y) = \alpha  +\beta \mu_z$, $\sigma^2 = Var(Y) = \beta^2(1-\mu_z^2 )$ and $\gamma =  \frac{4 - \pi}{2} \frac{(b \delta)^3}{(1 - b^2 \delta^2)^{3/2}}$, where $\gamma$ denotes the Pearson's skewness coefficient. The centered parameterization of the SN distribution will be denoted by $Y \sim CSN(\mu,\sigma^2,\gamma)$.

Using the Jacobian transformation, the density of  (\ref{centred_sn}), after some algebra, is given by $f(y\rvert \mu,\sigma^2,\gamma) = 2\omega^{-1}\phi(\omega^{-1}(y-\xi))\Phi\left( \lambda \left( \frac{y-\xi}{\omega}\right) \right)$, where 
\begin{align}
\begin{split}
s &= \left(\frac{2}{4-\pi}\right)^{1/3}, \quad \xi = \mu - \sigma\gamma^{1/3}s,\\
\omega &= \sigma \sqrt{1+s^2\gamma^{2/3}} \quad \text{and} \quad
\lambda = \frac{s\gamma^{1/3}}{\sqrt{b^2 + s^2\gamma^{2/3}(b^2-1)}}.
\end{split}   
\label{parameters_density_centred_sn}
\end{align}    

\cite{Henze1986} introduced a useful stochastic representation of the SN, which is given by $Y \das \alpha + \beta (\delta H + \sqrt{1-\delta^2} T)$, where $\das$ means ``distributed as'' and $H \sim HN(0,1) \bot T \sim N(0,1) $, where $HN(.)$ denotes the half-normal distribution.
Therefore, using the Henze's stochastic representation and the SN under the centered parameterization as described in (\ref{centred_sn}), we have that the stochastic representation of the SN under the centered parameterization $Y \sim CSN(\mu,\sigma^2,\gamma)$ is
\begin{align}
Y \das \xi + \omega (\delta H + \sqrt{1-\delta^2} T),
\label{stochastic_centred_sn}
\end{align} 
where $\xi$ and $\omega$ are defined in (\ref{parameters_density_centred_sn}). 


\subsection{Scale mixtures of skew-normal distribution under the centered parameterization}   \label{ss1: smsn centered}
The scale mixtures of normal distributions (SMN) were first introduced by \cite{Andrews_Mallows} and it is often used to model symmetrical data~\citep{Clecio_VH_2011}. A wide class of unimodal and symmetrical distributions can be written as a SMN distribution, as Student-t, contaminated normal, slash, among others~\citep{West}. In \cite{Marcia} the authors proposed a general class of multivariate skew-elliptical distribution that included the SMN distributions as special case. \cite{Clecio_VH_2011} introduced an easy representation of the SMSN distribution class and presented some of its probability and inferential properties, considering the EM algorithm for parameter estimation. As discussed in the previous section, the use of the direct parameterization of SN distribution can lead to some inferential problems. Then, since the SMSN distribution is defined through the direct parameterization of the SN distribution, we now define the CSMSN distribution.


\begin{dfn}
	A random variable Y follows a CSMSN distribution under the centered parameterization, if Y can be stochastically represented by $Y \das \mu + k(U)^{1/2}Z$, where $Z \sim CSN(0,\sigma^2,\gamma)$ and U is a scale random variable with CDF  $G(.\rvert  \bm{\nu})$. 
	\label{def_smsn_cp}
\end{dfn}     
We use the notation $Y \sim CSMSN(\mu,\sigma^2,\gamma,G,\nub)$ for a random variable represented as in Definition \ref{def_smsn_cp}.
From Definition \ref{def_smsn_cp}, it follows that $E(Y)=\mu$, since $E(Z)=0$, and $Var(Y)= \sigma^2 E(k(U))$.
It also can be noted that when $\gamma=0$ we get the corresponding SMN distribution family, since when the skewness parameter is equal to zero, $Z \sim N(0,\sigma^2)$. From Definition \ref{def_smsn_cp}, the hierarchical representation of Y is given by
\begin{align}
\begin{split}
Y\rvert U=u  &\sim CSN(\mu,\sigma^2 k(u), \gamma),\\
U &\sim G(.\rvert  \bm{\nu}).
\label{hierarch_smsn_cp1}
\end{split}
\end{align}


\subsection{Examples of scale mixtures of skew-normal distribution under the centered parameterization}  \label{sss1: ex smsn}
In this section we will present some members of the SMSN family under the centered parameterization. For this work, we will restrict this family considering $k(u)=\frac{1}{u}$. Then, it follows that $Var(Y) =  \sigma^2 E(U^{-1})$.
Before introducing some members of this family, let us first define $d = \frac{(y-\mu)}{\sigma \omega_1}$, $\omega_1 = \sqrt{1+s^2\gamma^{\frac{2}{3}}}$ and $\xi_1 = -\gamma^{1/3}s$. In the section 1 of the supplementary material there are figures and more comments for all distributions described below.

\subsubsection{Skew-t distribution under the centered parameterization:} 

Considering $U \sim \mbox{gamma}(\nu/2,\nu/2)$, where $h(u\rvert \nu) = \frac{\frac{\nu}{2}^{\frac{\nu}{2}}}{\Gamma(\frac{\nu}{2})}  u^{\frac{\nu}{2} - 1} e^{-\frac{\nu}{2} u} I_{(0,\infty)}(u)$, we have the skew-t distribution under the centered parameterization, denoted by $CST(\mu,\sigma^2,\gamma,\nu)$, whose density is given by $f(y\rvert \mu,\sigma^2,\gamma,\nu)= \frac{2\frac{\nu}{2}^{\frac{\nu}{2}}}{\sigma \omega_1 \sqrt{2\pi} \Gamma(\frac{\nu}{2})} e^{\frac{-\xi_1 ^2}{2\omega_1 ^2}} \int_{0}^{\infty} u ^{\frac{\nu +1}{2} -1}e^{-\frac{1}{2}\left[ u (d^2 + \nu) - 2\sqrt{u}d\frac{\xi_1}{\omega_1}\right]}\Phi\left(\lambda\left(\sqrt{u}d - \frac{\xi_1}{\omega_1}\right)\right)du$, where $\nu$ is the degree of freedom and $Var(Y) = \sigma^2 \frac{\nu}{\nu -2} $, since $E(U^{-1}) = \frac{\nu}{\nu -2}$.

\subsubsection{Skew-slash distribution under the centered parameterization:} 

If we consider $U \sim beta(\nu,1)$ then $f(y\rvert \mu,\sigma^2,\gamma,\nu)= \frac{2\nu}{\sigma \omega_1 \sqrt{2\pi}} e^{\frac{-\xi_1 ^2}{2\omega_1 ^2}} \int_{0}^{1} u ^{ \nu + \frac{1}{2} -1}e^{-\frac{1}{2}\left[ u d^2 - 2\sqrt{u}d\frac{\xi_1}{\omega_1}\right]}\Phi\left(\lambda\left(\sqrt{u}d - \frac{\xi_1}{\omega_1}\right)\right)du$, where $\nu$ is the degree of freedom and $Var(Y) = \sigma^2 \frac{\nu}{\nu -1}$. This distribution is denoted by $CSS(\mu,\sigma^2,\gamma,\nu)$.

\subsubsection{Skew-contaminated normal distribution under the centered parameterization:} 

Considering U a discrete random variable assuming only two values, with the following probability function $h(u\rvert \bm{\nu}) = \nu_1 I(u=\nu_2) + (1-\nu_1)I(u=1)$ and $E(U^{-1}) =\frac{\nu_1 + \nu_2(1-\nu_1)}{\nu_2} $. Then the respective density is given by: $f(y\rvert \mu,\sigma^2,\gamma,\nu_1,\nu_2)= 2\biggl\{  \nu_1 \frac{\sqrt{\nu_2}}{\sigma \omega_1 \sqrt{2\pi}} e^{ -\frac{1}{2}\left(\sqrt{\nu_2}d - \frac{\xi_1}{\omega_1} \right)^2}  \Phi\left(\lambda\left(\sqrt{\nu_2}d - \frac{\xi_1}{\omega_1}\right)\right) + (1-\nu_1) \frac{1}{\sigma \omega_1 \sqrt{2\pi}} e^{ -\frac{1}{2}\left(d - \frac{\xi_1}{\omega_1} \right)^2}  \Phi\left(\lambda\left(d - \frac{\xi_1}{\omega_1}\right)\right)  \biggl\}$, denoted by $CSCN(\mu,\sigma^2,\gamma,\bm{\nu})$. According to \cite{Aldo} the parameters $\nu_1$ and $\nu_2$ can be interpreted as the proportion of outliers and a scale factor,
respectively. For this distribution, the variance of Y is equal to $\sigma^2 \frac{\nu_1 + \nu_2(1-\nu_1)}{\nu_2} $.

\subsubsection{Skew generalized t distribution under the centered parameterization:} 

Skew generalized t distribution is obtained by considering $U \sim \mbox{gamma}(\nu_1/2, \nu_2/2)$ as the scale random variable, where $h(u\rvert \bm{\nu}) = \frac{\frac{\nu_2}{2}^{\frac{\nu_1}{2}}}{\Gamma(\frac{\nu_1}{2})}  u^{\frac{\nu_1}{2} - 1} e^{-\frac{\nu_2}{2} u} I_{(0,\infty)}(u)$. However, before provide further details, we will discuss about a identifiability problem. Without loss of generality, consider the case when $\gamma=0$ and $\mu=0$, that is, $Z \sim N(0,\sigma^2)$. Then, we have that $f(y\rvert \mu,\sigma^2,\gamma,\nu_1,\nu_2) = \int_{0}^{\infty} \phi(y\rvert 0,\frac{\sigma^2}{u}) h(u\rvert \bm{\nu}) du$, which leads, after some algebra, to
\begin{align}
f(y\rvert \mu,\sigma^2,\gamma,\nu_1,\nu_2) = \frac{\Gamma(\frac{\nu_1+ 1}{2})}{\sqrt{\nu_2 \sigma^2 \pi} \Gamma(\frac{\nu_1}{2})} \left( 1 + \frac{y^2}{\nu_2\sigma^2}\right)^{\frac{\nu_1 + 1}{2}}.
\label{sgtn_ident}
\end{align}

\begin{figure}[h!]
	\centering
	\includegraphics[width=0.5\textwidth]{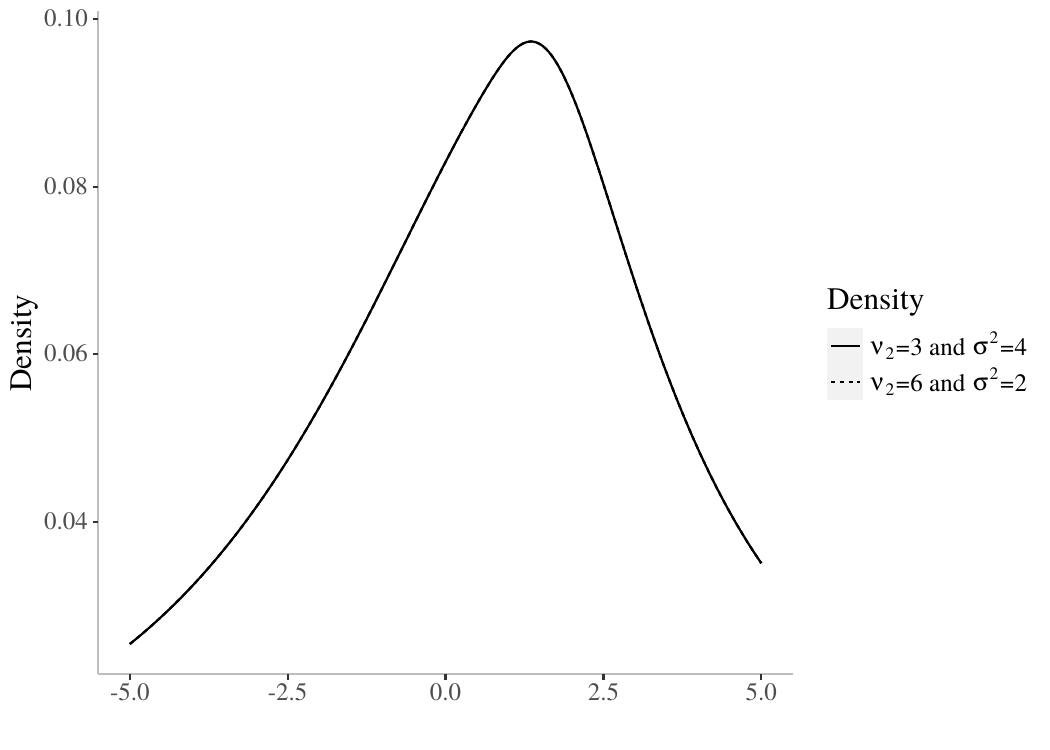}
	\caption{Probability density function of CSGT distribution under the centered parameterization for two parameter sets of $\nu_2$ and $\sigma^2$.}
	\label{fig:sgtn_problema}
\end{figure}

From equation (\ref{sgtn_ident}) it is evident that different values of $\sigma^2$ and $\nu_2$ can produce the same value for the likelihood. Identifiability problems can also be noted in skew generalized t distribution under the centered parameterization. To illustrate this, consider the plot in Figure \ref{fig:sgtn_problema}. It is clear that the two curves overlap each other. Then, to avoid this problem, we will fix $\sigma = 1$. In this way, we will have the skew-t with $\sigma^2=1$ and the SN with $\sigma^2=1$, as special cases of the skew generalized t distribution. 

The density of the skew generalized t distribution under the centered parameterization, denoted by $CSGT(\mu,\gamma,\nu_1,\nu_2)$, is given by $f(y\rvert \mu,\gamma,\nu_1,\nu_2)= \frac{2\frac{\nu_2}{2}^{\frac{\nu_1}{2}}}{\omega_1 \sqrt{2\pi} \Gamma(\frac{\nu_1}{2})} e^{\frac{-\xi_1 ^2}{2\omega_1 ^2}} \int_{0}^{\infty} u ^{\frac{\nu_1 +1}{2} -1} e^{-\frac{1}{2}\left[ u \left(\frac{(y-\mu)^2}{\omega_1^2} + \nu_2\right) - 2\sqrt{u}(y-\mu)\frac{\xi_1}{\omega_1^2}\right]}\times\Phi\left(\lambda\left(\sqrt{u}\frac{(y-\mu)}{\omega_1} - \frac{\xi_1}{\omega_1}\right)\right)du$, where $\nu_1$ and $\nu_2$ are shape parameters that control the variance of Y, since $Var(Y) = \frac{\nu_2}{\nu_1 -2} $.

\section{Profiled log-likelihood for skewness parameter of the SMSN distributions} \label{s1: lvpr}

By introducing the SN distribution, we pointed out some inferential problems related to its direct parameterization. Since the SMSN distribution depends on the direct parameterization (\cite{Clecio_VH_2011}), this family may heritage the same problems, then the respective centered parameterization of this family can circumvents these problems. In this section, our objective is to show that under the direct parameterization for the SMSN distributions, the log-likelihood for $\lambda$ presents problems. In addition, we will show that under the centered parameterization, this problem is no longer observed.


Consider a random sample $\bm{y}=(y_1,y_1,\dots,y_n)^\top$ from the SMSN family under the direct and centered parameterizations and $\thetab = (\mu,\sigma,\lambda,\bm{\nu})$ the parameter vector. We shall denote the log-likelihood under the centered parameterization by $l(\thetab\rvert y)_{CP}$. Under direct parameterization we shall use the density functions as described in \cite{Clecio_VH_2011} and denote the respective log-likelihoods by $l(\thetab\rvert y)_{DP}$.

Following the ideas of \cite{Nathalia} and \cite{ZeRoberto}, the profiled log-likelihood for $\lambda$ is calculated by obtaining the maximum-likelihood estimator for the parameters $\mu, \sigma^2,\bm{\nu}$, for a specific value of $\lambda$ and then plugging these estimates in the log-likelihood. for each $\lambda$ defined in its parametric space we repeat the former steps. 

We denote the profiled log-likelihood of the direct parameterization as $l_{DP}(\hat{\mu}(\lambda), \hat{\bm{\nu}}(\lambda),\lambda)$ for the skew generalized t distribution or $l_{DP}(\hat{\mu}(\lambda),\hat{\sigma^2}(\lambda),\hat{\bm{\nu}}(\lambda),\lambda)$ for the others distributions, where $\hat{\mu}(\lambda), \hat{\sigma^2}(\lambda), \hat{\bm{\nu}}(\lambda) $ stand for values of $\mu, \sigma^2, \bm{\nu}$ that maximizes $l_{DP}$ for a specif value of $\lambda$. 
To make interpretations easier, we have calculated the relative profiled log-likelihood, as suggested by \cite{ArellanoAzzalini2008} obtained by subtracting  $l_{DP}(\hat{\mu}(\lambda),\hat{\sigma^2}(\lambda),\hat{\bm{\nu}}(\lambda),\lambda)$ of $l_{DP}(\hat{\mu}(\lambda),\hat{\sigma^2}(\lambda),\hat{\bm{\nu}}(\lambda),\hat{\lambda})$. Similarly to the direct parameterization, the relative profiled log-likelihood under the centered parameterization is obtained by subtracting  $l_{CP}(\hat{\mu}(\gamma),\hat{\sigma^2}(\gamma), \hat{\bm{\nu}}(\gamma),\gamma)$ of $l_{CP}(\hat{\mu}(\gamma),\hat{\sigma^2}(\gamma),\hat{\bm{\nu}}(\gamma),\hat{\gamma})$.

\begin{figure}[h!]
     \centering
     \begin{subfigure}[b]{0.45\textwidth}
         \centering
         \includegraphics[width=\textwidth]{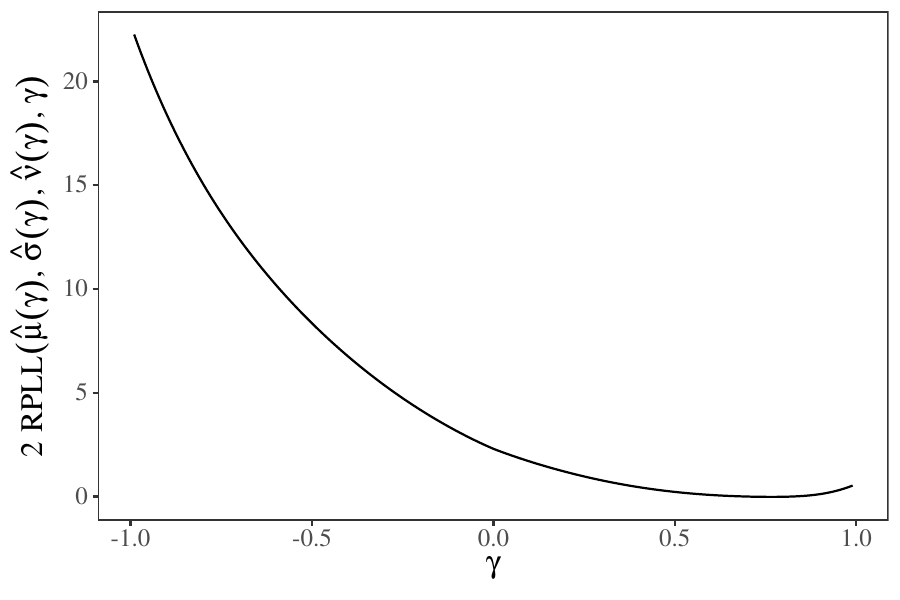}
         \caption{}
     \end{subfigure}
     \hfill
     \begin{subfigure}[b]{0.45\textwidth}
         \centering
         \includegraphics[width=\textwidth]{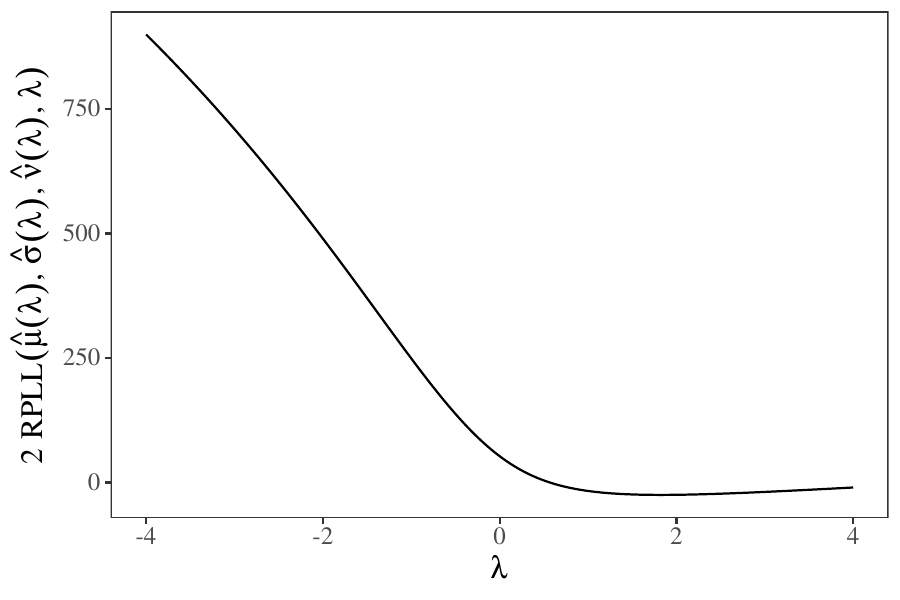}
         \caption{}
     \end{subfigure}
        \caption{Profile twice the relative log-likelihood for $\gamma$ in the centered parameterization (left panel) and for $\lambda$ in the direct parameterization (right panel) for the CST model}
        \label{fig:lvpr_stn}
\end{figure}

\begin{figure}[h!]
     \centering
     \begin{subfigure}[b]{0.45\textwidth}
         \centering
         \includegraphics[width=\textwidth]{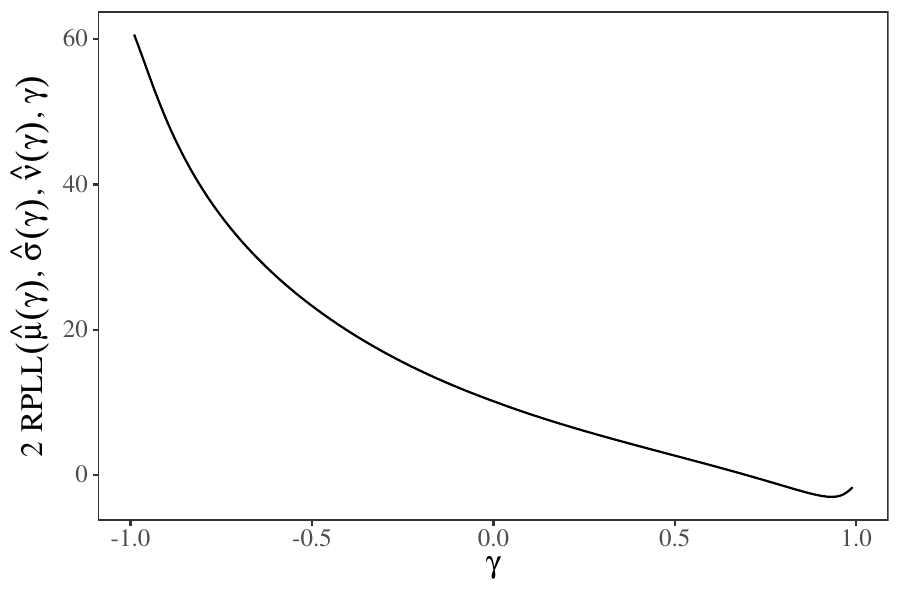}
         \caption{}
     \end{subfigure}
     \hfill
     \begin{subfigure}[b]{0.45\textwidth}
         \centering
         \includegraphics[width=\textwidth]{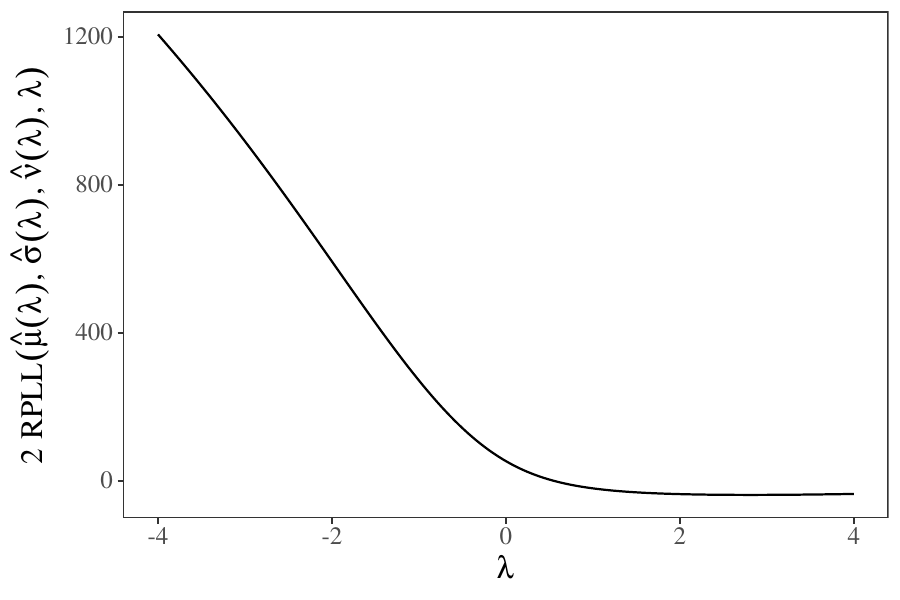}
         \caption{}
     \end{subfigure}
        \caption{Profile twice the relative log-likelihood for $\gamma$ in the centered parameterization (left panel) and for $\lambda$ in the direct parameterization (right panel) for the CSS model}
        \label{fig:lvpr_ssl}
\end{figure}

\begin{figure}[h!]
     \centering
     \begin{subfigure}[b]{0.45\textwidth}
         \centering
         \includegraphics[width=\textwidth]{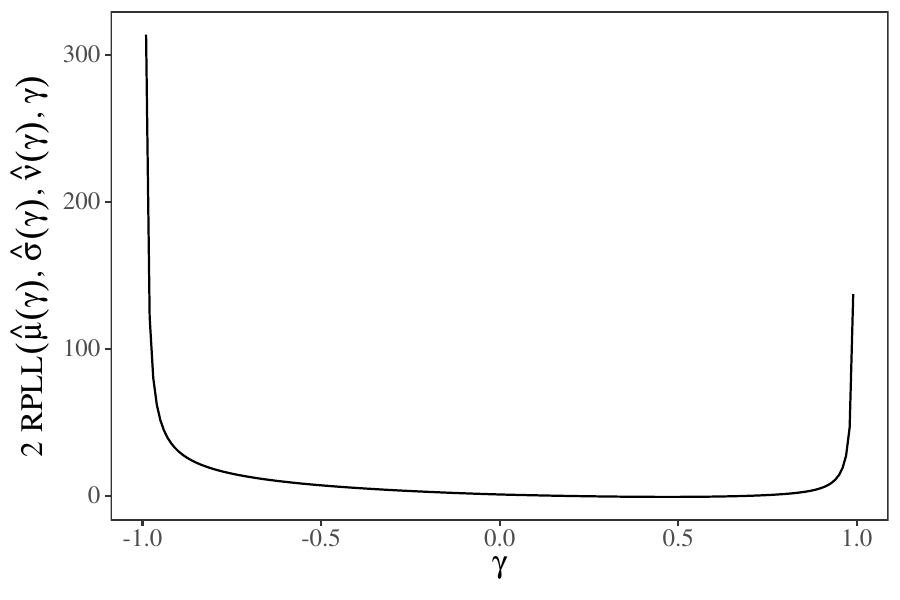}
         \caption{}
     \end{subfigure}
     \hfill
     \begin{subfigure}[b]{0.45\textwidth}
         \centering
         \includegraphics[width=\textwidth]{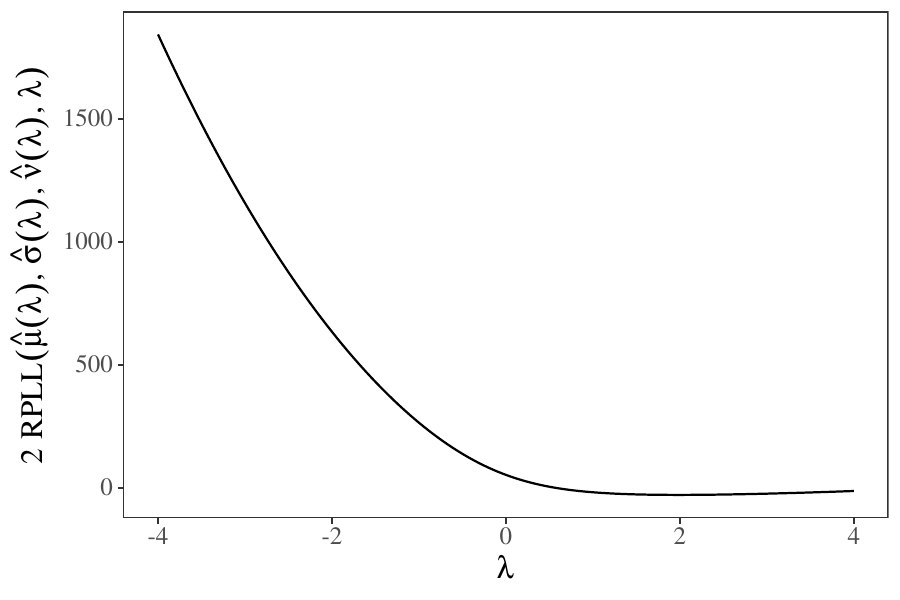}
         \caption{}
     \end{subfigure}
        \caption{Profile twice the relative log-likelihood for $\gamma$ in the centered parameterization (left panel) and for $\lambda$ in the direct parameterization (right panel) for the CSCN model}
        \label{fig:lvpr_scn}
\end{figure}

\begin{figure}[h!]
     \centering
     \begin{subfigure}[b]{0.45\textwidth}
         \centering
         \includegraphics[width=\textwidth]{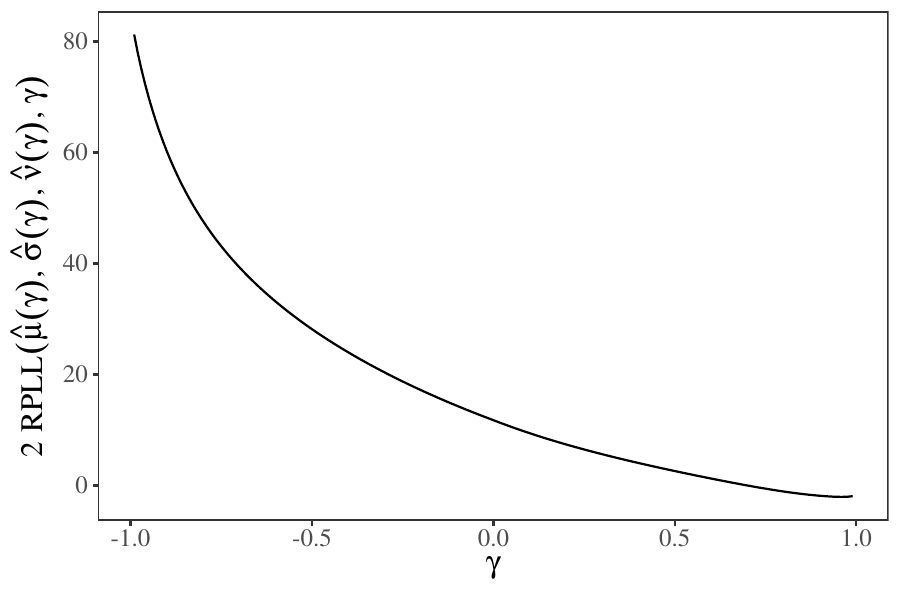}
         \caption{}
     \end{subfigure}
     \hfill
     \begin{subfigure}[b]{0.45\textwidth}
         \centering
         \includegraphics[width=\textwidth]{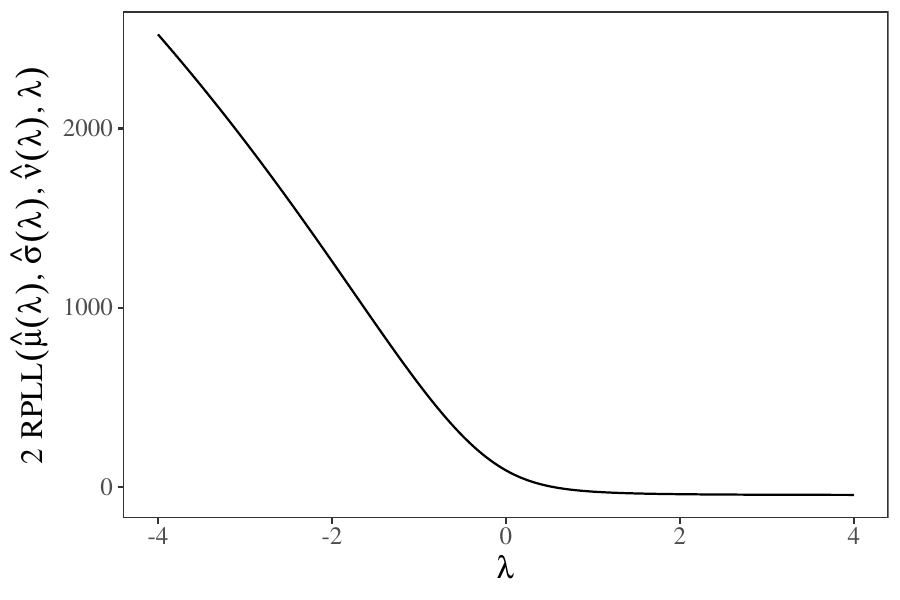}
         \caption{}
     \end{subfigure}
        \caption{Profile twice the relative log-likelihood for $\gamma$ in the centered parameterization (left panel) and for $\lambda$ in the direct parameterization (right panel) for the CSGT model}
        \label{fig:lvpr_stngen}
\end{figure}

We generated a random sample of size 100 from a SMN distribution under the centered parameterization with $\mu=0$, $\sigma^2=1$, $\gamma=0.7$ and $\nu = 4$ for skew slash, $\nu_1=0.4$ and $\nu_2=0.6$ for the skew contaminated normal, $\nu=5$ for skew-t and $\nu_1 = 15$ $\nu_2=5$ for the skew generalized-t. From Figures \ref{fig:lvpr_stn} - \ref{fig:lvpr_stngen}, it is possible to see that under the centered parameterization twice the relative log-likelihood presents a concave behavior while in direct parameterization the relative log-likelihood presents a non-quadratic shape around point zero. We can notice that for values of $\lambda >0$ the function is almost linear. These figures illustrate that under the centered parameterization we have better behavior of  log-likelihood for the parameter $\gamma$ than for the parameter $\lambda$ in direct parameterization. 

We are also interest in analyze the log-likelihood for $\nu$, since depending on its behavior, the estimation procedure can be a complicated task. As noted in \cite{Fili}, for the sinh-contaminated normal distribution for example, large posterior standard deviation and large length of credibility intervals for $\nu_1$ and $\nu_2$ may be explained by the ill-behavior of the profiled log-likelihoods for these parameters. In the section 2 of the supplementary material there is a study of the profiled log-likelihood for $\nu$, in which we can assess the behavior of the likelihood under different values of $\nu$ considering all the CSMSN distributions in this paper.

\section{Regression Model} \label{s1: reg}

Let $\bm{X} = (1,\bm{X}_1,\bm{X}_2,\dots,\bm{X}_{p-1})^\top$ a $p\times n$ design matrix of fixed covariates, $\bm{Y}=(Y_1,\dots,Y_n)^\top$ a $n \times 1$ vector of response variables and $\bm{\beta}= (\beta_0,\beta_1,\dots,\beta_{p-1})^\top$ a $p \times 1$ vector of regression coefficients. The regression model is given by ${Y_i} = \bm{X}_i^\top\bm{\beta} + {\varepsilon_i}, i=1,\dots, n$, where $\varepsilon_i$ are independent random variables identically distributed as a member of CSMSN family. In this work, we say that $\varepsilon_i \stackrel{iid}{\sim} CSS(0,\sigma^2,\gamma,\nu)$ or $\varepsilon_i \stackrel{iid}{\sim} CST(0,\sigma^2,\gamma,\nu)$ or $\varepsilon_i \stackrel{iid}{\sim} CSCN(0,\sigma^2,\gamma,\nu_1,\nu_2)$ or $\varepsilon_i \stackrel{iid}{\sim} CSGT(0,1,\gamma,\nu_1,\nu_2)$, where $CSS,\mbox{ }CST,\mbox{ }CSCN \mbox{ and }CSGT$ denote, respectively, the skew slash, skew-t, skew contaminated and skew generalized t distributions under the centered parameterization.
From the CSMSN it follows that $E({Y_i})=\mu_i= \bm{X}_{i}^{\top}\bm{\beta}$, where $\bm{X}_i$ is the i-th row of matrix $\bm{X}$.
From the hierarchical representation described in (\ref{hierarch_smsn_cp1}), we have that $Y_i\rvert U_i=u_i \sim CSN\left(\bm{X}_{i}^{\top}\bm{\beta}, \frac{\sigma^2}{u_i},\gamma\right)$ and $U_i \sim G(.\rvert  \bm{\nu})$.

We also can represent $Y_i$ as $Y_i\rvert U_i=u_i \das \bm{X}_{i}^{\top}\bm{\beta} + \frac{\sigma}{\sqrt{u_i}}\left(\frac{V_i - \mu_v}{\sigma_v}\right)$,
where $V_i \stackrel{iid}{\sim} SN(0,1,\lambda)$, $\mu_v$ and $\sigma_v$ are the mean and the variance of $V_i$, respectively. Using Henze's stochastic representation for $V_i$, then
\begin{align}
Y_i\rvert U_i=u_i \das \bm{X}_{i}^{\top}\bm{\beta} - \frac{\sigma}{\sqrt{u_i}} \frac{\mu_v}{\sigma_v} +  \frac{\sigma}{\sigma_v \sqrt{u_i}}(\delta H + \sqrt{1-\delta^2}Z). 
\label{stoc_y_reg}
\end{align}

Since $\mu_v =\delta b$ and $\sigma_v = \sqrt{1-b^2\delta^2}$ then (\ref{stoc_y_reg}) becomes
\begin{align*}
Y_i\rvert U_i=u_i  & \das \bm{X}_{i}^{\top}\bm{\beta} - \frac{\sigma \delta b}{\sqrt{u_i}\sqrt{1-b^2\delta^2}}  +  \frac{\sigma \delta}{\sqrt{u_i}\sqrt{1-b^2\delta^2}} H_i + \frac{\sigma \sqrt{1-\delta^2}}{\sqrt{u_i}\sqrt{1-b^2\delta^2}}Z_i \nonumber \\
&= \bm{X}_{i}^{\top}\bm{\beta} + \frac{\sigma \delta }{\sqrt{u_i}\sqrt{1-b^2\delta^2}} \left(H_i -b  \right) +  \frac{\sigma \sqrt{1-\delta^2}}{\sqrt{u_i}\sqrt{1-b^2\delta^2}}Z_i.
\end{align*}

Setting $\Delta = \frac{\sigma \delta}{\sqrt{1-b^2\delta^2}}$ and $\tau = \frac{\sigma^2 ({1-\delta^2})}{{1-b^2\delta^2}}$, we have an one by one transformation such that it is possible to recover $\sigma$ and $\delta$ through: $\delta = \frac{\Delta}{\sqrt{\tau + \Delta^2}}$ and $\sigma^2 = \tau + \Delta^2(1-b^2)$. Then
\begin{align}
Y_i\rvert U_i=u_i \das \bm{X}_{i}^{\top}\bm{\beta} + \frac{\Delta}{\sqrt{u_i}}  (h_i - b) + \frac{\sqrt{\tau}}{\sqrt{u_i}}Z_i.
\label{stoc_y_reg3}
\end{align}

Using (\ref{stoc_y_reg3}), the hierarchical representation of Y becomes
\begin{align*}
Y_i\rvert U_i=u_i,H_i=h_i &\sim N(\bm{X}_{i}^{\top}\bm{\beta} + \frac{\Delta}{\sqrt{u_i}}(h_i-b) ,\frac{\tau}{u_i}), \nonumber \\
H_i &\sim HN(0,1),\\
U_i &\sim G(.\rvert  \bm{\nu}). \nonumber
\end{align*}

\subsection{Bayesian Inference} \label{ss1: bayes inf}
To use the Bayesian paradigm, it is essential to obtain the joint posterior distribution. However, since the necessary integrals are not easy to calculate, it is not possible to obtain such distribution, analytically. However, it is possible to draw samples from the posterior distribution of interest by using Markov Chain Monte Carlo (MCMC) algorithms and use appropriate statistics to obtain estimates for the parameters, see \cite{Geman1984} and \cite{Hasting}, for example.

To obtain the posterior distribution we need to consider the complete likelihood: $L_c(\bm{\thetab}\rvert y,u,h) \propto \prod_{i=1}^{n} \phi\left(y_i\rvert \mu_i, \tau u_i^{-1}\right) f(h_i)h(u_i\rvert \bm{\nu})$
where $\mu_i = \bm{X}_{i}^{\top}\bm{\beta}+ \frac{\Delta}{\sqrt{u_i}}(h_i-b)$ and $\bm{\thetab} = (\bm{\beta},\Delta,\tau,\bm{\nu})$. Since we set $\sigma^2=1$ for the CSGT model, then we have that $\bm{\thetab} = (\bm{\beta},\delta,\bm{\nu})$, which implies that $\Delta$ and $\tau$ are functions of only $\delta$. Then, for the CSGT model, is preferable to sample directly from $\delta$, instead of sampling for $\Delta$ and $\tau$.

We need to consider a prior distribution for $\bm{\thetab}$. We will assume an independence structure, that is $\pi(\bm{\thetab}) = \pi(\bm{\beta}) \pi(\Delta)\pi(\tau)\pi(\bm{\nu})$ for the CST, CSS and CSCN models and $\pi(\bm{\thetab}) = \pi(\bm{\beta}) \pi(\delta)\pi(\bm{\nu})$, for the CSGT model. Furthermore, we will assume conditional conjugate prior distributions, as in \cite{Gelman06}, for $\bm{\beta}$, $\tau^{-1}$, $\Delta$ and $\delta \in U(-1,1)$. On the other hand, for $\bm{\nu}$, the choice of the prior distribution will depend on the model. Therefore, the priors for  $\bm{\beta}$, $\tau^{-1}$ and $\Delta$ are $\bm{\beta} \sim N(\mu_\beta, \Sigma_\beta)$, $\tau^{-1} \sim \mbox{gamma}(c,d)$ and $\Delta \sim N(\mu_\Delta, \sigma_\Delta^2)$. From Bayesian theory, the posterior distribution of $\bm{\thetab}$ can be calculated as $\pi(\bm{\thetab}\rvert \bm{y}) \propto \pi(\bm{\thetab}) f(\bm{y}\rvert \bm{\thetab}) $ where $f(\bm{y}\rvert \bm{\thetab})$ is the joint distribution of the data $\bm{y} = (y_1,y_2, \dots, y_n)$. More details about the full conditional and prior distributions can be found in the section 4 of the supplementary material.


\subsection{Residual analysis} \label{sss1: res}
Residual analysis is an important tool for model fit assessment, including detection of departing from model assumptions as well as the presence of outliers. For this section we consider a residual analysis based on Bayesian estimate for each unknown parameter: $R_i = \frac{Y_i - \bm{X}_{i}^{\top}\bm{\hat{\beta}}}{\hat{\sigma}}, \quad i=1,\dots, n$.

We expected that the residuals $R_i$ approximately follows, under the good fit of the model, a $CST(0,1,\gamma,\nu)$, $CSS(0,1,\gamma,\nu)$, $CSCN(0,1,\gamma,\nu_1, \nu_2)$ or $CSGT(0,1,\gamma,\nu_1,\nu_2)$ distribution, according to the respective adopted distribution, with $\bm{\nu}$ and $\gamma$ equal to the Bayesian estimates. For checking goodness of fit, we can build envelope plot, using the above mentioned distributions to simulate the envelopes. For outliers detection, we can graph the residual versus the index of the observations and the residuals against the fitted values.

Model fit assessment and model comparison criteria can be found in details in the section 3 of the supplementary material. We showed the expected Akaike information criterion (EAIC), expected Bayesian information criterion (EBIC), deviance information criterion (DIC) and log pseudo-marginal likelihood (LPML) criterion to compare competing models for a given data set. Also, we provided the Kullback-Leibler (KL) divergence to identify influential observations.

\section{Simulation study} \label{s1: simulation}
We performed simulation studies in order to evaluate the performance of the model and the estimation method proposed in this work. All these models were implemented in Nimble (\cite{nimble}) through the interface provided by the nimble package (\cite{nimblepackage}) available in R program (\cite{R}). The codes are available from the authors upon request.

\subsection{Parameter recovery} \label{ss1: par recov}

We considered different scenarios based on the crossing of the levels of some factors of interest. For the four CSMSN distribution examples exposed in this work, we simulate from samples of size n=50, 100, 250 and 500 and R=50 replicas were made. A sample of the regression model were simulated considering $Y_i = \beta_0 + \beta_1 x_i + \varepsilon_i \quad i=1,\dots,n$, where $\beta_0=1$, $\beta_1=2$, $\varepsilon_i$ belongs to the CSMSN family with $\sigma^2=1$ and $\gamma$ to have strong asymmetry. Also, we set $\nu = 5$ for the CST distribution; $\nu = 3$ for the CSS distribution; $\bm{\nu} = (\nu_1,\nu_2) = (0.5, 0.5)$ for the CSCN distribution and $\bm{\nu} = (\nu_1,\nu_2) = (15,5)$ for the CSGT distribution. These values for $\bm{\nu}$ were chosen in order to have distributions with heavy tails than the SN distribution. The covariates were simulated from a standard normal distribution and centered in their respective means. To eliminate the effect of the initial values and to avoid correlations problems, we run a MCMC chain of size 60,000 with a burn-in of 20,000 and thin 40, so we retain a valid MCMC chain of size 1000. The analyses of traceplots and autocorrelation plots indicated that the MCMC algorithm converged and the autocorrelation were negligible. 

To compare the performance of the estimation methods we considered some appropriate statistics. Let $\vartheta$ be an element of  $\bm{\thetab}=(\bm{\beta},\gamma,\sigma^2,\bm{\nu})$ and $\hat{\vartheta_r}$ the respective posterior estimate from the r-th  replica. These statistics are: mean of the estimates of parameter $\vartheta$ (Est) $\bar{\hat{\vartheta}} =\frac{\sum_{r=1}^{R} \hat{\vartheta_r} }{R}$, variance of the estimates $Var_{\vartheta} =\frac{\sum_{r=1}^{R} ( \hat{\vartheta_r}  -\bar{\hat{\vartheta}} )^2  }{R-1}$, bias of the estimates $\bar{\hat{\vartheta}} -\vartheta$, relative bias $\frac{\rvert  Bias_{\vartheta}\rvert }{\vartheta}$, square root of the mean square error $RMSE_{\vartheta} =\sqrt{ Bias_{\vartheta}^2 + Var_{\vartheta} }$, the length of the credibility interval and the Coverage Ratio of the 95\% credibility interval of parameter $\vartheta$. To obtain it, we have calculated the numbers of intervals which contain the true value of the parameter, and then divide it by the total of replicas. 

Posterior mean, median and mode were calculated for each parameter $\vartheta$, in each replica. The choice of one of these three statistics was made analyzing overall bias and variance, considering a high sample size of $n=500$ and the true value of $\nu$. We chose for each distribution that posteriori statistic that balanced bias and variance, where these choices are in the Table \ref{statpost}. In section 5.1 of the supplementary material we have the bias and variance under different posterior statistics for each distribution.

\begin{table}[h!]
    \centering
    \caption{Posterior statistics chosen to estimate each parameter of each distribution.}
    \label{statpost}
    \begin{tabular}{c|ccccc}
    \hline
    Distribution & $\beta$ & $\sigma^2$ & $\delta$ & $\nu (\text{or }\nu_1)$& $\nu_2$\\
    \hline
    Skew-t & Median & Mean & Median & Mode & -\\
    Skew slash & Median & Mean & Mean & Mode& - \\
    Skew contaminated normal  & Median & Mean & Mode & Mean & Mode\\
    Skew generalized t & Median & - & Median & Median & Median\\
    \hline
    \end{tabular}
\end{table}




We adopted weakly informative priors for all parameters, that is: $\Delta \sim N(0,100)$,  $\tau \sim \mbox{gamma}(0.01, 0.01)$, $\beta_0 \sim N(0,100)$ and $\beta_1 \sim N(0,100)$. For the CST model we set $\nu \sim exp(\thetab)T(2,)$ and $\thetab \sim unif(0.02, .49)$; for the CSCN model we used $\nu_1 \sim beta(1,1)$ and $\nu_2\sim beta(1,1)$, for the CSGT model we set $\nu_1 \sim exp(\thetab_1)T(2,)$ and $\thetab_1 \sim unif(0.02, .49)$ and $\nu_2 \sim exp(\thetab_2)$ and $\thetab_2 \sim unif(0.02, .49)$, and for the CSS model we set $\nu \sim exp(\thetab)T(1,)$ and $\thetab \sim unif(0.02, .99)$.

All results for the parameter recovery study for all distributions can be found in the section 5.1 of the supplementary material. For the CST and CSS models, we can notice that for all sample sizes, the estimates of all parameters were close to the true values, and the length of the credibility intervals decreases as the sample size increases. Only for $\nu$ we noticed that this parameter can present a somewhat high variance in small samples, but that it decreases considerably with increasing of the sample size.

For the CSCN distribution we can see that in general, for all sample sizes, the estimates of all parameters were close to the true values as well as its variances are low, and the length of the credibility intervals decreases as the sample size increases. For the CSGT distribution, the estimates of all parameters become more precise, as the sample size increases. Also the length of the credibility intervals decreases. We only make an observation that for very small samples ($n=50$) the estimates for $\delta$, $\nu_1$ and $\nu_2$ may not be good, and for small samples ($n=100$) the estimates for $\nu_1$ and $\nu_2$ may still be biased. Also, for the shape parameters, its variances can be high in all scenarios.

Comparing all the results obtained in the simulations, it is observed that the estimates, in general, are more accurate as the sample size increases. For small sample sizes ($< 250$), the CST, CSS and CSCN models performs very well with estimates very close to actual values and in many cases very accurately. For the CSGT we notice that this model is only reliable in large sample sizes ($n>=250$).

\subsection{Residual analysis} \label{ss1: simulation residual}

In this section we analyzed the behavior of the residuals, presented in Section \ref{sss1: res}, under some conditions of interest. We have conducted a simulation study, considering a sample size of 500. We simulated data sets, for each regression model, considering $\beta_0=1$, $\beta_1=2$, $\sigma^2=1$ and $\gamma=-0.9$. Also, we considered: $\nu=5$ for CST and $\nu=3$ for the CSS distributions; $\bm{\nu}=(15,5)$ for CSGT and $\bm{\nu}=(0.5, 0.5)$ for CSCN distribution. 

For each simulated data, we fitted the SN, CST, CSS, CSGT and CSCN models using the priors described in Section \ref{ss1: par recov}. We built suitable quantile-quantile plots for all models, where the confidence bands were made considering the underlying distribution, with $\mu=0$ and $\sigma^2=1$. All figures for the residual simulation study for all distributions can be found in the section 5.2 of the supplementary material.

For all simulated data, we can see that when data exhibits heavy tails, residuals obtained from the SN fit indicated that this model did not fit well to the data, since the residuals lying outside the confidence bands. For all models, when we fitted the correct model to the data, there is no residual lying outside the confidence bands, indicating that the model is well adjusted. Comparing the fit by another members of CSMSN class, we notice that when observations are generated by CST and CSS distributions and we fitted with the CSGT and CSCN models, the fit using these distributions are not as good as the rest.

\section{Application} \label{s1: application}
The application is made on Australian Athletes dataset described in \cite{weisberg}. This data consist on a sample of 202 elite athletes who were in training at the Australian Institute of Sport. We consider the lean body mass (LBM) as our response and the height in cm (Ht), weight in kg (Wt), and the sex (0 = male and 1 = female) as our covariates for the regression model.
We consider a regression model of the form $Y_i = \beta_0 + \beta_1 x_1 + \beta_2 x_2 + \beta_3 x_3 + \varepsilon_i$, for $i=1,2,\dots, 202$, where $x_1 = 0$ if male and 1 if female, $x_2$ and $x_3$ are, respectively, the covariates Ht and Wt centered in their respective mean. 

We fitted five models, assuming that: $\varepsilon_i \stackrel{iid}{\sim} CST(0,\sigma^2,\gamma,\nu)$, or , $\varepsilon_i \stackrel{iid}{\sim} CSCN(0,\sigma^2,\gamma,\nu_1,\nu_2)$, or $\varepsilon_i \stackrel{iid}{\sim} CSGT(0,1,\gamma,\nu_1,\nu_2)$, or $\varepsilon_i \stackrel{iid}{\sim} CSS(0,\sigma^2,\gamma,\nu)$, or, $\varepsilon_i \stackrel{iid}{\sim} CSN(0,\sigma^2,\gamma)$ that we denote, respectively, by CST, CSCN, CSGT, CSS and CSN. The values for the MCMC algorithm were the same used in the simulation study. The priors for all other parameters were chosen according to the simulation study available in Section \ref{ss1: par recov}. 

\begin{figure}[h!]
	\begin{subfigure}{\linewidth}
		\begin{subfigure}{\linewidth/3-1em}
			\includegraphics[width=\linewidth]{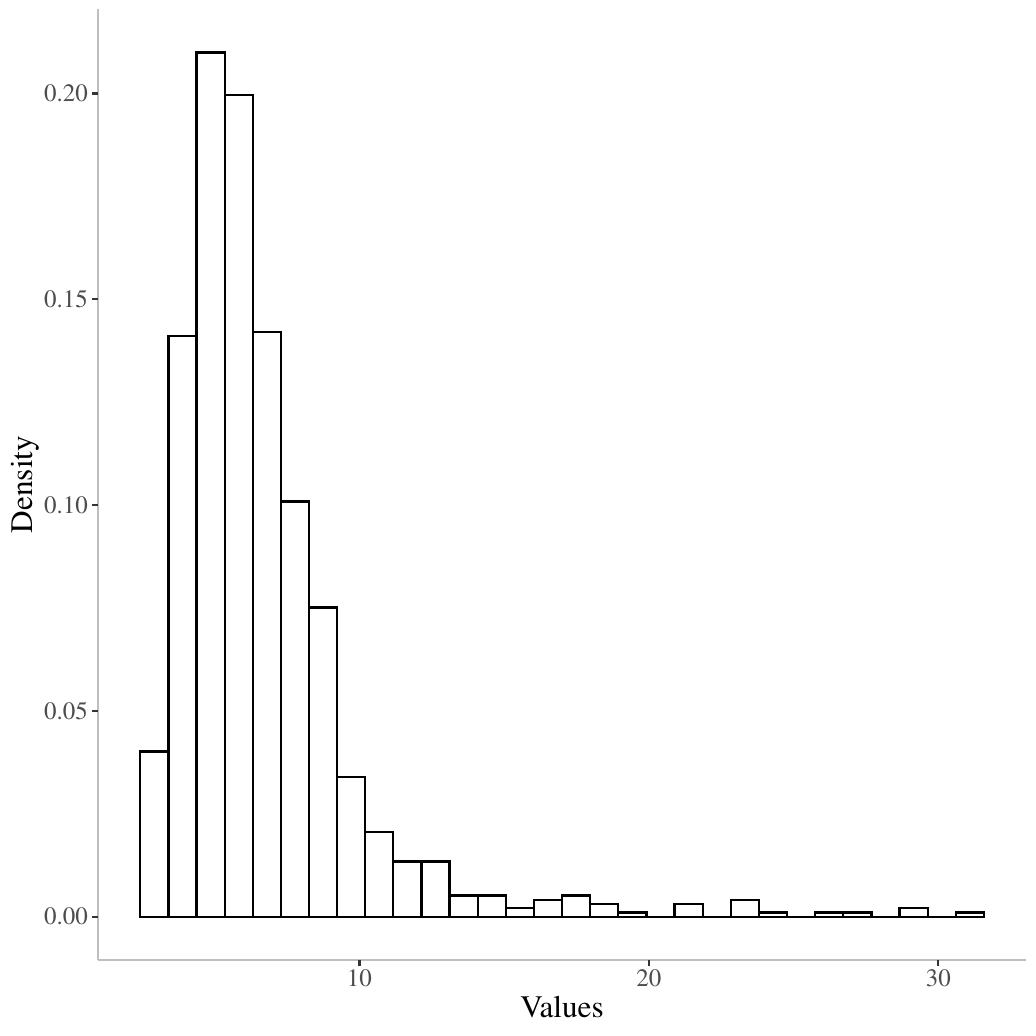}
			\caption{}
			\label{fig:nu_cst}
		\end{subfigure}\hfill
		\begin{subfigure}{\linewidth/3-1em}
			\includegraphics[width=\linewidth]{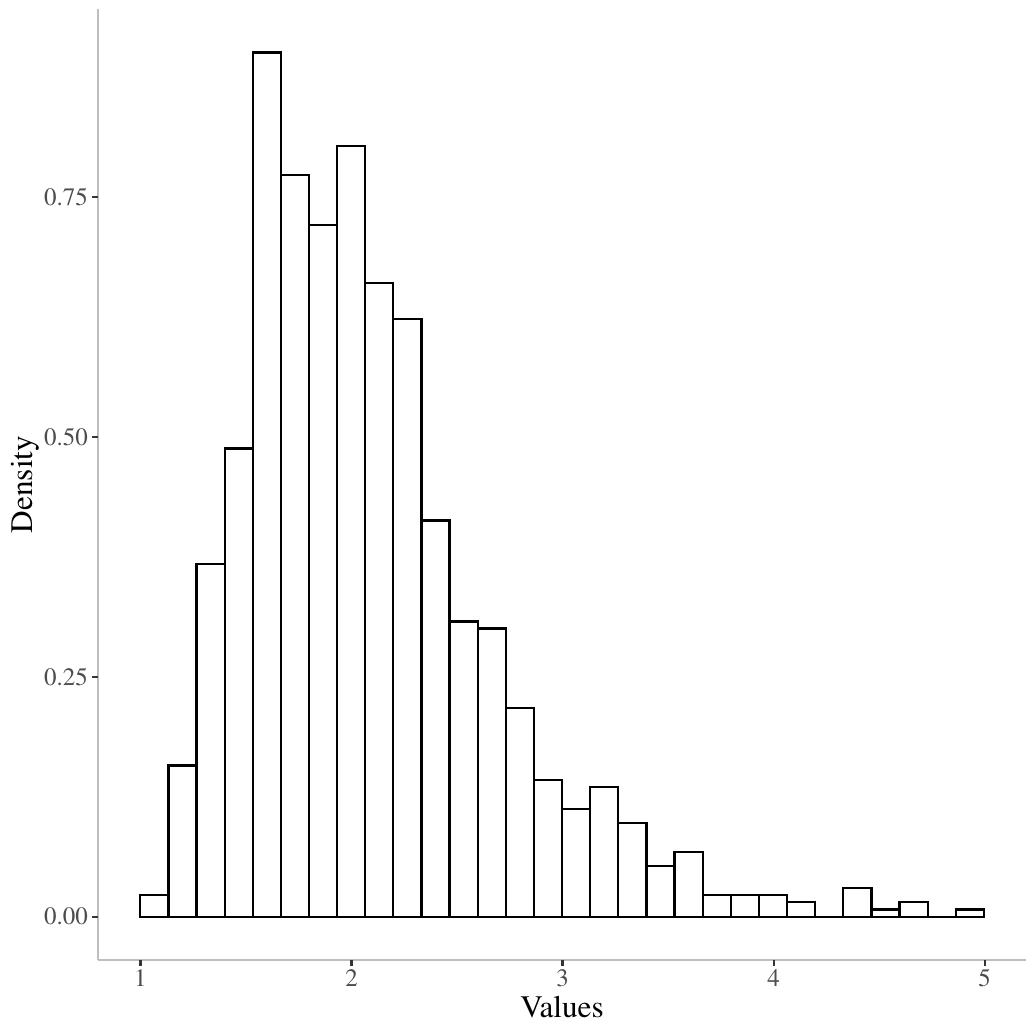}
			\caption{}
			\label{fig:nu_css}
		\end{subfigure}\hfill
		\begin{subfigure}{\linewidth/3-1em}
			\includegraphics[width=\linewidth]{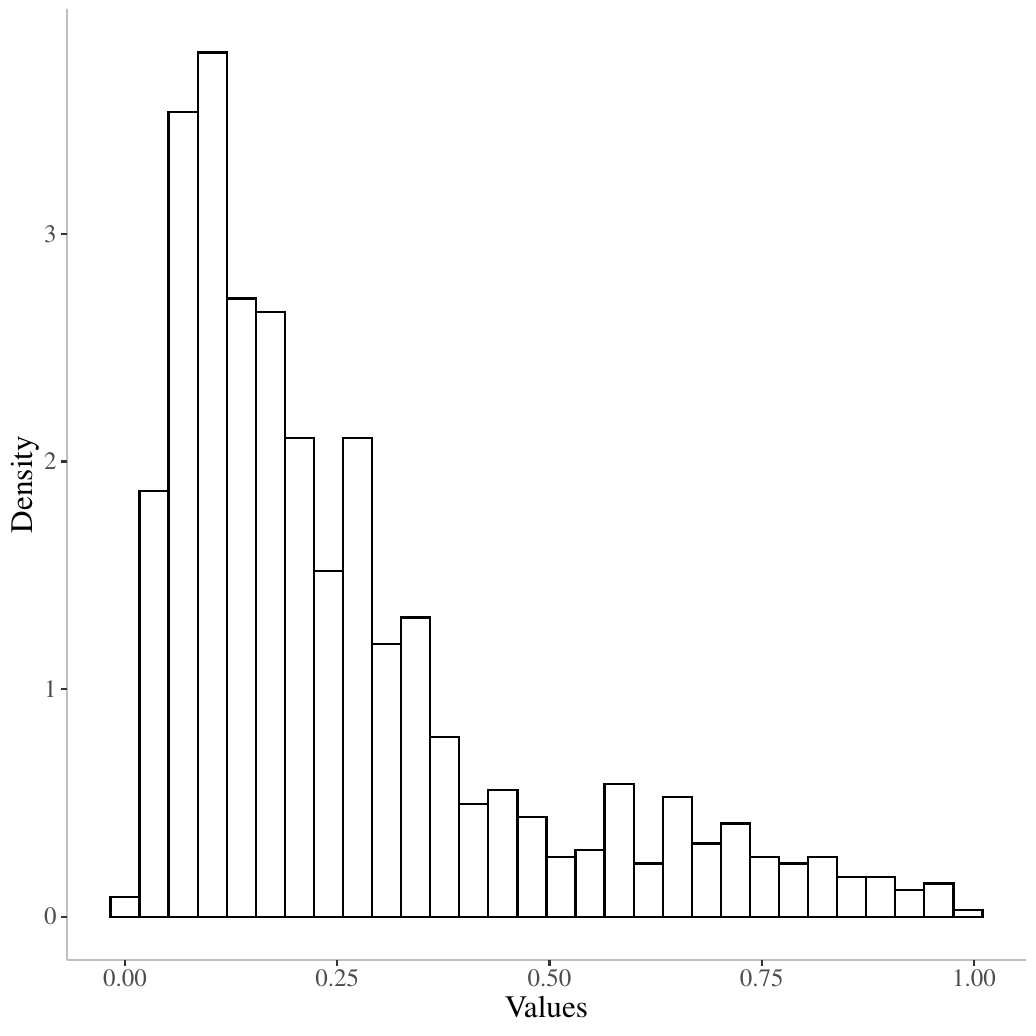}
			\caption{}
			\label{fig:nu1_cscn}
		\end{subfigure}
	\end{subfigure}
	\par\bigskip
	\begin{subfigure}{\linewidth}
		\begin{subfigure}{\linewidth/3-1em}
			\includegraphics[width=\linewidth]{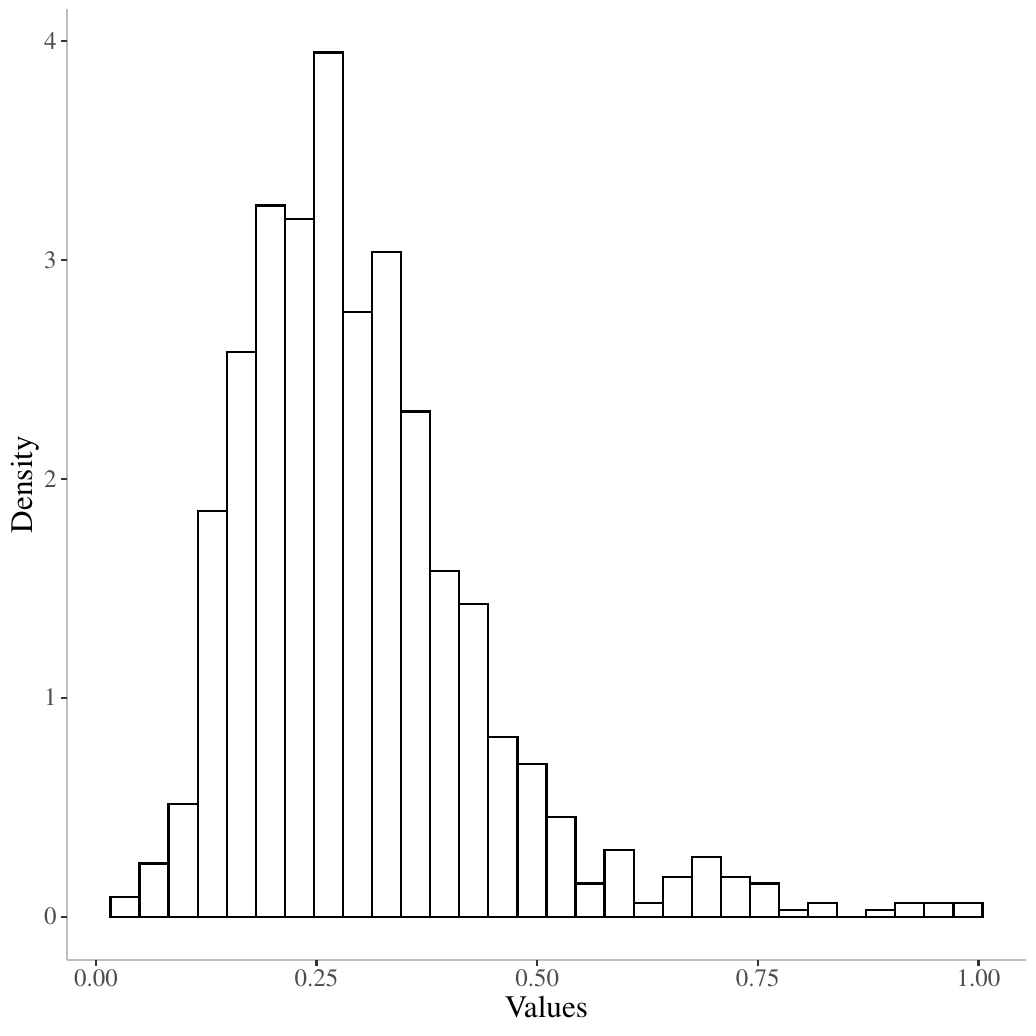}
			\caption{}
			\label{fig:nu2_cscn}
		\end{subfigure}\hfill
		\begin{subfigure}{\linewidth/3-1em}
			\includegraphics[width=\linewidth]{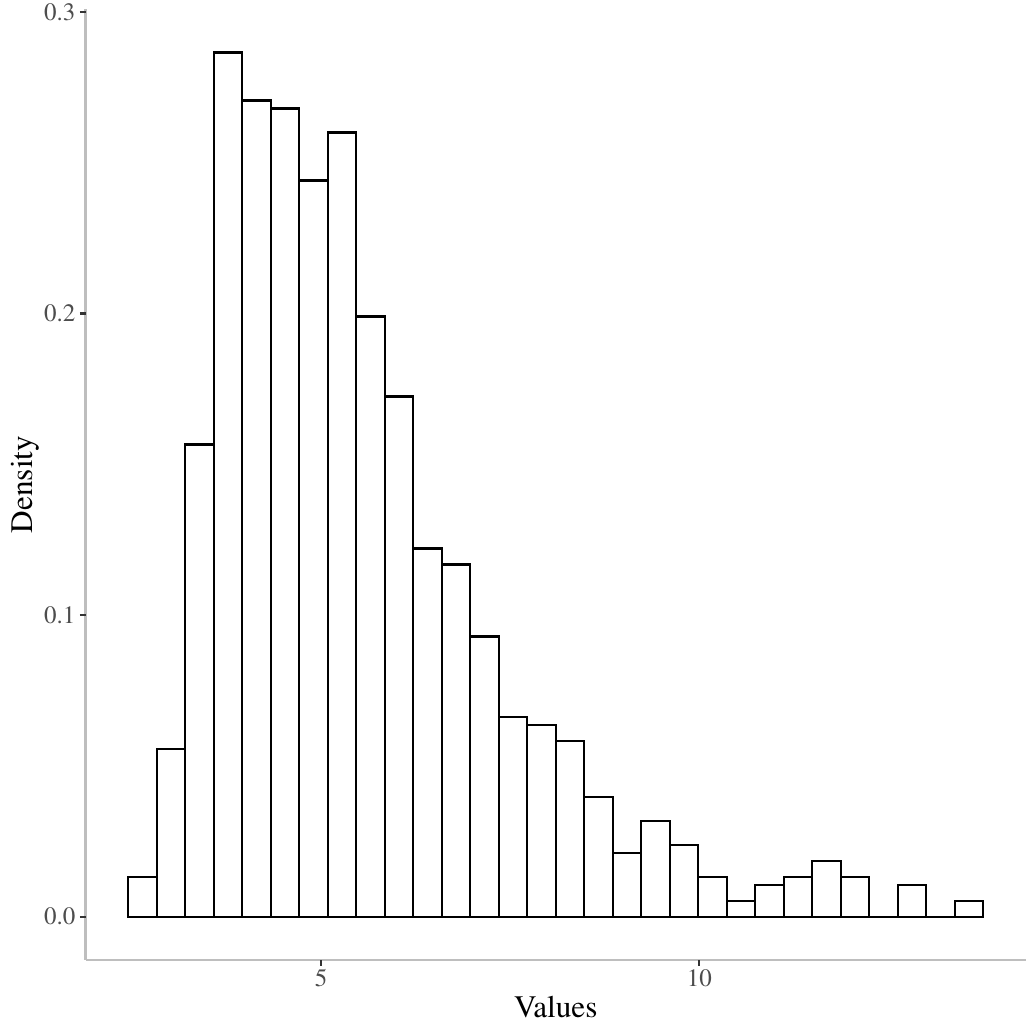}
			\caption{}
			\label{fig:nu1_csgt}
		\end{subfigure}\hfill
		\begin{subfigure}{\linewidth/3-1em}
			\includegraphics[width=\linewidth]{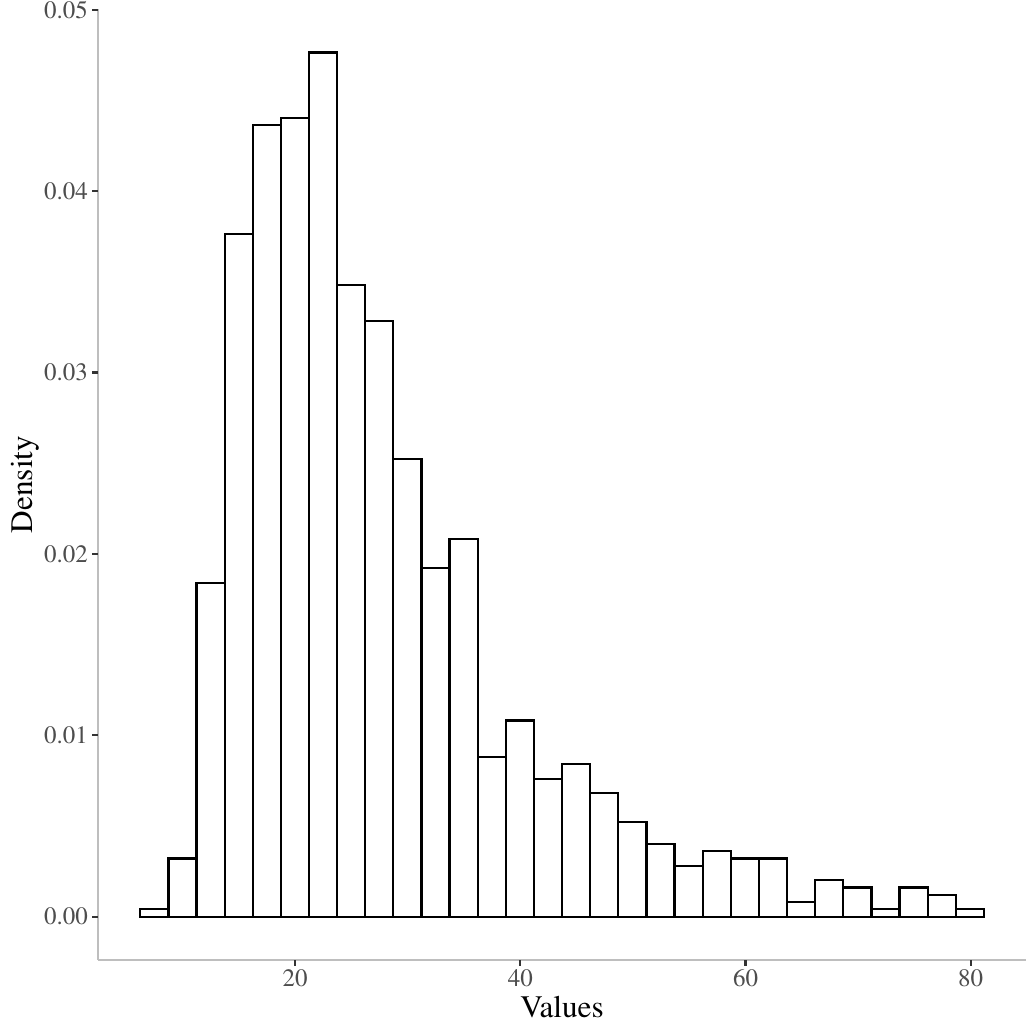}
			\caption{}
			\label{fig:nu2_csgt}
		\end{subfigure}
	\end{subfigure}
	\caption{\textbf{AIS dataset}: Posterior distribution of $\nu$ for the CST (\subref{fig:nu_cst}) and CSS (\subref{fig:nu_css}), $\nu_1$ (\subref{fig:nu1_cscn}) and $\nu_2$ (\subref{fig:nu2_cscn}) for the CSCN model and posterior distributions of $\nu_1$ (\subref{fig:nu1_csgt}) and $\nu_2$ (\subref{fig:nu2_csgt}) for the CSGT model.}
	\label{fig:ais_nu}
\end{figure}

\begin{figure}[h!]
	\begin{center}
		\begin{subfigure}{\linewidth/3-1em}
			\includegraphics[width=\linewidth]{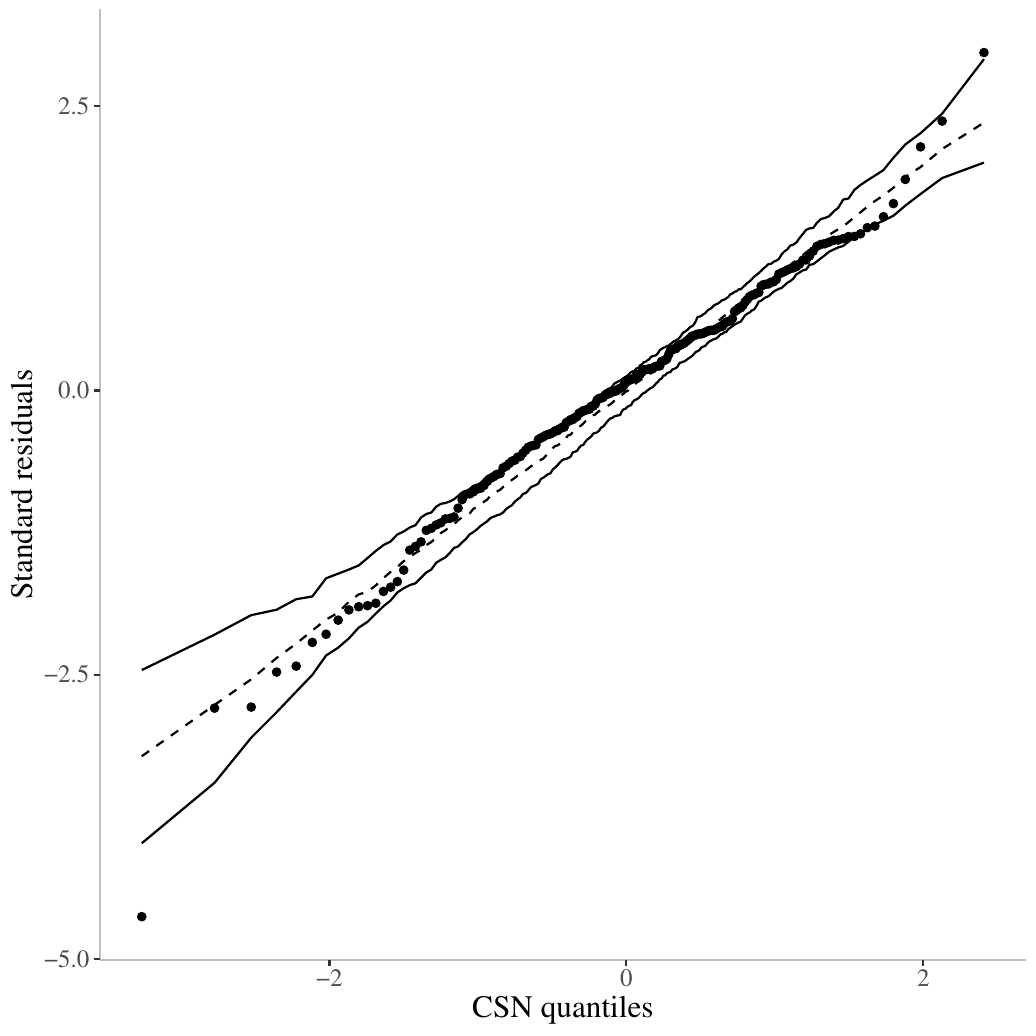}
			\caption{}
			\label{fig:res_csn}
		\end{subfigure}
		\begin{subfigure}{\linewidth/3-1em}
			\includegraphics[width=\linewidth]{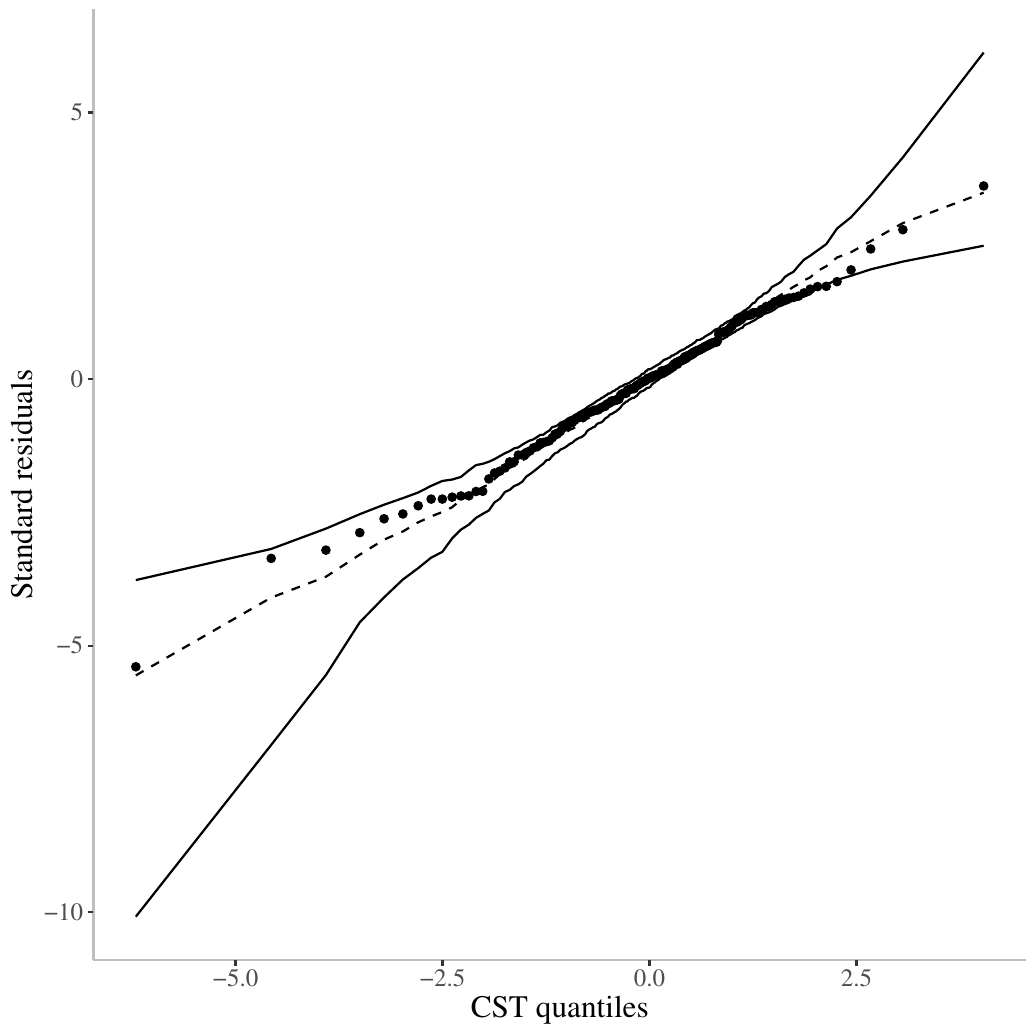}
			\caption{}
			\label{fig:res_cst}
		\end{subfigure}
		\begin{subfigure}{\linewidth/3-1em}
			\includegraphics[width=\linewidth]{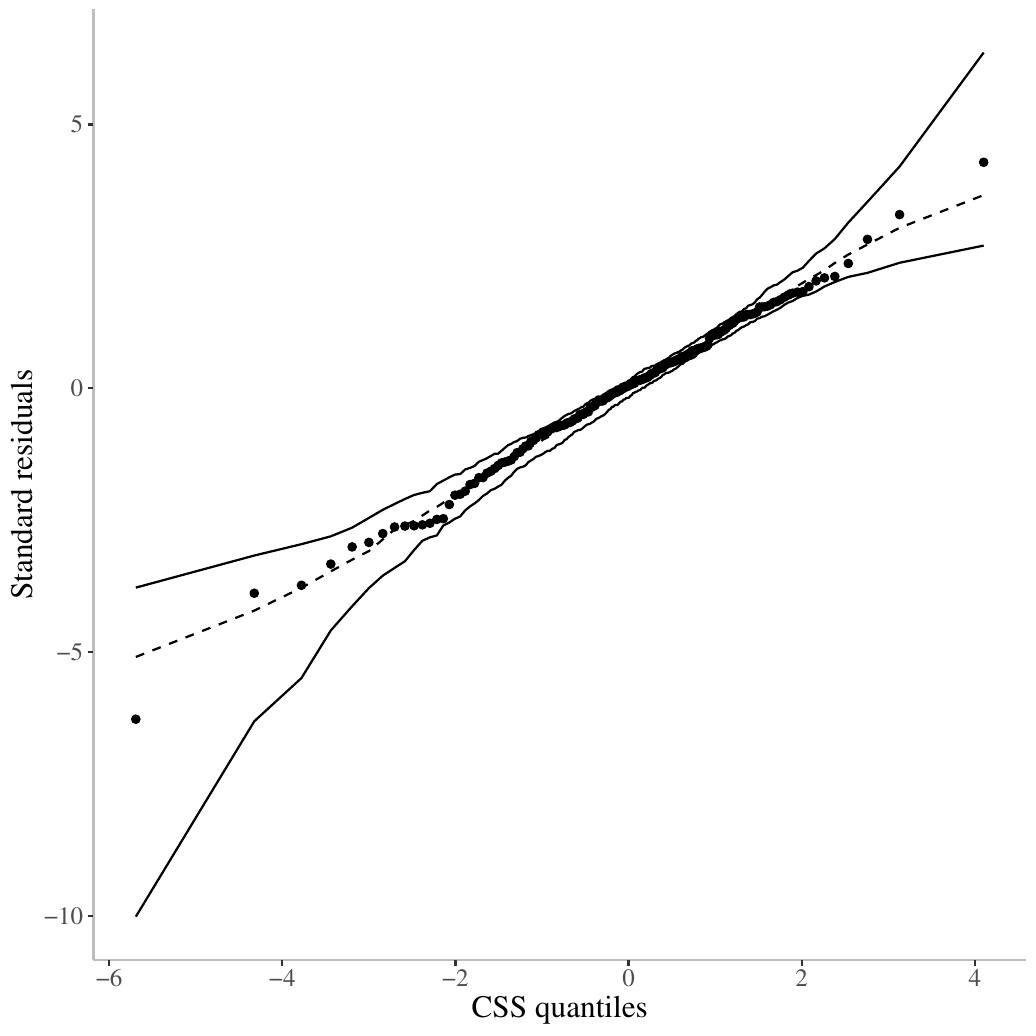}
			\caption{}
			\label{fig:res_css}
		\end{subfigure}
		\begin{subfigure}{\linewidth/3-1em}
			\includegraphics[width=\linewidth]{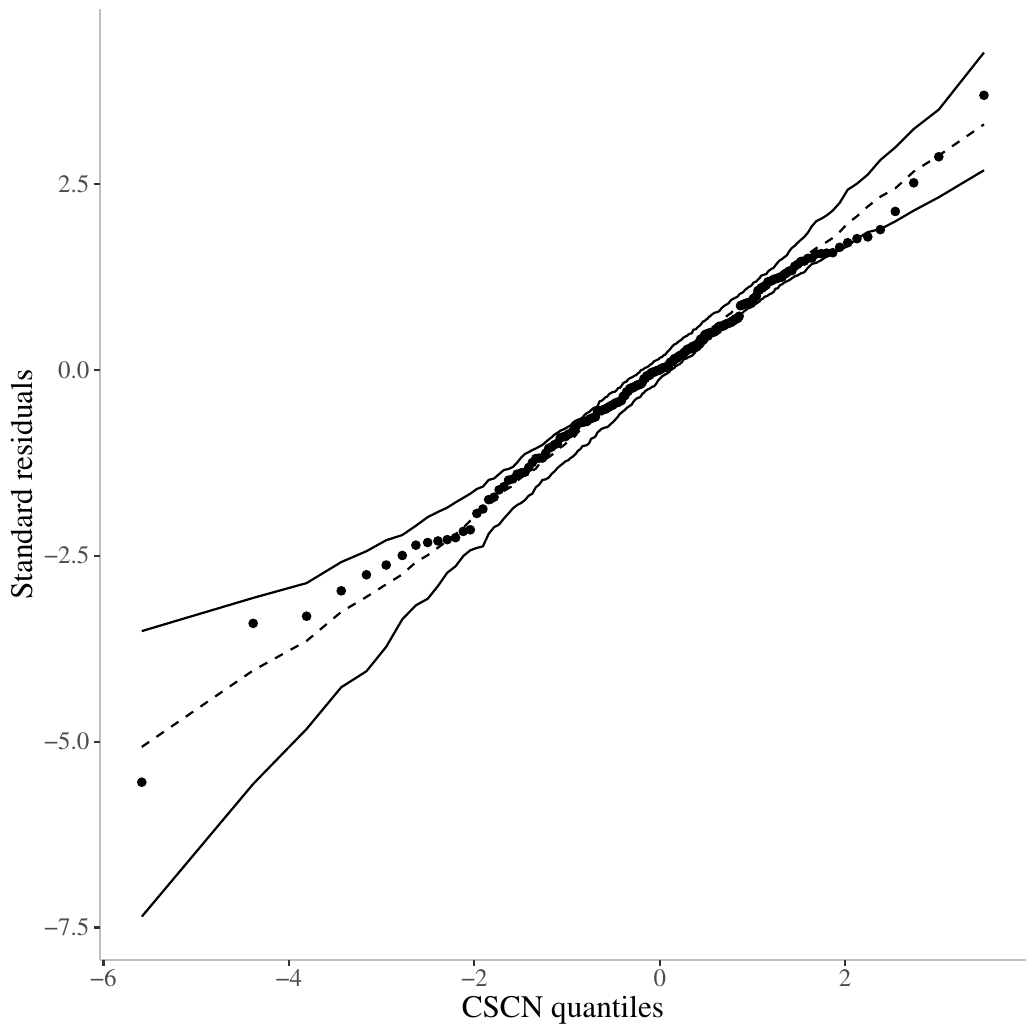}
			\caption{}
			\label{fig:res_cscn}
		\end{subfigure}
		\begin{subfigure}{\linewidth/3-1em}
			\includegraphics[width=\linewidth]{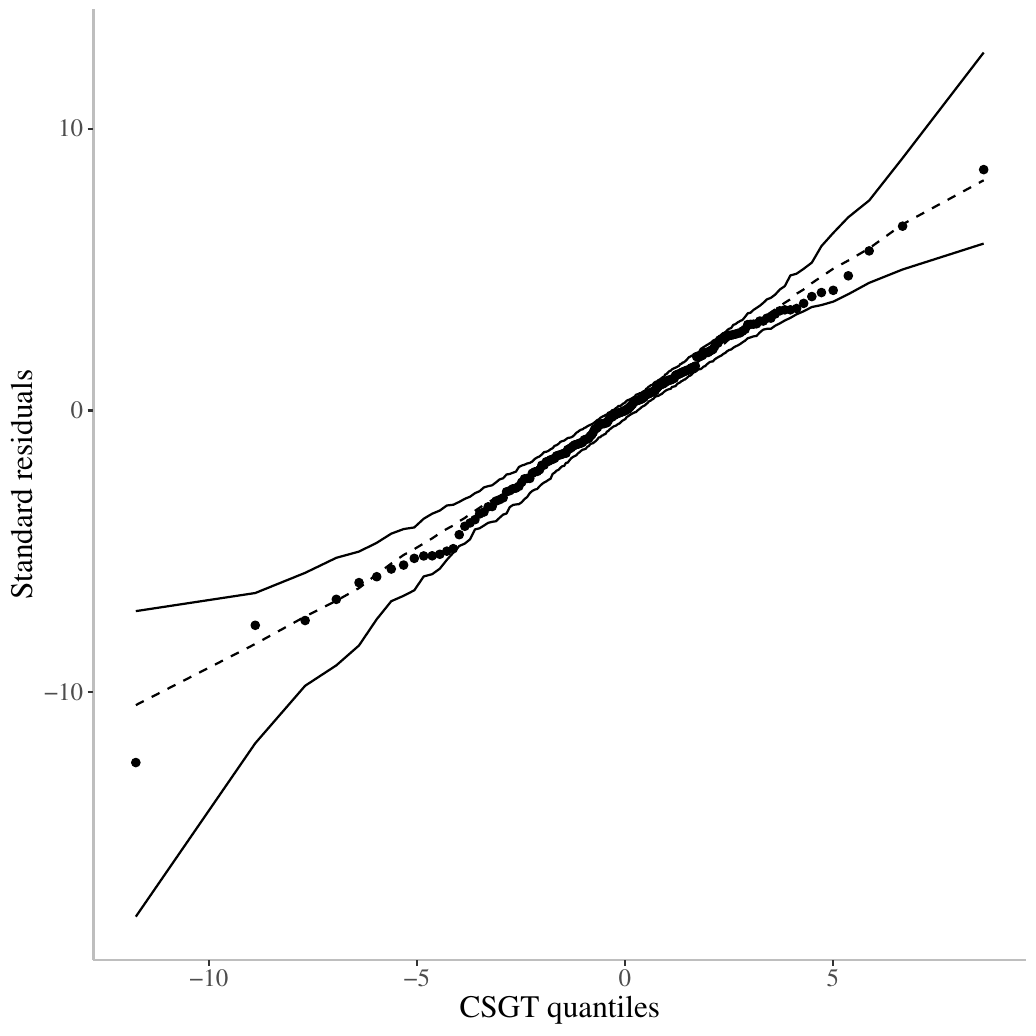}
			\caption{}
			\label{fig:res_csgt}
		\end{subfigure}
	\end{center}
	\caption{\textbf{AIS dataset}: QQ plots with envelopes for the ais dataset using the models: CSN (\subref{fig:res_csn}), CST (\subref{fig:res_cst}), CSS (\subref{fig:res_css}), CSCN (\subref{fig:res_cscn}), and CSGT (\subref{fig:res_csgt}).}
	\label{fig:ais_res}
\end{figure}

\begin{figure}[h!]
\begin{center}
		\begin{subfigure}{\linewidth/3-1em}
			\includegraphics[width=\linewidth]{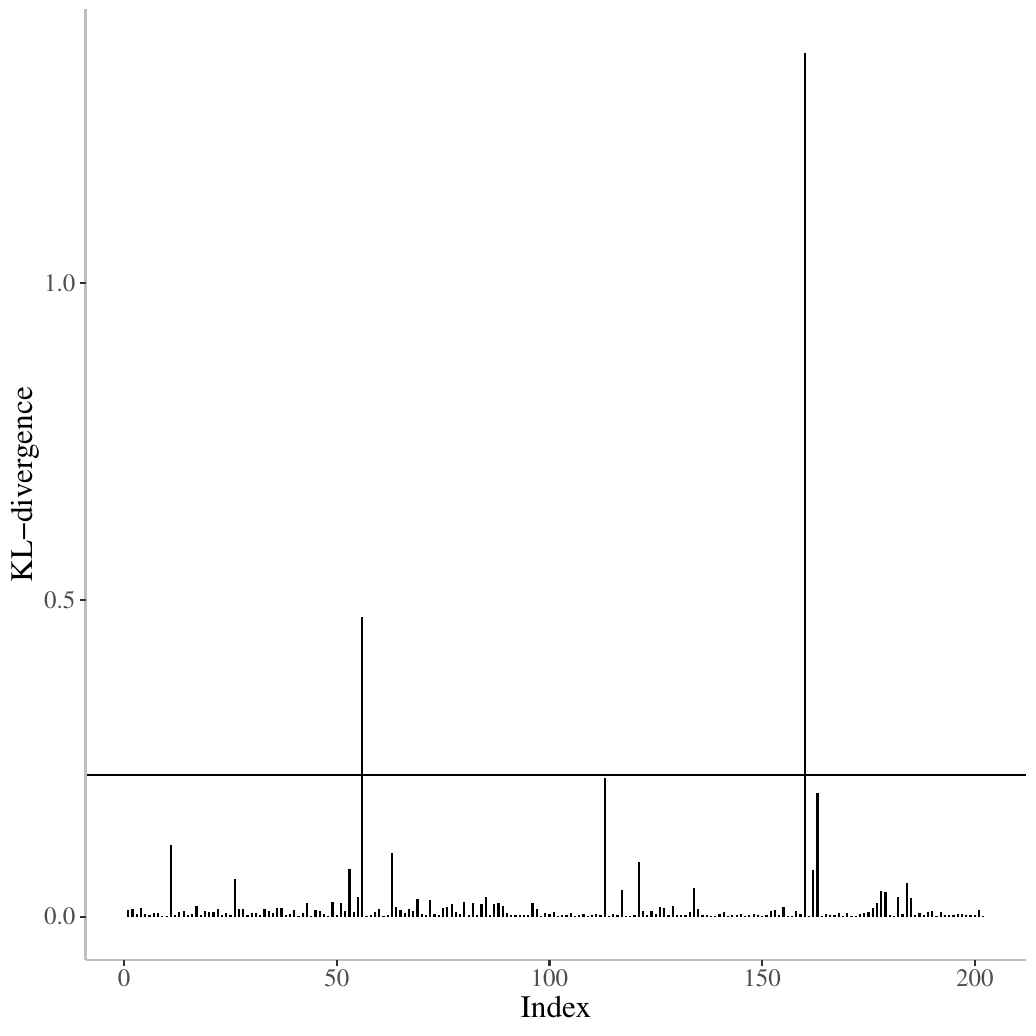}
			\caption{}
			\label{fig:kl_csn}
		\end{subfigure}
		\begin{subfigure}{\linewidth/3-1em}
			\includegraphics[width=\linewidth]{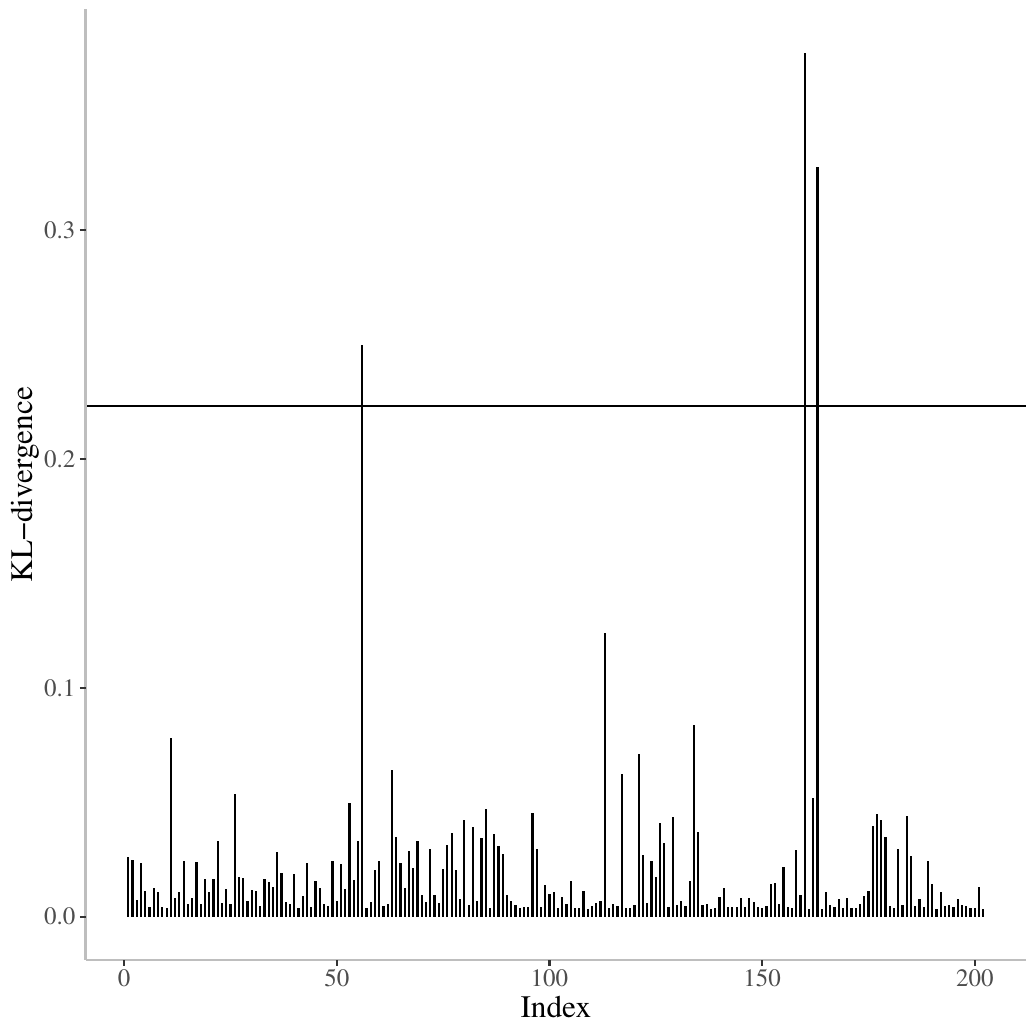}
			\caption{}
			\label{fig:kl_cst}
		\end{subfigure}
		\begin{subfigure}{\linewidth/3-1em}
			\includegraphics[width=\linewidth]{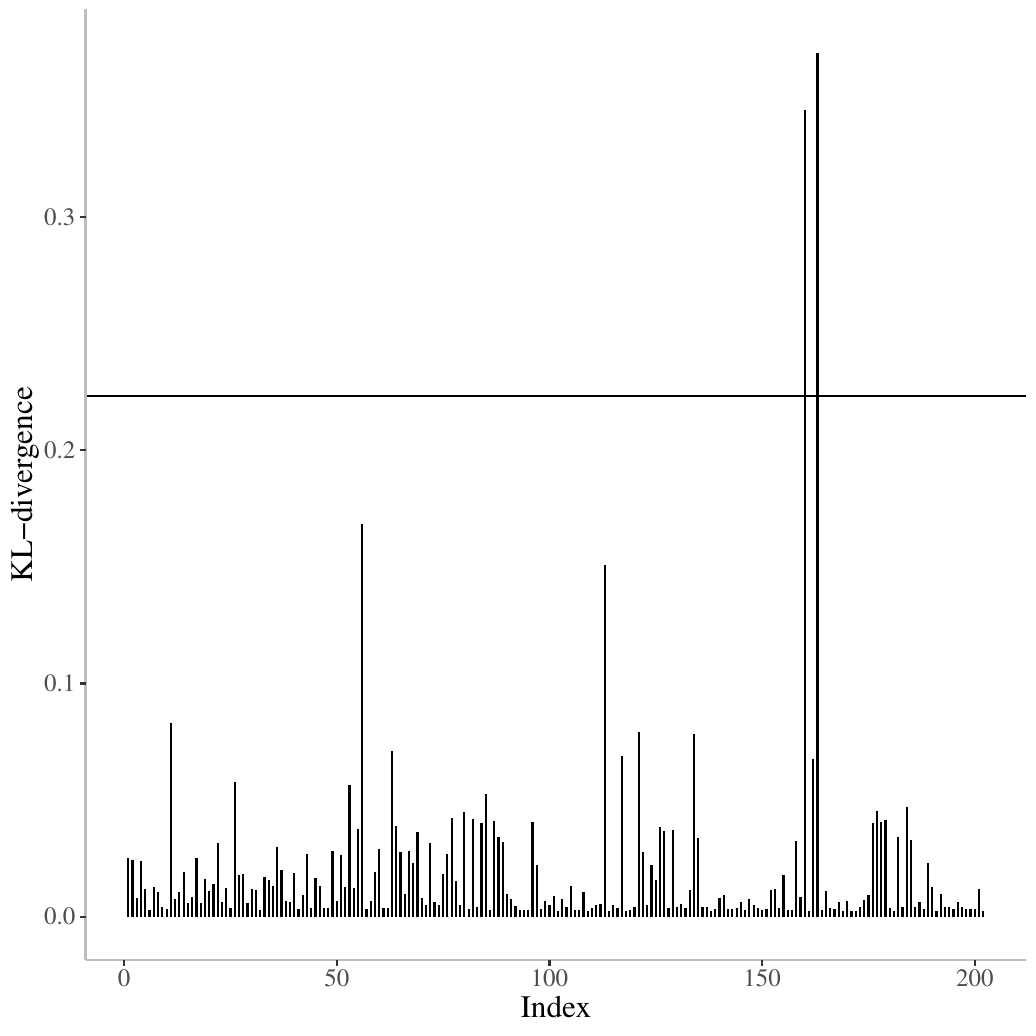}
			\caption{}
			\label{fig:kl_css}
		\end{subfigure}
	\par\bigskip
		\begin{subfigure}{\linewidth/3-1em}
			\includegraphics[width=\linewidth]{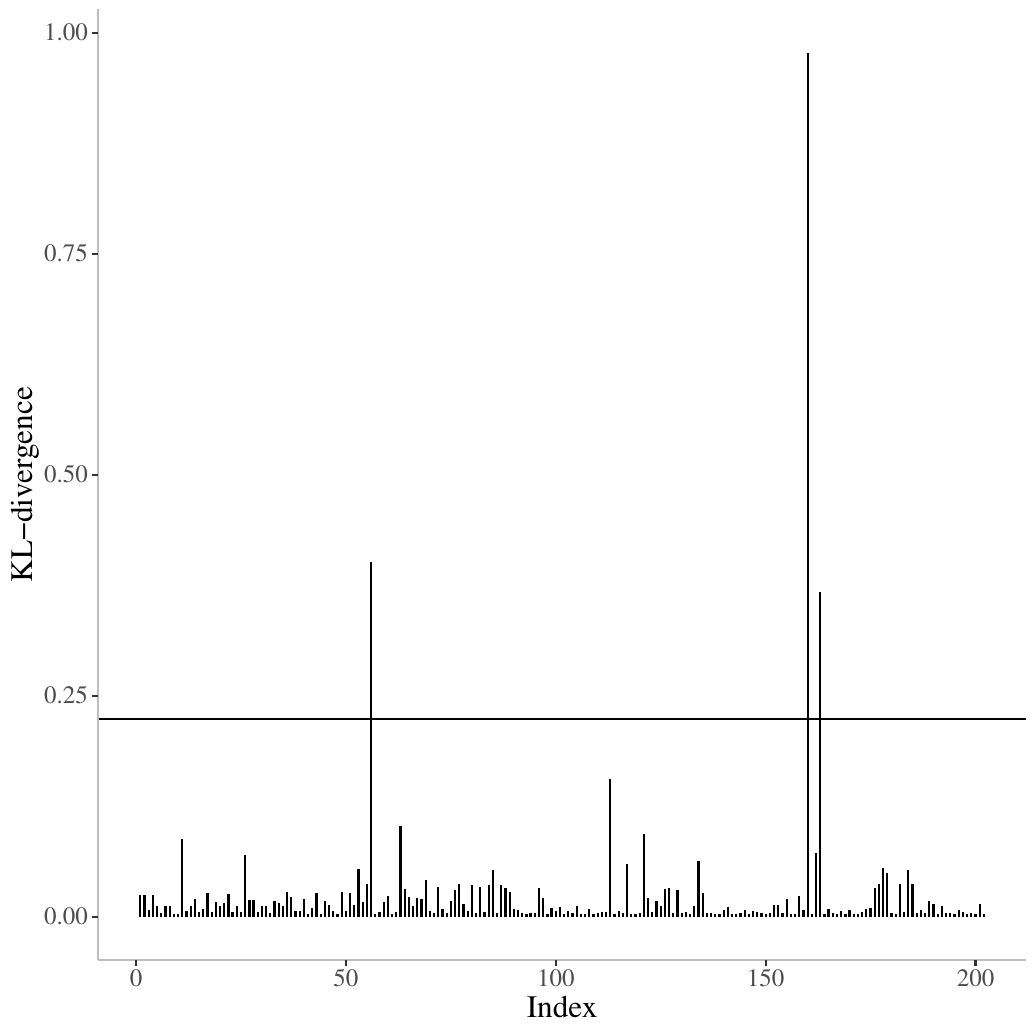}
			\caption{}
			\label{fig:kl_cscn}
		\end{subfigure}
		\begin{subfigure}{\linewidth/3-1em}
			\includegraphics[width=\linewidth]{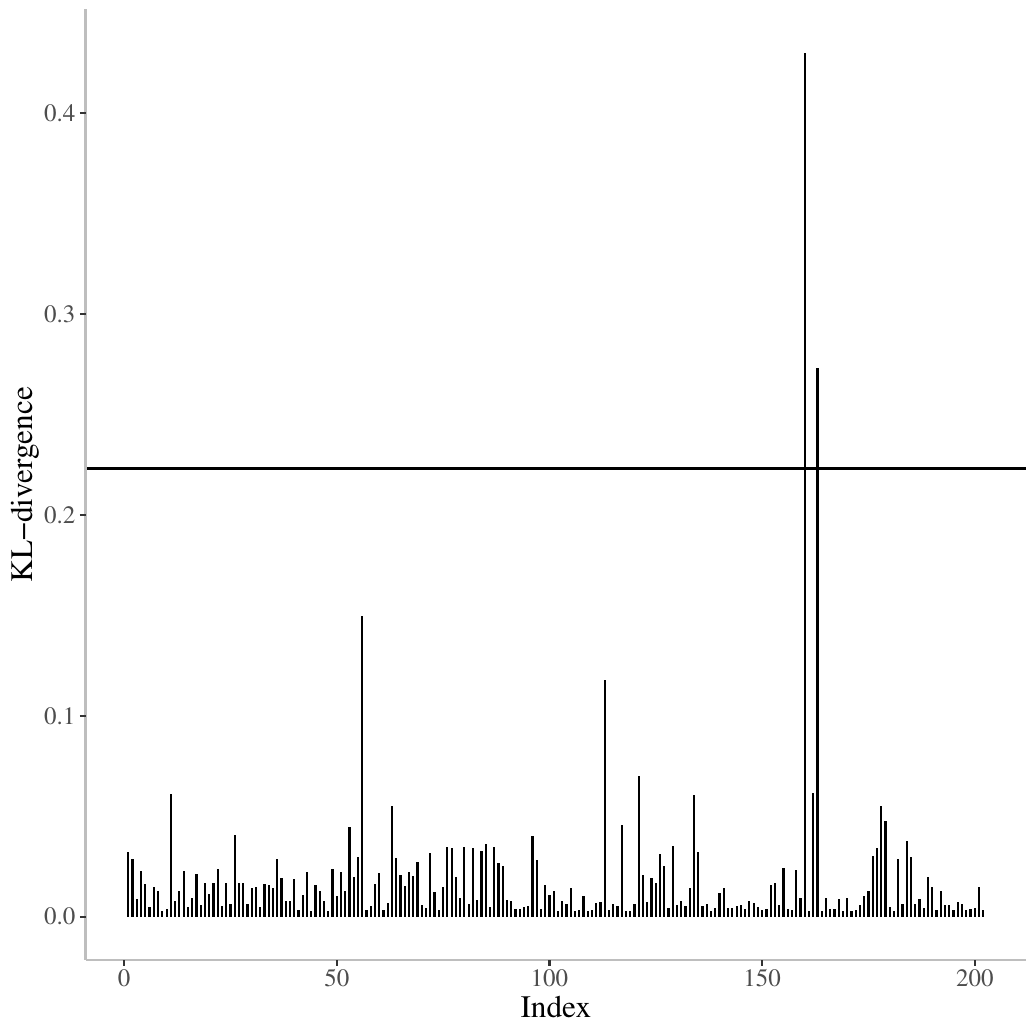}
			\caption{}
			\label{fig:kl_csgt}
		\end{subfigure}
		\end{center}
	\caption{\textbf{AIS dataset}: Index plots of $K(P,P_{(i)})$ for the ais dataset: CSN (\subref{fig:kl_csn}), CST (\subref{fig:kl_cst}), CSS (\subref{fig:kl_css}), CSCN (\subref{fig:kl_cscn}) and CSGT (\subref{fig:kl_csgt}) models.}
	\label{fig:ais_kl}
\end{figure}

Table \ref{tab1: AISdata_selection} presents the statistics for model comparison. The CSS model was selected by EAIC, EBIC and LPML. The CST model was select by DIC criterion. For the CSCN model, it is not clear if the CSN model is preferable to the CSCN model, since the width of the credibility intervals are large. For CST and CSS models, the credibility intervals do not include values of $\nu > 30$. Analyzing the posterior distributions in Figure \ref{fig:ais_nu}, we can notice that $\nu_1$ for the CSCN model is concentrated toward .2 and for $\nu_2$, the histogram shows that it is concentrated toward .25, which indicates that CSCN model is preferable to CSN. From the posterior distribution of $\nu_1$ for the CSGT model, we have the indicative that this model is preferred to the CSN model. From Figure \ref{fig:ais_res}, QQ plot with envelopes for all fitted models are shown. It is possible to see that for CST, CSCN and CSN models, there are some points lying outside the confidence bands, which do not happen with CSS and CSGT models. The analysis of influential observations, presented in Figure \ref{fig:ais_kl}, indicated that there are two influential observations for the CSN, CSS and CSGT models and three for the CST and CSCN models.

Finally, under the residual analysis, KL-divergence, and the EAIC, EBIC and LPML we have that the CSS model outperform the others models. From Table \ref{tab1: AISdata_summ_stat}, it is possible to see that through the credibility intervals all regression parameters are different from zero. Also, the model indicated that the height and weight have a positive influence in the lean body mass, and females presented lean body mass higher than males.

\begin{table}[h!]
	\centering
	\caption{\textbf{AIS dataset}: EAIC, EBIC, DIC and LPML for model comparison.}
	\begin{tabular}{r|rrrrr}
		\hline
		\multicolumn{1}{l}{} & \multicolumn{5}{c}{Model} \\
		\hline
		criterion &  CST & SSl & CSCN & CSGT & CSN \\ 
		\hline
		EAIC & 972.23 & \textbf{969.44}   & 973.93  & 970.41      & 976.28 \\ 
		EBIC & 995.39 & \textbf{992.60}   & 1000.40 & 993.57     & 996.13 \\ 
		DIC  & \textbf{967.57}          & 969.19   & 968.61  & 969.94      & 976.30  \\ 
		LPML & -485.60 & \textbf{-485.22} & -486.56 & -485.47     & -489.61  \\ 
		\hline		
	\end{tabular}
	\label{tab1: AISdata_selection}
\end{table}

\begin{table}[ht!]
	\centering
	\caption{\textbf{AIS dataset}: Posterior parameter estimates for the CSN, CST, CSS, CSCN and CSGT models.}
	\begin{tabular}{ccccc}
		\hline
		Model & parameter & Est & SD & CI (95\%) \\ 
		\hline
		\multirow{5}{*}{CST} 
		& $\beta_0$ & 60.78 & 0.35 & (60.13, 61.49) \\ 
		& $\beta_1$ & 8.00  & 0.54 & (7.02, 9.05) \\ 
		& $\beta_2$ & 0.10  & 0.03 & (0.03, 0.16) \\ 
		& $\beta_3$ & 0.67  & 0.02 & (0.62, 0.71) \\ 
		& $\gamma$  & -0.83 & 0.24 & (-0.99, -0.21) \\ 
		& $\nu$     & 4.43  & 3.30 & (2.57, 12.24) \\ 
		& $\sigma^2$ & 5.42 & 0.94 & (3.59, 7.21) \\ 
		\hline
		\multirow{5}{*}{CSS} 
		& $\beta_0$ & 60.81 & 0.35 & (60.20, 61.52) \\
		& $\beta_1$ & 8.01  & 0.54 & (7.01, 9.09) \\ 
		& $\beta_2$ & 0.10  & 0.03 & (0.05, 0.17) \\ 
		& $\beta_3$	& 0.66  & 0.02 & (0.61, 0.71) \\ 	
		& $\gamma$  & -0.63 & 0.20 & (-0.99, -0.32) \\ 
		& $\nu$     & 2.03  & 0.59 & (1.20, 3.32) \\ 
		& $\sigma^2$ & 4.01 & 0.80 & (2.51, 5.59) \\ 
		\hline
		\multirow{6}{*}{CSCN} 
		& $\beta_0$ & 60.74 & 0.34 & (60.08, 61.43) \\ 
		& $\beta_1$ & 8.17  & 0.57 & (6.99, 9.21) \\ 
		& $\beta_2$ & 0.09  & 0.03 & (0.02, 0.15) \\ 
		& $\beta_3$	& 0.67  & 0.02 & (0.62, 0.72) \\ 
		& $\gamma$ & -0.72  & 0.20 & (-0.92, -0.10) \\
		& $\nu_1$ & 0.25    & 0.21 & (0.02, 0.74) \\ 
		& $\nu_2$ & 0.26    & 0.14 & (0.07, 0.59) \\ 
		& $\sigma^2$ & 5.11 & 1.37 & (1.80, 7.41) \\ 
		\hline
		\multirow{5}{*}{CSGT} 
		& $\beta_0$ & 60.76   & 0.31  & (60.18, 61.37) \\ 
		& $\beta_1$ & 8.20    & 0.49  & (7.23, 9.20) \\ 
		& $\beta_2$  & 0.10   & 0.03  & (0.04, 0.16) \\ 
		& $\beta_3$	 & 0.66   & 0.02  & (0.62, .71) \\ 
		& $\gamma$ & -0.58    & 0.24  & (-0.88, 0) \\ 
		& $\nu_1$ & 5.14      & 1.89  & (2.90, 9.55) \\ 
		& $\nu_2$ & 24.10     & 12.65 & (9.64, 54.26) \\ 
		\hline
		\multirow{6}{*}{CSN} 
		& $\beta_0$ & 60.58 & .32 & (59.96, 61.20) \\ 
		& $\beta_1$ & 8.38 & .53 & (7.33, 9.45) \\ 
		& $\beta_2$ & 0.07 & .03 & (0.01, .13) \\ 
		& $\beta_3$	& 0.68 & .02 & (0.63, .73) \\ 
		& $\gamma$ & -0.46 & .14 & (-0.74, -0.20) \\ 
		& $\sigma^2$ & 7.48 & .82 & (6.09, 9.25) \\ 
		\hline		
	\end{tabular}
	\label{tab1: AISdata_summ_stat}
\end{table}

\section{Conclusions}
In this paper we developed a scale mixture of skew-normal distribution under the centered parameterization class of probability distributions as an alternative to the parameterization used in \cite{Clecio_VH_2011}. 
It was decided to use a new parameterization for this class for several reasons, among them, the simplicity of parameter interpretation compared to the parameterization used in \cite{Clecio_VH_2011}. Another motivation was the issues related to the estimation process of parameter $\lambda$ in the direct parameterization. We have showed, through profiled log-likelihood, that the SMSN class under direct parameterization can heritage the problem caused by the non quadratic likelihood shape of the direct parameterization. 

A class of linear regression models based on the SMSN family under the centered parameterization was introduced, and we developed the Bayesian estimation approach. Also, we described model comparison criteria, and we developed analysis of influential observations and residual analysis. Simulation studies were performed in order to evaluate the parameter recovery under different scenarios. We concluded that for values of $\nub$ that generate distributions with heavy tails the estimates are accurate and improved as the sample size increases. An application of the proposed model in a real dataset was performed in order to show that CSMSN models provide better fits than the CSN linear regression.

\backmatter

\bmhead{Acknowledgements}

We gratefully acknowledge São Paulo Research Foundation (FAPESP), for the financial support of this project, through a Master’s scholarship, grant number \#2015/25867-2. Also, the first author acknowledge master's scholarship from São Paulo Research Foundation (FAPESP), grant number \#2018/26780-6.

\section*{Declarations}

Not applicable


\bibliography{sn-bibliography.bib}

\end{document}


\title[Bayesian inference for SMSN linear models under the CP]{Bayesian inference for scale mixtures of skew-normal linear models under the centered parameterization - supplementary material}


\author*[1]{\fnm{João} \sur{Victor B. de Freitas}}\email{jvbfreitas@ime.unicamp.br}\equalcont{These authors contributed equally to this work.}

\author[1]{\fnm{Caio} \sur{L. N. Azevedo}}\email{cnaber@ime.unicamp.br}
\equalcont{These authors contributed equally to this work.}

\affil[1]{\orgdiv{Departamento de Estatística, Instituto de Matemática, Estatística e Computação Científica}, \orgname{Universidade Estadual de Campinas}, \orgaddress{\city{Campinas}, \state{São Paulo}, \country{Brazil}}}




\maketitle

\section{Figures for the densities}

\subsection{Centered skew-t}

\begin{figure}[h!]
	\begin{subfigure}{\linewidth}
		\begin{subfigure}{0.5\textwidth}
			\includegraphics[width=\linewidth]{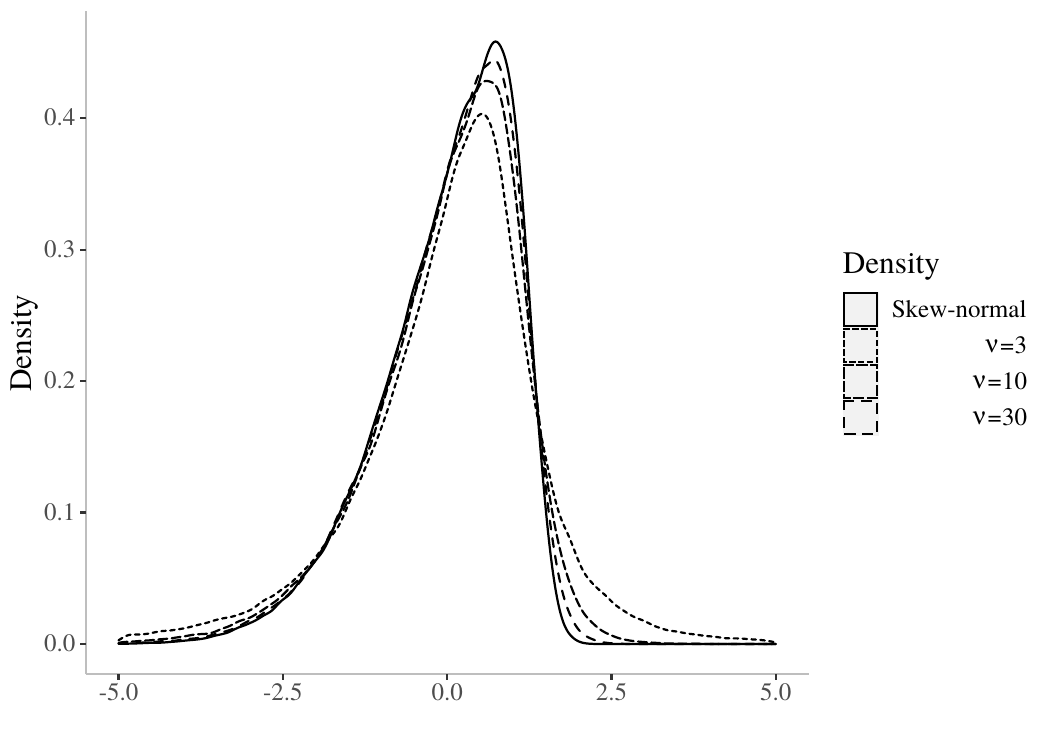}
			\caption{}
			\label{fig:stn_1}
		\end{subfigure}\hfill
		\begin{subfigure}{0.5\textwidth}
			\includegraphics[width=\linewidth]{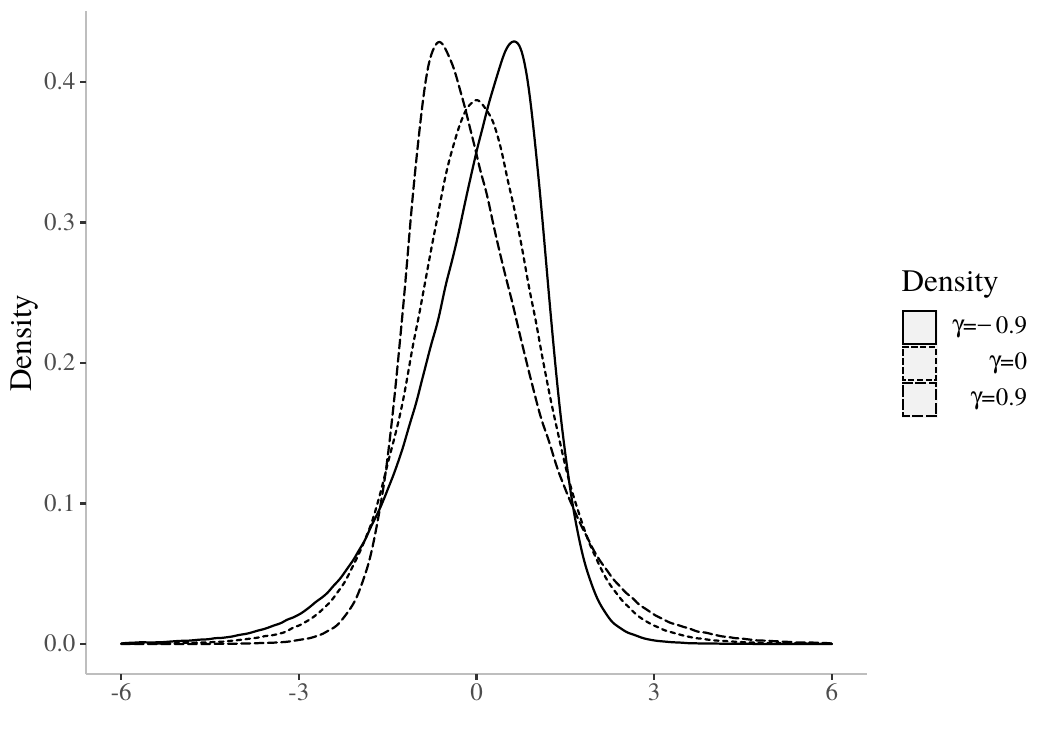}
			\caption{}
			\label{fig:stn_2}
		\end{subfigure}\hfill
	\end{subfigure}
	\caption{Probability density function of skew-t distribution under the centered parameterization  (\subref{fig:stn_1}) varying the values of $\nu$, with $\gamma=-0.9$, $\sigma^2=1$, $\mu=0$ and (\subref{fig:stn_2}) varying the parameter $\gamma$, with $\nu=8$, $\sigma^2=1$, $\mu=0$.}
	\label{fig:dens_stn}
\end{figure}

Figure \ref{fig:stn_1} presents the density of the CST distribution for three values of $\nu$ ($\nu =3$ is the dashed line, $\nu=10$, dotted line and $\nu = 30$, dot-dashed line) and the CSN (solid line) for $\sigma^2 =1$, $\gamma=-0.9$ and $\mu=0$. As $\nu$ increases the CST approaches to the CSN distribution. For $\nu=30$ it is possible to see that the CST and CSN curves almost overlap. In Figure \ref{fig:stn_2}, the density of CST distribution is drawn for $\gamma=0$ (solid line), $\gamma=0.9$ (dashed line) and $\gamma=-0.9$ (dotted line). It is possible to see right/left asymmetry as $\gamma$ assumes positive/negative values, respectively.

\subsection{Centered skew slash}

\begin{figure}[h!]
	\begin{subfigure}{\linewidth}
		\begin{subfigure}{0.5\textwidth}
			\includegraphics[width=\linewidth]{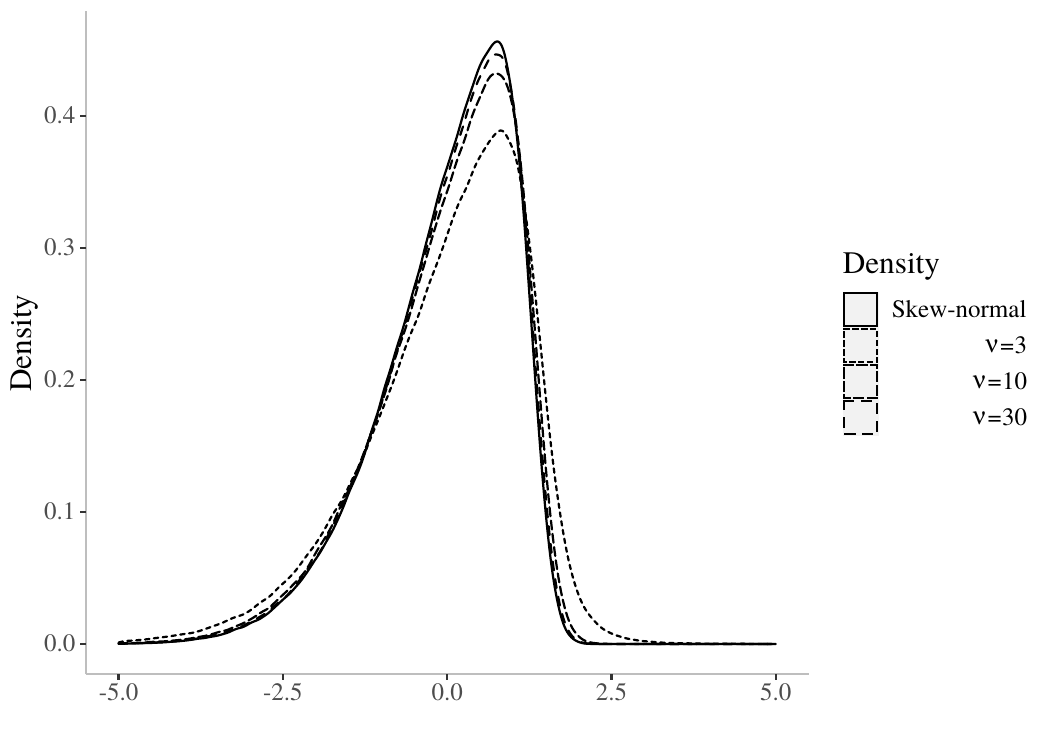}
			\caption{}
			\label{fig:ssl_1}
		\end{subfigure}\hfill
		\begin{subfigure}{0.5\textwidth}
			\includegraphics[width=\linewidth]{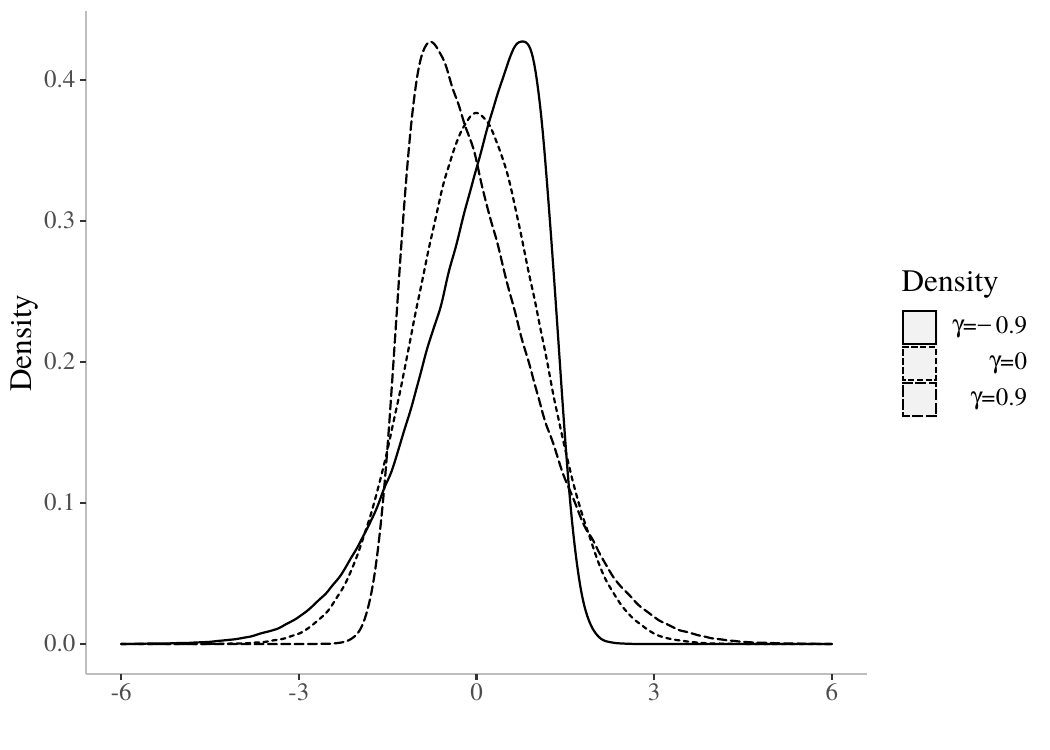}
			\caption{}
			\label{fig:ssl_2}
		\end{subfigure}\hfill
	\end{subfigure}
	\caption{Probability density function of skew-slash distribution under the centered parameterization (\subref{fig:ssl_1}) varying the values of $\nu$, with $\gamma=-0.9$, $\sigma^2=1$, $\mu=0$ and (\subref{fig:ssl_2}) varying the parameter $\gamma$, with $\nu=8$, $\sigma^2=1$, $\mu=0$.}
	\label{fig:dens_ssl}
\end{figure}

Again, from Figures \ref{fig:ssl_1} and \ref{fig:ssl_2} it is possible to see a similar behavior as observed in the CST distribution.

\subsection{Skew contaminated normal}
\begin{figure}
	\begin{subfigure}{\linewidth}
		\begin{subfigure}{0.5\textwidth}
			\includegraphics[width=\linewidth]{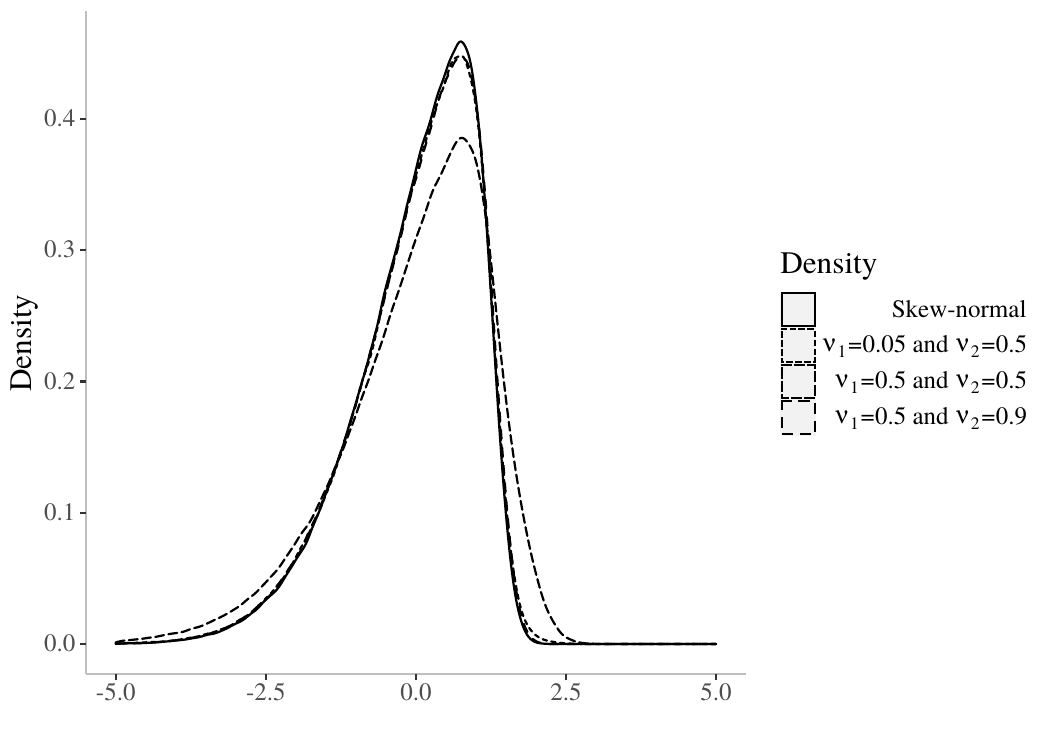}
			\caption{}
			\label{fig:scn_1}
		\end{subfigure}\hfill
		\begin{subfigure}{0.5\textwidth}
			\includegraphics[width=\linewidth]{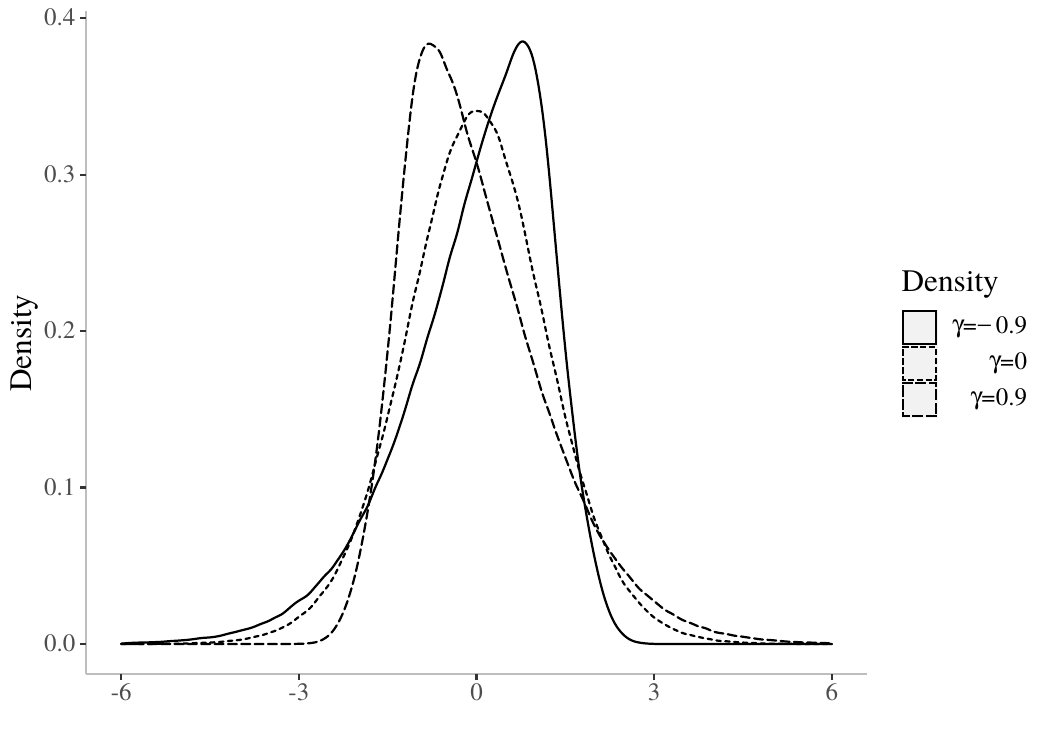}
			\caption{}
			\label{fig:scn_2}
		\end{subfigure}\hfill
	\end{subfigure}
	\caption{Probability density function of skew contaminated normal distribution under the centered parameterization (\subref{fig:scn_1}) varying the values of $\nu_1$ and $\nu_2$, with $\sigma^2 = 1$,  $\gamma = -0.9$ and $\mu=0$ and (\subref{fig:scn_2}) varying the parameter $\gamma$, with $\sigma^2=1$, $\mu=0$, $\nu_1= 0.5$ and $\nu_2=0.5$.}
	\label{fig:dens_scn}
\end{figure}

From Figure \ref{fig:scn_1} it is possible to see that as $\nu_2$ approaches 1 and/or $\nu_1$ approaches 0, the CSCN density tends to the CSN distribution. 

\subsection{Skew generalized t}

\begin{figure}[!h]
	\centering
	\includegraphics[width=0.5\textwidth]{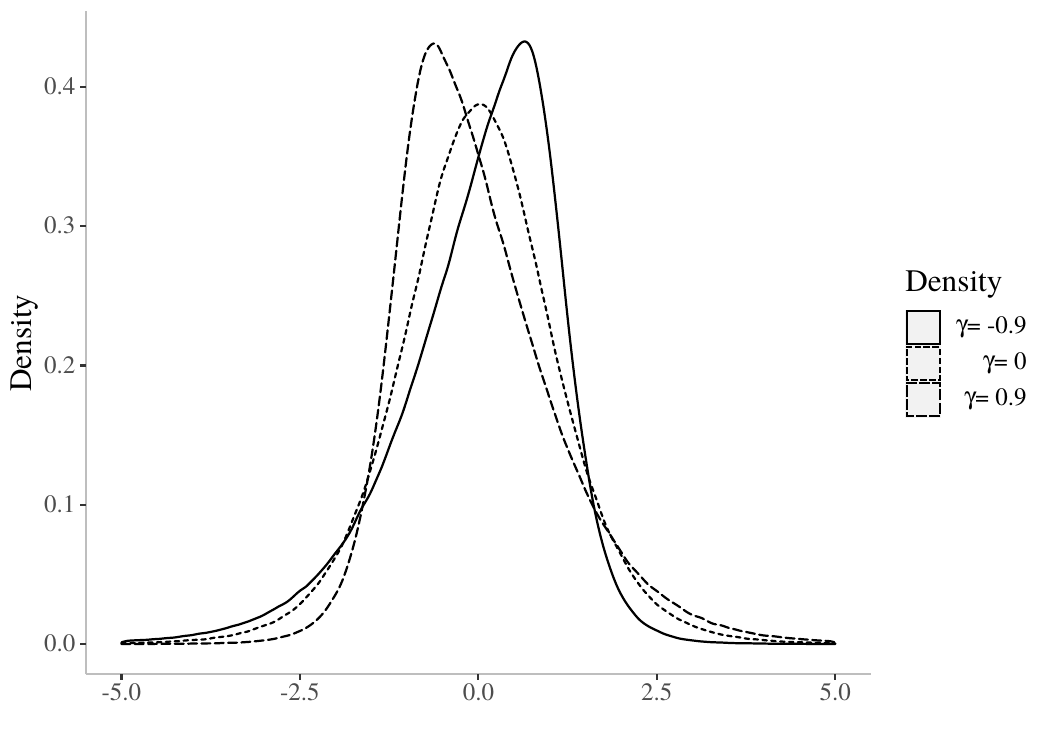}
	\caption{Probability density function of skew generalized t distribution under the centered parameterization for some values of $\gamma$, $\sigma^2=5$, $\mu=0$, $\nu_1= 8$ and $\nu_2=8$.}
	\label{fig:sgtn_2}
\end{figure}

In Figure \ref{fig:sgtn_2}, the density of CSGT distribution is drawn to $\gamma=0$ (solid line), $\gamma=0.9$ (dashed line) and $\gamma=-0.9$ (dotted line). 

	
\section{Profiled log-likelihood for parameter $\nu$} \label{ss1: lvpr nu}
The profiled log-likelihood for parameter $\bm{\nu}$ was calculated in the same way as described in section 3 of the paper. The only difference from the former procedure is that now we replaced the maximum-likelihood estimator of parameters $\mu$,$\sigma^2$ and $\gamma$ in the log-likelihood function, calculated by fixing values of $\bm{\nu}$.

\begin{figure}
	\begin{subfigure}{\linewidth}
		\begin{subfigure}{0.5\textwidth}
			\includegraphics[width=\linewidth]{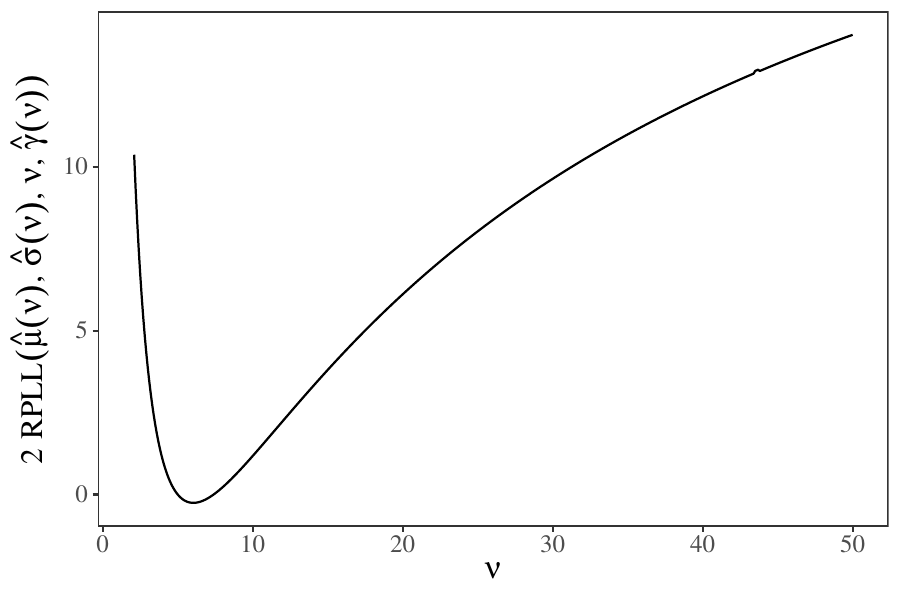}
			\caption{}
			\label{fig:lvpr_stn_nu}
		\end{subfigure}\hfill
		\begin{subfigure}{0.5\textwidth}
			\includegraphics[width=\linewidth]{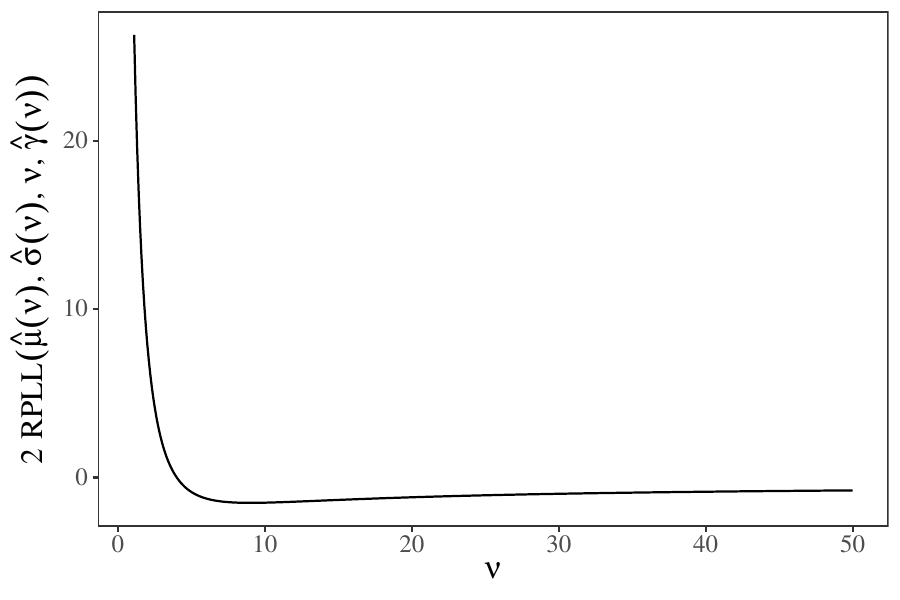}
			\caption{}
			\label{fig:lvpr_ssl_nu}
		\end{subfigure}\hfill
	\end{subfigure}
	\caption{Profiled relative log-likelihood for $\nu$ for the: skew-t (\subref{fig:lvpr_stn_nu}) and skew slash  (\subref{fig:lvpr_ssl_nu})distributions.} 
	\label{fig:lvpr1}
\end{figure}

\begin{figure}
	\begin{subfigure}{\linewidth}
		\begin{subfigure}{0.5\textwidth}
			\includegraphics[width=\linewidth]{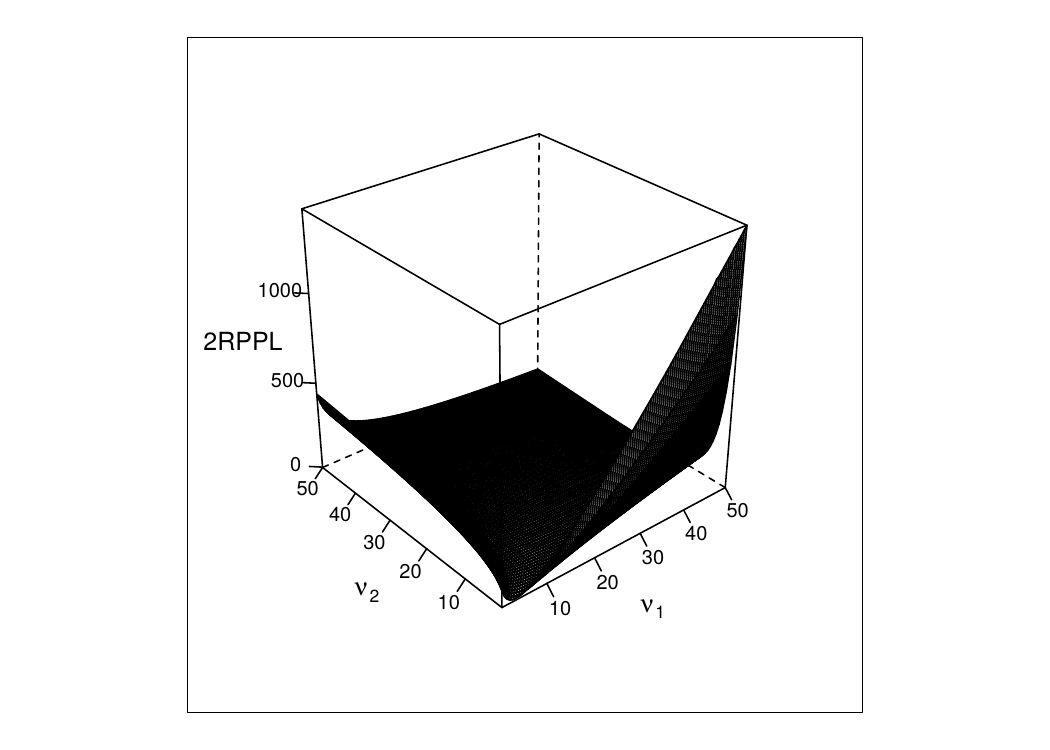}
			\caption{}
			\label{fig:lvpr_stngen_nu}
		\end{subfigure}\hfill
		\begin{subfigure}{0.5\textwidth}
			\includegraphics[width=\linewidth]{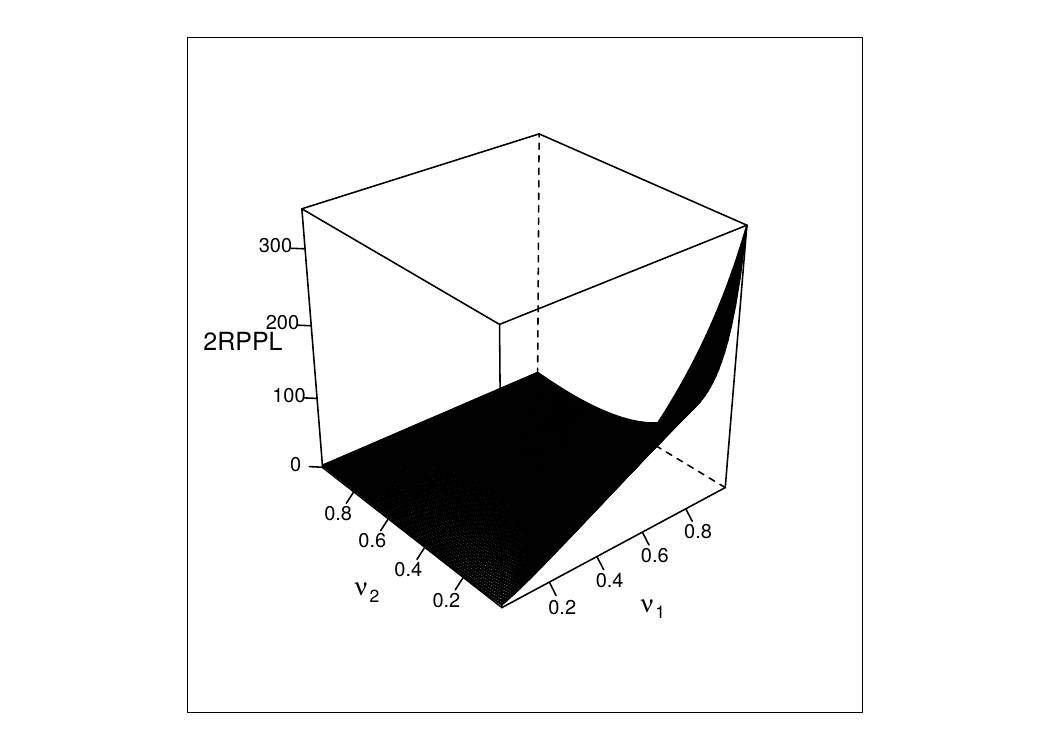}
			\caption{}
			\label{fig:lvpr_scn_nu}
		\end{subfigure}\hfill
	\end{subfigure}
	\caption{Profiled relative log-likelihood for $\nu_1$ and $\nu_2$ for the: skew generalized t (\subref{fig:lvpr_stngen_nu}) and skew contaminated normal (\subref{fig:lvpr_scn_nu}) distributions.}
	\label{fig:lvpr2}
\end{figure}

From Figures \ref{fig:lvpr1} and \ref{fig:lvpr2}, it is possible to observe that as $\nu$ increases, the profiled relative log-likelihood tends to be a straight line, i.e., for larger values of $\nu$, the estimates tends to range within such interval with almost the same probability.
Similar behavior can be noticed for the CSGT and CSCN distributions, as we can see from Figures \ref{fig:lvpr_stngen_nu} and \ref{fig:lvpr_scn_nu}. As long as at least $\nu_1$ or $\nu_2$ increases, flatter is the surface.
Analysis of these figures indicates that the profiled log-likelihoods are ill-behaved, which can result in large posterior standard deviations and large credibility intervals for $\bm{\nu}$.

\section{Model fit assessment and model comparison} \label{ss1: diag and sel}

\subsection{Model comparison} \label{sss1: model sel}
In Bayesian inference, there are several methodologies to compare a set of competing models for a given data set. According to \cite{Spiegelhalter_2002} under the Bayesian inference, the selection criteria are obtained through the parameters posterior distribution. Moreover, when MCMC algorithms are used to approximate the posterior distribution, such criteria can be easily calculated. In this work we will consider the expected Akaike information criterion (EAIC), expected Bayesian information criterion (EBIC), deviance information criterion (DIC) and log pseudo-marginal likelihood (LPML) criterion.

The LPML criterion is defined based on the conditional predictive ordinate (CPO), which is based on the cross validation criterion. Let $\bm{\thetab} = (\bm{\beta},\sigma^2,\gamma,\bm{\nu})$ and $D(\bm{\thetab}) = -2 l(\bm{\thetab}\rvert \bm{y})$, where $l(\bm{\thetab}\rvert \bm{y})$ is the log-likelihood function, according to the fitted model.
Also, let $\bm{\thetab}^{(m)}$, $m= 1, \dots, M$ a valid MCMC sample (after discharging the burn-in and the spacing between the values), $\bm{\bar{\thetab}}$ the vector with the posterior expectation, and $\overline{D(\bm{\thetab})} = (1/M)\sum_{m=1}^{M} D(\bm{\thetab}^{(m)})$.
According to \cite{Aldo}, using the MCMC sample, the approximation for the $CPO_i$ for the i-th observation, is defined as
\begin{equation}
\widehat{CPO_i} = \left\{ \frac{1}{M} \sum_{m=1}^{M} \frac{1}{L(\bm{\thetab}^{(m)}\rvert y_i)} \right\}^{-1}. \label{CPO_exp}
\end{equation}

A summary statistic of the CPO is the LPML criterion, which is defined as $LPML = \sum_{i=1}^{n}ln(\widehat{CPO_i})$. For calculating DIC, EAIC and EBIC criteria, we need to define $D(\bar{\bm{\thetab}}) = -2 l(\bar{\bm{\thetab}}\rvert \bm{y})$, therefore $EAIC=  D(\bar{\bm{\thetab}}) +2k \quad \text{and} \quad EBIC=  D(\bar{\bm{\thetab}}) +klog(n)$ and DIC is defined as $DIC=  D(\bar{\bm{\thetab}}) + 2 p_D$ where $p_D = \overline{D(\bm{\thetab})} - D(\bar{\bm{\thetab}})$.
Smaller values of DIC, EAIC and EBIC indicate the best model, occurring the opposite with the LPLM. More discussions about these criteria can be found in \cite{Ando}.

\subsection{Influential observations} \label{sss1: infl}
As noted in \cite{Cho}, influential observations in a data set can have a strong impact in statistical inference and the related conclusions. Computation of divergence measures between posterior distributions with and without a given subset of the data is a useful way of quantifying influence, and the most popular Bayesian case deletion influence diagnostics is based on the Kullback-Leibler divergence, which is a type of q-divergence measure~\citep{AldoeVH}. 

Let $K(P, P_{(-i)})$ denote the KL divergence between two densities P and $P_{(-i)}$ for $\bm{\thetab}$, which is defined as 
\begin{align}
K(P, P_{(-i)}) = \int \pi(\bm{\thetab}\rvert \bm{y}) log\left( \frac{\pi(\bm{\thetab}\rvert \bm{y})}{\pi(\bm{\thetab}\rvert \bm{y}_{(-i)})}\right) d\bm{\thetab},
\label{KL divergence}
\end{align}
where P denotes the posterior distribution of $\bm{\thetab}$ based on the full data $\bm{y}$, and $P_{(-i)}$ denotes the posterior distribution obtained from the data $\bm{y}$ without the i-th observation. 
The $K(P, P_{(-i)})$ measures the effect of deleting the i-th observation from the full data on the joint posterior distribution of $\bm{\thetab}$.
From expression (\ref{KL divergence}) it is possible to rewrite the KL divergence as a posterior expectation, that is
\begin{align}
K(P, P_{(-i)}) = E_{\bm{\thetab}\rvert \bm{y}}\left(log\left( \frac{\pi(\bm{\thetab}\rvert \bm{y})}{\pi(\bm{\thetab}\rvert \bm{y}_{(-i)})}\right)\right).
\label{Expc_KL divergence}
\end{align}

From (\ref{Expc_KL divergence}), the computation of this measure can be approximated by using the MCMC posterior samples. Also, \cite{Cho} developed a simplified expressions for computing the KL divergence, through the CPO statistic and the log-likelihood as $K(P, P_{(-i)}) = -log(\widehat{CPO_i}) + \frac{1}{M}\sum_{m=1}^{M} l(\bm{\thetab}^{(m)}\rvert y_i), i=1, \dots,n$, where $\widehat{CPO_i}$ is expressed as in (\ref{CPO_exp}).
As usual, we need to establish a cut-off point, in order to determine whether an observation is influent or not.
As pointed by \cite{Cho}, the calibration of KL divergence can be done by solving for $p_i$ the equation:
\begin{align}
K(P, P_{(-i)}) = K(Ber(1/2),Ber(p_i)) = - \frac{1}{2} log(4p_i(1-p_i)),
\label{KL_Ber}
\end{align}
where $Ber(p_i)$ is the Bernoulli distribution with success probability $p_i$. The equality $K(P, P_{(-i)}) = K(Ber(1/2),Ber(p_i))$ we have that describing outcomes using $\pi(\bm{\thetab}\rvert \bm{y})$ instead of $\pi(\bm{\thetab}\rvert \bm{y}_{(-i)})$ is compatible with describing an unobserved event as having probability $p_i$ when correct probability is .5~\citep{Cho}.
Solving (\ref{KL_Ber}), the calibration of the KL divergence is $p_i = .5[1 + \sqrt{1-exp\{-2K(P, P_{(-i)})\}}]$.

This equation implies that $0.5 \le p_i \le 1$. For $p_i$ much greater than .5 implies that the i-th observation is influential. In this work we are going to consider a influential observation if $p_i \ge .8$, as used in \cite{Nathalia} and \cite{AldoeVH}. So, for KL divergence measure greater than $K(Ber(1/2),Ber(0.8)) \approx .2231436$, the observation is considered influential.

\section{Full conditional distributions} \label{ss1: gibs}
In order to implement the MCMC algorithm, we have to simulate iteratively from the full conditionals described bellow. Denoting by $\bm{\thetab}_{-\thetab_i}$ the parameter vector $\bm{\thetab}$ without the component $\thetab_i$, the full conditional distributions are:
\paragraph*{For $\beta$:}
\begin{align*}
\pi(\bm{\beta}\rvert \bm{\thetab}_{-\bm{\beta}}, \bm{y}, \bm{u},\bm{h}) &\propto \exp\left\{ -\frac{1}{2} \left( \bm{\beta}^t \Sigma_{*}^{-1} \bm{\beta} - 2\mu_{*}^t\Sigma_{*}^{-1}\bm{\beta}\right)  \right\} I_{\mathbb{R}^p}(\bm{\beta}),
\end{align*}
which can be recognized as the kernel of p-variate normal distribution with variance $$\bm{\Sigma_{*}} = \left(\frac{\sum_{i=1}^{n} u_i x_ix_i^t}{\tau} + \bm{\Sigma_{\beta}}^{-1}\right)^{-1},$$ and mean $$\bm{\mu_{*}} = \left( \sum_{i=1}^{n} \frac{u_i}{\tau} \left(y_i - \frac{\Delta}{\sqrt{u_i}}(h_i - b)\right)\bm{x_i}^t + \bm{\mu_{\beta}}^t \bm{\Sigma_{\beta}}^{-1} \right)\bm{\Sigma_{*}}.$$

\paragraph*{For $h_i$:}
\begin{align*}
f(\thetab,u_i,h_i\rvert y_i) &\propto \exp\left\{ -\frac{1}{2} \left(\frac{\Delta^2+\tau}{\tau}\right) \left[ h_i^2 - 2h_i\left(\frac{\Delta^2b + \Delta\sqrt{u_i}(y_i - \bm{X}_{i}^{t}\bm{\beta} )}{\Delta^2+\tau}\right) \right]  \right\}\times \\
&\times I_{(0,\infty)}(h_i),
\end{align*}
which can be recognized as the kernel of a truncated normal distribution, so
$$h_i\rvert \thetab,u_i,y_i \sim TN\left( \frac{\Delta^2b + \Delta\sqrt{u_i}(y_i - \bm{X}_{i}^{t}\bm{\beta} )}{\Delta^2+\tau},\frac{\tau}{\Delta^2+\tau}\right)I_{(0,\infty)}(h_i).$$

\paragraph*{For $u_i$:}

\begin{itemize}
	\item Skew slash: 
	\begin{align*}
	f(\thetab,u_i,h_i\rvert y_i) &\propto u_i^{\nu+ 1/2 - 1} \exp\biggl\{ -\frac{u_i}{2\tau} \biggl[ (y_i - \bm{X}_{i}^{t}\bm{\beta} )^2 - 2\frac{\Delta}{\sqrt{u_i}} (h_i - b)(y_i -\\
	&\bm{X}_{i}^{t}\bm{\beta})\biggl]\biggl\} I_{(0,1)}(u_i).
	\end{align*}
	\item Skew-t: 
	\begin{align*}
	f(\thetab,u_i,h_i\rvert y_i) &\propto u_i^{\frac{\nu+1}{2} - 1} \exp\biggl\{ -\frac{u_i}{2} \biggl[\frac{(y_i - \bm{X}_{i}^{t}\bm{\beta})^2}{\tau} + \nu \biggl]  + \frac{\Delta\sqrt{u_i}}{\tau} (h_i - b)\times\\
	&\times(y_i - \bm{X}_{i}^{t}\bm{\beta})\biggl\} I_{(0,\infty)}(u_i).
	\end{align*}    
	\item Skew generalized t: 
	\begin{align*}
	f(\thetab,u_i,h_i\rvert y_i) &\propto u_i^{\frac{\nu_1+1}{2} - 1} \exp\biggl\{ -\frac{u_i}{2} \biggl[\frac{(y_i - \bm{X}_{i}^{t}\bm{\beta})^2}{\tau} + \nu_2 \biggl]  + \frac{\Delta\sqrt{u_i}}{\tau}\times \\
	&\times(h_i - b)(y_i - \bm{X}_{i}^{t}\bm{\beta})\biggl\} I_{(0,\infty)}(u_i).
	\end{align*}
	\item Skew-contaminated normal: the discrete conditional distribution of $u_i$ assumes $\nu_2$ with probability $\frac{p_i}{p_i+q_i}$ and 1 with probability $\frac{q_i}{p_i+q_i}$ where
	\begin{align*}
	p_i &= \nu_1 \sqrt{\nu_2} \exp\left\{ -\frac{\nu_2}{2\tau} \left[ (y_i - \bm{X}_{i}^{t}\bm{\beta} )^2 - 2\frac{\Delta}{\sqrt{\nu_2}} (h_i - b)(y_i - \bm{X}_{i}^{t}\bm{\beta})\right]\right\}, \\
	q_i &=(1-\nu_1) \exp\left\{ -\frac{1}{2\tau} \left[ (y_i - \bm{X}_{i}^{t}\bm{\beta} )^2 - 2\Delta (h_i - b)(y_i - \bm{X}_{i}^{t}\bm{\beta})\right]\right\}.
	\end{align*}   
\end{itemize}

\paragraph*{For $\nu$:}

\begin{itemize}
	\item Skew slash: Considering a gamma distribution left truncated at 1 as prior with mean $\frac{\alpha_1}{\alpha_2}$ and variance $\frac{\alpha_1}{\alpha_2^2}$, it follows that 
	\begin{align*}
	\pi(\nu\rvert \bm{\thetab}_{-\nu}, \bm{y}, \bm{u},\bm{h}) &\propto \nu^{n+\alpha_1 -1} \exp\left\{ -\nu (\alpha_2 - \sum_{i=1}^{n}ln(u_i))\right\} I_{(1,\infty)}(\nu),
	\end{align*}
	that is, $\nu\rvert \bm{\thetab}_{-\nu}, \bm{y}, \bm{u},\bm{h} \sim TG(n+\alpha_1,\alpha_2 - \sum_{i=1}^{n}ln(u_i) )I(1,\infty)$, where TG denotes the Truncated Gamma distribution.
	\item Skew-t: For Skew-t, we have adopted a very useful hierarchical prior distribution as noted in \cite{Cabral_VH}, which consists in $\nu\rvert \lambda \sim exp(\lambda)I(\nu)_{(2,\infty)}$ and $\lambda \sim U(\rho_0,\rho_1)$ where $0<\rho_0<\rho_1$ are known. The exponential distribution is left truncated at 2 to insure finite variance. Then 
	\begin{align*}
	\pi(\nu\rvert \bm{\thetab}_{-\nu},\lambda, \bm{y}, \bm{u},\bm{h}) &\propto \frac{\frac{\nu}{2}^{\frac{n\nu}{2}}}{\Gamma(\nu/2)^n} \left(\prod_{i=1}^{n}u_i\right)^{\nu/2 - 1} \exp\left\{-\nu \left(\frac{\sum_{i=1}^{n}u_i}{2} + \lambda\right) \right\}\times\\
	&\times I_{(2,\infty)}(\nu),\\
	\pi(\lambda\rvert \bm{\thetab}, \bm{y}, \bm{u},\bm{h}) &\propto \lambda \exp{-\lambda(\nu-2)} I_{(\rho_0,\rho_1)}(\lambda),
	\end{align*}  	 
	that is, $ \lambda\rvert \bm{\thetab}, \bm{y}, \bm{u},\bm{h} \sim TG(2,\nu-2)I(\rho_0,\rho_1)$.
	\item Skew generalized t: Assuming $\nu_1\rvert \lambda_1 \sim exp(\lambda_1)I(\nu_1)_{(2,\infty)}$ and $\lambda_1 \sim U(\rho_0,\rho_1)$ where $0<\rho_0<\rho_1$ are known and $\nu_2\rvert \lambda_2 \sim exp(\lambda_2)$ and $\lambda_2 \sim U(\psi_0,\psi_1)$ where $0<\psi_0<\psi_1$ are known, we have
	\begin{align*}
	\pi(\nu_1\rvert \bm{\thetab}_{-\nu_1},\lambda_1, \bm{y}, \bm{u},\bm{h}) &\propto \frac{\nu_2/2 ^{n\nu_1/2}}{\Gamma(\nu_1/2)^n} \left(\prod_{i=1}^{n}u_i\right)^{\nu_1/2 - 1} \exp\left\{-\lambda_1 (\nu_1 - 2)\right\}\times\\
	&\times I_{(2,\infty)}(\nu_1),\\
	\pi(\lambda_1\rvert \bm{\thetab}, \bm{y}, \bm{u},\bm{h}) &\propto \lambda_1 \exp \left\{ {-\lambda_1(\nu_1-2)} \right\} I_{(\rho_0,\rho_1)}(\lambda_1),
	\end{align*}  	
	that is, $ \lambda_1\rvert \bm{\thetab}, \bm{y}, \bm{u},\bm{h} \sim TG(2,\nu_1-2)I(\rho_0,\rho_1)$. Also, we have that
	\begin{align*}
	\pi(\nu_2\rvert \bm{\thetab}_{-\nu_2},\lambda_2, \bm{y}, \bm{u},\bm{h}) &\propto \nu_2/2 ^{n\nu_1/2}\exp\left\{-\nu_2\left( \frac{\sum_{i=1}^{n}u_i}{2} + \lambda_2\right)\right\}I_{(0,\infty)}(\nu_2),
	\end{align*}  	
	and 
	\begin{align*}
	\pi(\lambda_2\rvert \bm{\thetab}, \bm{y}, \bm{u},\bm{h}) &\propto \lambda_2 \exp\{-\lambda_2(\nu_2)\} I_{(\psi_0,\psi_1)}(\lambda_2),
	\end{align*} 
	that is, $\nu_2\rvert \bm{\thetab}_{-\nu_2},\lambda_2, \bm{y}, \bm{u},\bm{h} \sim \mbox{gamma}(\frac{n\nu_1}{2}+1,\frac{\sum_{i=1}^{n}u_i}{2} + \lambda_2)$ and $ \lambda_2\rvert \bm{\thetab}, \bm{y}, \bm{u},\bm{h} \sim TG(2,\nu_2-2)I(\xi_0,\xi_1)$
	\item Skew-contaminated normal: Observe that distribution of $U$ can be written as 
	\begin{align*}
	h(u\rvert \bm{\nu}) = \nu_1^{\frac{1-u}{1-\nu_2}}(1-\nu_1)^{\frac{u-\nu_2}{1-\nu_2}}I_{\{\nu_2,1\}}(u).
	\end{align*}
	
	Setting as prior distributions $\nu_1 \sim beta(\alpha_1,\beta_1)$, $\nu_2 \sim beta(\alpha_2,\beta_2)$, it follows that the conditional distributions of $\nu_1$ and $\nu_2$ are 
	\begin{align*}
	\pi(\nu_1\rvert \bm{\thetab}_{-\nu_1}, \bm{y}, \bm{u},\bm{h}) &\propto \nu_1^{\frac{n-\sum_{i=1}^{n}u_i}{1-\nu_2} + \alpha_1 - 1}(1-\nu_1)^{\frac{\sum_{i=1}^{n}u_i - n\nu_2}{1-\nu_2} + \beta_1 - 1} I_{(0,1)}(\nu_1),
	\end{align*}
	which can be recognized as the kernel of a beta distribution. So,
	
	$\nu_1\rvert \bm{\thetab}_{-\nu_1}, \bm{y}, \bm{u},\bm{h} \sim beta\left(\frac{n-\sum_{i=1}^{n}u_i}{1-\nu_2} + \alpha_1,\frac{\sum_{i=1}^{n}u_i - n\nu_2}{1-\nu_2} + \beta_1  \right)$
	and
	\begin{align*}
	\pi(\nu_2\rvert \bm{\thetab}_{-\nu_2}, \bm{y}, \bm{u},\bm{h}) &\propto \nu_1^{\frac{n-\sum_{i=1}^{n}u_i}{1-\nu_2}}(1-\nu_1)^{\frac{\sum_{i=1}^{n}u_i - n\nu_2}{1-\nu_2} } \nu_2^{\alpha_2 -1} (1-\nu2)^{(\beta_2 -1)}\times\\
	&\times I_{(0,1)}(\nu_2).
	\end{align*}	
\end{itemize}

Finally, for the skew generalized t distribution the conditional distribution of $\delta$ is 
\paragraph*{For $\delta$:}
\begin{align*}
\begin{split}
\pi(\delta\rvert \bm{\thetab}_{-\delta}, \bm{y}, \bm{u},\bm{h}) &\propto \biggl( \frac{\sqrt{1-b^2\delta^2}}{\sqrt{1-\delta^2}}\biggl)^n \exp\biggl\{ -\frac{1-b^2\delta^2}{2 (1-\delta^2)} \sum_{i=1}^{n} u_i \biggl( y_i - \biggl(\bm{X}_{i}^{t}\bm{\beta} +\\
&\frac{\delta}{\sqrt{u_i}\sqrt{1-b^2\delta^2}} (h_i-b)\biggl) \biggl) \biggl\} I_{(-1,1)}(\delta),
\end{split}
\end{align*}
and for the other distributions: 
\paragraph*{For $\Delta$:}
\begin{align*}
	\begin{split}
	\pi(\Delta\rvert \bm{\thetab}_{-\Delta}, \bm{y}, \bm{u},\bm{h}) &\propto \exp\biggl\{ -\frac{1}{2} \biggl(\frac{\sigma^2_\Delta \sum_{i=1}^{n}(h_i - b)^2 + \tau}{\tau\sigma^2_\Delta}\biggl) \biggl[ \Delta^2 -\\
	&2\Delta \frac{\sigma^2_\Delta\sum_{i=1}^{n}(h_i-b)\sqrt{u_i}(y_i - \bm{X}_{i}^{t}\bm{\beta}) + \mu_\Delta\tau}{\sigma^2_\Delta \sum_{i=1}^{n}(h_i - b)^2 + \tau}\biggl] \biggl\}I_{\mathbb{R}}(\Delta),
	\end{split}
	\end{align*}
which can be recognized as the kernel of normal distribution with mean $\frac{\sigma^2_\Delta\sum_{i=1}^{n}(h_i-b)\sqrt{u_i}(y_i - \bm{X}_{i}^{t}\bm{\beta}) + \mu_\Delta\tau}{\sigma^2_\Delta \sum_{i=1}^{n}(h_i - b)^2 + \tau}$ and variance $\frac{\tau\sigma^2_\Delta}{\sigma^2_\Delta \sum_{i=1}^{n}(h_i - b)^2 + \tau}$.

\paragraph*{For $\tau^{-1}$:}
\begin{align*}
\pi(\tau^{-1}\rvert \bm{\thetab}_{-\tau^{-1}}, \bm{y}, \bm{u},\bm{h}) &\propto (\tau^{-1})^{n/2 + c-1} \exp\biggl\{ - \tau^{-1} \biggl\{ d+\sum_{i=1}^{n}\frac{u_i}{2}\biggl( y_i - \biggl(\bm{X}_{i}^{t}\bm{\beta} +\\
&\frac{\Delta}{\sqrt{u_i}}(h_i-b)\biggl) \biggl)^2 \biggl\}  \biggl\}I_{(0,\infty)}(\tau^{-1}),
\end{align*}
that can be recognized as the kernel of a gamma distribution. So, 
$\tau^{-1}\rvert \bm{\thetab}_{-\tau^{-1}}, \bm{y},\\ \bm{u},\bm{h} \sim \mbox{gamma}\left(n/2 + c,d+\sum_{i=1}^{n}\frac{u_i}{2}\left( y_i - \left(\bm{X}_{i}^{t}\bm{\beta} + \frac{\Delta}{\sqrt{u_i}}(h_i-b)\right) \right)^2 \right)$.

\section{Simulation study results}

\subsection{Parameter recovery study}

	\begin{table}[h!]
	    \centering
	    	    \caption{Bias and variance under different posterior statistics for centered skew-t regression model.}
	    \begin{tabular}{ccccccc}
	    \hline
	    &\multicolumn{2}{c}{Mean}&\multicolumn{2}{c}{Median}&\multicolumn{2}{c}{Mode}\\
	    Paramater & Bias & Var& Bias & Var& Bias & Var\\
	    \hline
	    $\beta_0$ & -0.0022 & 0.0040 & -0.0011 & 0.0040 & -0.0007 & 0.0042\\
	    $\beta_1$ & -0.0115 & 0.0023 & -0.0116 & 0.0024 & -0.0120 & 0.0030\\
	    $\sigma^2$ & 0.0009 & 0.0125 & -0.0071 & 0.0118 & -0.0263 & 0.0129\\
	    $\delta$ & 0.0020 & 0.0009 & -0.0010 & 0.0005 & -0.0057 & 0.0003\\
	    $\nu$ & 0.3960 & 1.9702 & 0.2032 & 1.4093 & -0.0529 & 1.1254\\
	    \hline
	    \end{tabular}
	\end{table}
	
	\begin{table}[h!]
	    \centering
	    	    \caption{Bias and variance under different posterior statistics for centered skew slash regression model.}
	    \begin{tabular}{ccccccc}
	    \hline
	    &\multicolumn{2}{c}{Mean}&\multicolumn{2}{c}{Median}&\multicolumn{2}{c}{Mode}\\
	    Paramater & Bias & Var& Bias & Var& Bias & Var\\
	    \hline
	    $\beta_0$ & 0.0036 & 0.0025 & 0.0024 & 0.0025 & -0.0021 & 0.0029\\
	    $\beta_1$ & 0.0040 & 0.0025 & 0.0039 & 0.0024 & 0.0030 & 0.0024\\
	    $\sigma^2$ & 0.0005 & 0.0176 & -0.0070 & 0.0175 & -0.0271 & 0.0173\\
	    $\delta$ & 0.0010 & 0.0001 & 0.0022 & 0.0001 & 0.0058 & 0.0001\\
	    $\nu$ & 0.7786 & 2.8018 & 0.3055 & 1.0913 & -0.1174 & 0.3698\\
	    \hline
	    \end{tabular}
	\end{table}
	
	\begin{table}[h!]
	    \centering
	    	    \caption{Bias and variance under different posterior statistics for centered skew contaminated normal regression model.}
	    \begin{tabular}{ccccccc}
	    \hline
	    &\multicolumn{2}{c}{Mean}&\multicolumn{2}{c}{Median}&\multicolumn{2}{c}{Mode}\\
	    Paramater & Bias & Var& Bias & Var& Bias & Var\\
	    \hline
	    $\beta_0$ & -0.0116 & 0.0035 & -0.0104 & 0.0036 & -0.0062 & 0.0036\\
	    $\beta_1$ & 0.0100 & 0.0018 & 0.0103 & 0.0018 & 0.0125 & 0.0019\\
	    $\sigma^2$ & 0.1300 & 0.0187 & 0.1531 & 0.0190 & 0.1941 & 0.0271\\
	    $\delta$ & 0.0079 & $<$0.0001 & 0.0061 & $<$0.0001 & -0.0003 & $<$0.0001\\
	    $\nu_1$ & -0.0251 & 0.0110 & -0.0272 & 0.0167 & -0.0912 & 0.0777\\
	    $\nu_2$ & 0.0999 & 0.0093 & 0.0821 & 0.0127 & 0.0258 & 0.0322\\
	    \hline
	    \end{tabular}
	\end{table}
	
	\begin{table}[h!]
	    \centering
	    	    \caption{Bias and variance under different posterior statistics for centered skew generalized-t regression model.}
	    \begin{tabular}{ccccccc}
	    \hline
	    &\multicolumn{2}{c}{Mean}&\multicolumn{2}{c}{Median}&\multicolumn{2}{c}{Mode}\\
	    Paramater & Bias & Var& Bias & Var& Bias & Var\\
	    \hline
	    $\beta_0$ & -0.0030 & 0.0010 & 0.0027 & 0.0010 & 0.0020 & 0.0010\\
	    $\beta_1$ & -0.0036 & 0.0004 & -0.0035 & 0.0004 & -0.0045 & 0.0005\\
	    $\delta$ & -0.0021 & $<0.0001$ & -0.0005 & $<0.0001$ & 0.0040 & $<0.0001$\\
	    $\nu_1$ & 1.4777 & 62.0683 & -0.0667 & 40.2128 & -1.8664 & 42.5915\\
	    $\nu_2$ & 0.5929 & 10.1638 & 0.0312 & 7.0368 & -0.7113 & 6.6483\\
	    \hline
	    \end{tabular}
	\end{table}

\begin{table}[h!]
\centering
	\caption{Results of the simulation study for the skew-t model with $\nu=5$ and $\gamma=-0.9$}
	\begin{tabular}{rrr|rrrrrrr}
			\hline
			n	&  Parameter & Real & Est & Var & Bias &  Rel Bias & RMSE & CR & length CI\\ 
			\hline
			\multirow{5}{*}{50}
			& $\beta_0$  & 1.0000  &  1.0359  & 0.0320  & 0.0359 & 0.0359   & 0.1826 & 0.82 & 0.5792\\
			& $\beta_1$  & 2.0000  & 1.9879  & 0.0376  & -0.0121 & 0.0060   & 0.1943 & 0.84 & 0.5276\\
			& $\delta$   & -0.9876 & -0.8129 & 0.2322  & 0.1747  & 0.1801   & 0.5193 & 0.62 & 0.7746\\
			& $\sigma^2$ &1.0000   & 0.9202  & 0.1266  & -0.0798  & 0.0798  & 0.3647 & 0.86 & 1.0430   \\
			& $\nu$      &5.0000   & 4.4932  & 3.5373 & -0.5068  & 0.1013   & 1.9479 & 0.94 & 18.1367  \\
			\hline	
			\multirow{5}{*}{100}
			& $\beta_0$  & 1.0000  & 0.9737  & 0.0185 & -0.0262 & 0.0262 & 0.1386 & 0.76 & 0.3952 \\
			& $\beta_1$  & 2.0000  & 1.9949  & 0.0127 & -0.0051 & 0.0025 & 0.1122 & 0.84 & 0.3560\\
			& $\delta$   & -0.9876 & -0.9658 & 0.0118 & 0.0218  & 0.0221 & 0.1108 & 0.46 & 0.2379 \\
			& $\sigma^2$ & 1.0000  & 1.0333  & 0.0848 & 0.0333  & 0.0333  & 0.2930 & 0.82 & 0.7703  \\
			& $\nu$      & 5.0000 & 5.0743  & 3.6676 & 0.0743   & 0.0149  & 1.9165 & 0.92 & 11.3941 \\
			\hline	
			\multirow{5}{*}{250}
			& $\beta_0$  & 1.0000  & 1.0037  & 0.0079 & 0.0037  & 0.0037  &0.0891  & 0.82 &  0.2840 \\
			& $\beta_1$  & 2.0000  & 1.9797  & 0.0075 & -0.0203 & 0.0101  & 0.0892 & 0.86 &0.2547  \\
			& $\delta$   & -0.9876 & -0.9778 & 0.0014 & 0.0098  & 0.0099  & 0.0392 & 0.76 & 0.1352  \\
			& $\sigma^2$ & 1.0000  & 1.0627  & 0.0333 & 0.0627  & 0.0627  & 0.1929 & 0.80 & 0.5645 \\
			& $\nu$      &  5.0000 &  5.3274 & 2.6146 & 0.3274  & 0.0655  & 1.6498 & 0.94  & 10.2040\\
			\hline	
			\multirow{5}{*}{500}
			& $\beta_0$  & 1.0000  & 0.9989  & 0.0039 & -0.0012 & 0.0012   & 0.0632 & 0.82 & 0.1896 \\
			& $\beta_1$  & 2.0000  & 1.9884  & 0.0024 & -0.1161    & 0.0058 & 0.0506 & 0.96 & 0.1723 \\
			& $\delta$   & -0.9876 & -0.9887 & 0.0005 & -0.0010   & 0.0010  & 0.0233 & 0.50 & 0.0544 \\
			& $\sigma^2$ & 1.0000  & 1.0009  & 0.0125 & 0.0009   & 0.0009   & 0.1119 & 0.94 & 0.3680  \\
			& $\nu$      & 3.0000  & 4.9471  & 1.1254 & -0.0529    & 0.0106 & 1.0621 & 0.92 & 4.0947\\
			\hline			
	\end{tabular}
	\label{recor:skewt}
\end{table}

\begin{table}[!htbp]
\centering
	\caption{Results of the simulation study for the skew slash model with $\nu=3$ and $\gamma=0.9$}
	\begin{tabular}{rrr|rrrrrrr}
			\hline
			n	&  Parameter & Real & Est & Var & Bias &  Rel Bias & RMSE & CR & length CI\\ 
			\hline
			\multirow{5}{*}{50}
			& $\beta_0$  & 1.0000 & 1.0093 & 0.0275  & 0.0094  & 0.0094  & 0.1661 & 0.74 & 0.5256 \\ 
			& $\beta_1$  & 2.0000 & 2.0146 & 0.0192  & 0.0146 & 0.0073 & 0.1395 & 0.82 & 0.4819 \\ 
			& $\delta$   & 0.9876  & 0.9486 & 0.0149  & -0.0390 & 0.0395 & 0.1283 & 0.54 & 0.3499 \\ 	
			& $\sigma^2$ & 1.0000 & 1.0019 & 0.1008  & 0.0019  & 0.0019  & 0.3176 & 0.80 & 1.0458 \\ 
			& $\nu$      & 3.0000 & 2.4790 & 4.9842 & -0.5210  & 0.1737  & 2.2925 & 0.90 & 13.7703 \\ 
			\hline	
			\multirow{5}{*}{100}
			& $\beta_0$ & 1.0000  & 1.0193 & 0.0156   & 0.0193  & 0.0193 & 0.1263 & 0.86 & 0.3975 \\ 
			& $\beta_1$ & 2.0000 & 2.0043  & 0.0123   & 0.0043  & 0.0021 & 0.1112 & 0.88 & 0.3452 \\ 
			& $\delta$ & 0.9876   & 0.9720 & 0.0074   & -0.0156 & 0.0158 & 0.0877 & 0.54 & 0.1708 \\ 
			& $\sigma^2$ &1.0000  & 1.0604 & 0.0826   & 0.0603  & 0.0604 & 0.2938 & 0.76 & 0.7932 \\ 
			& $\nu$ &3.0000     & 3.0461  & 2.9236   & 0.0461  & 0.0154 & 1.7105 & 0.88 & 10.0941 \\  
			\hline	
			\multirow{5}{*}{250}
			& $\beta_0$ & 1.0000  & 1.0082 & 0.0045 & 0.0082   & 0.0082 & 0.0676 & 0.94 & 0.2757 \\ 
			& $\beta_1$ & 2.0000  & 1.9835 & 0.0052 & -0.0165  & 0.0082 & 0.0737 & 0.88 & 0.2314 \\ 
			& $\delta$ & 0.9876   & 0.9890 & 0.0001 & 0.0014   & 0.0014 & 0.0114 & 0.78 & 0.0292 \\ 
			& $\sigma^2$ &1.0000  & 1.0608 & 0.0273 & 0.0608   & 0.0608 & 0.1761 & 0.88 & 0.5847 \\ 
			& $\nu$ &3.0000      & 2.8397 & 0.5433 & -0.1603  & 0.0534 & 0.7543 & 0.92 & 7.3272 \\ 
			\hline	
			\multirow{5}{*}{500}
			& $\beta_0$  & 1.0000 & 1.0024 & 0.0025    & 0.0041  & 0.0041  & 0.0501 & 0.92 & 0.1950 \\ 
			& $\beta_1$  & 2.0000 & 2.0039 & 0.0024    & 0.0040 & 0.0020   & 0.0493 & 0.88 & 0.1601 \\ 
			& $\delta$   & 0.9876  & 0.9883 & 0.0001    & 0.0007  & 0.0007  & 0.0110 & 0.66 & 0.0231 \\ 
			& $\sigma^2$ & 1.0000 & 1.0104 & 0.0202    & 0.0104  & 0.0104   & 0.1424 & 0.84 & 0.4172 \\ 
			& $\nu$      & 3.0000 & 2.9227 & 0.4277    & -0.0773  & 0.0358  & 0.6586 & 0.88 & 5.3177 \\ 
			\hline			
	\end{tabular}
	\label{recor:skewslash}
\end{table}

\begin{table}[!htbp]
\centering
	\caption{Results of the simulation study for the skew contaminated normal model with $\nu_1 = \nu_2 = 0.5$ and $\gamma=-0.9$}
	\begin{tabular}{rrr|rrrrrrr}
			\hline
			n	&  Parameter & Real & Est & Var & Bias &  Rel Bias & RMSE & CR & length CI\\ 
			\hline
			\multirow{5}{*}{50}
			& $\beta_0$  & 1.0000 & 0.9945    & 0.0318  & -0.0055  & 0.0055  & 0.1787 & 0.90 & 0.6646 \\ 
			& $\beta_1$  & 2.0000 & 1.9740    & 0.0220  & -0.0260  & 0.0130 & 0.1506 & 0.92 & 0.5551 \\ 
			& $\delta$   & -0.9876  & -0.9565 & 0.0675  & 0.0311   & 0.0315 & 0.2617 & 0.72 & 0.4444 \\ 
			& $\sigma^2$ & 1.0000 & 1.0698    & 0.1417  & 0.0694   & 0.0694  & 0.3827 & 0.96 & 1.6816 \\ 
			& $\nu_1$    & 0.5000 & 0.5149    & 0.0098 & 0.0149    & 0.0299  & 0.1004 & 0.98 & 0.8678 \\ 
			& $\nu_2$    & 0.5000 & 0.4731    & 0.0620 & -0.0269   & 0.0537  & 0.2505 & 0.96 & 0.7269 \\ 
			\hline	
			\multirow{5}{*}{100}
			& $\beta_0$ & 1.0000  & 0.9931    & 0.0238 & -0.0069 & 0.0069 & 0.1543 & 0.84 & 0.4385 \\ 
			& $\beta_1$ & 2.0000 & 2.0183     & 0.0144 & 0.0183  & 0.0091 & 0.1215 & 0.84 & 0.3728 \\ 
			& $\delta$ & -0.9876   & -0.9951   & 0.0001 & -0.0075 & 0.0076 & 0.0148 & 0.82 & 0.1751 \\ 
			& $\sigma^2$ &1.0000  & 1.1050    & 0.0876 & 0.1050  & 0.1050 & 0.3140 & 0.92 & 1.2254 \\ 
			& $\nu_1$    & 0.5000 & 0.4584    & 0.0151 & -0.0415 & 0.0831 & 0.1298 & 0.92 & 0.8408 \\ 
			& $\nu_2$    & 0.5000 & 0.5280    & 0.0553 & 0.0280  & 0.0560 & 0.2369 & 0.92 & 0.6743 \\ 
			\hline	
			\multirow{5}{*}{250}
			& $\beta_0$ & 1.0000 & 0.9990   & 0.0061 & -0.0019 & 0.0019 & 0.0783 & 0.90 & 0.2841 \\ 
			& $\beta_1$ & 2.0000 & 1.9961   & 0.0062 & -0.0043 & 0.0021 & 0.0790 & 0.82 & 0.2321 \\ 
			& $\delta$ & -0.9876   & -0.9932 & 0.0001 & -0.0056 & 0.0057 & 0.0115 & 0.82 & 0.0357 \\ 
			& $\sigma^2$ &1.0000  & 1.1093  & 0.0366 & 0.1093  & 0.1093 & 0.2204 & 0.90 & 0.9239 \\ 
			& $\nu_1$    & 0.5000 & 0.4506  & 0.0099 & -0.0494 & 0.0988 & 0.1114 & 0.90 & 0.7260 \\ 
			& $\nu_2$    & 0.5000 & 0.5051  & 0.0268 & 0.0051  & 0.0103 & 0.1639 & 0.92 & 0.5477 \\ 
			\hline	
			\multirow{5}{*}{500}
			& $\beta_0$  & 1.0000 & 0.9896   & 0.0036    & -0.0104 & 0.0104 & 0.0613 & 0.92 & 0.2108 \\ 
			& $\beta_1$  & 2.0000 & 2.0103   & 0.0018    & 0.0103  & 0.0052 & 0.0434 & 0.98 & 0.1738 \\ 
			& $\delta$   & -0.9876  & -0.9879 & $<0.0001$ & -0.0003 & 0.0003 & 0.0100 & 0.96 & 0.0340 \\
			& $\sigma^2$ & 1.0000 & 1.1300   & 0.0187    & 0.1300  & 0.1300 & 0.1886 & 0.94 & 0.8887 \\ 
			& $\nu_1$    & 0.5000 & 0.4749   & 0.0110    & -0.0251 & 0.0502 & 0.1079 & 0.96 & 0.8195 \\ 
			& $\nu_2$    & 0.5000 & 0.5258   & 0.0322    & 0.0258  & 0.0516 & 0.1813 & 0.98 & 0.5923 \\ 
			\hline			
	\end{tabular}
	\label{recor:contn}
\end{table}

\begin{table}[!htbp]
\centering
	\caption{Results of the simulation study for the skew generalized t model with $\nu_1 = 15$ and $\nu_2 = 5$ and $\gamma=0.9$}
	\begin{tabular}{rrr|rrrrrrr}
			\hline
			n	&  Parameter & Real & Est & Var & Bias &  Rel Bias & RMSE & CR & length CI\\ 
			\hline
			\multirow{5}{*}{50}
			& $\beta_0$  & 1.0000   & 0.9310    & 0.0049  & -0.0690  & 0.0690  & 0.0984 & 0.86 & 0.3344 \\ 
			& $\beta_1$  & 2.0000   & 1.9941    & 0.0062  & -0.0059  & 0.0030 & 0.0793 & 0.94 & 0.3305 \\ 
			& $\delta$   & 0.9876  & 0.5971   & 0.1021    & -0.3905  & 0.3954 & 0.5046 & 0.72 & 1.5492 \\ 
			& $\nu_1$    & 15.0000  & 7.3217    & 2.5234  & -7.6782  & 0.5119  & 7.8408 & 0.86 & 20.3674 \\ 
			& $\nu_2$    & 5.0000  & 2.0004     & 0.3646  & -2.9996  & 0.5999  & 3.0598 & 072 & 6.8833 \\ 
			\hline	
			\multirow{5}{*}{100}
			& $\beta_0$ & 1.0000   & 0.9957   & 0.0042  & -0.0043  & 0.0043 & 0.0652 & 0.90 & 0.2417 \\ 
			& $\beta_1$ & 2.0000   & 2.0082   & 0.0032  & 0.0082   & 0.0041 & 0.0571 & 0.94 & 0.2194 \\ 
			& $\delta$ & 0.9876    & 0.8807   & 0.0473  & -0.1069  & 0.1082 & 0.2423 & 0.90 & 0.5757 \\ 
			& $\nu_1$    & 15.0000 & 11.1107  & 13.4414 & -3.8893  & 0.2593 & 5.3449 & 0.88 & 25.2330 \\ 
			& $\nu_2$    & 5.0000  & 3.4820   & 1.8910  & -1.5179  & 0.3036 & 2.0482 & 0.86 & 9.1148 \\ 
			\hline	
			\multirow{5}{*}{250}
			& $\beta_0$ & 1.0000 & 0.9967   & 0.0014  & -0.0034 & 0.0034 & 0.0383 & 0.98 & 0.1482 \\ 
			& $\beta_1$ & 2.0000 & 1.9944   & 0.0015  & -0.0056 & 0.0028 & 0.0387 & 0.98 & 0.1330 \\ 
			& $\delta$ & 0.9876   & 0.9740  & 0.0003  & -0.0136 & 0.0137 & 0.0234 & 0.98 & 0.0896 \\ 
			& $\nu_1$    & 15.0000 & 15.8465& 98.8371 & 0.8465  & 0.0564 & 9.9777 & 0.94 & 26.7851 \\ 
			& $\nu_2$    & 5.0000 & 5.2933  & 16.5690 & 0.2933  & 0.0587 & 4.0810 & 0.90 & 9.9441 \\ 
			\hline	
			\multirow{5}{*}{500}
			& $\beta_0$  & 1.0000 & 1.0027   & 0.0010    & 0.0027  & 0.0027 & 0.0323 & 0.88 & 0.1062 \\ 
			& $\beta_1$  & 2.0000 & 1.9964   & 0.0004    & -0.0035 & 0.0018 & 0.0217 & 0.96 & 0.0899 \\ 
			& $\delta$   & 0.9876  & 0.9871  & $<0.0001$ & -0.0005 & 0.0005 & 0.0077 & 0.96 & 0.0298 \\
			& $\nu_1$    & 15.0000 & 14.9332 & 40.2128   & -0.0667 & 0.0044 & 6.3417 & 0.90 & 22.1373 \\ 
			& $\nu_2$    & 5.0000 & 5.0312   & 7.0368    & 0.0312  & 0.0062 & 2.6529 & 0.86 & 8.3443 \\ 
			\hline			
	\end{tabular}
	\label{recor:skewgt}
\end{table}

\subsection{Residuals simulation study}
	
	\begin{figure}[h!]
		\begin{center}
		\begin{subfigure}{.37\linewidth}
			\includegraphics[width=\linewidth]{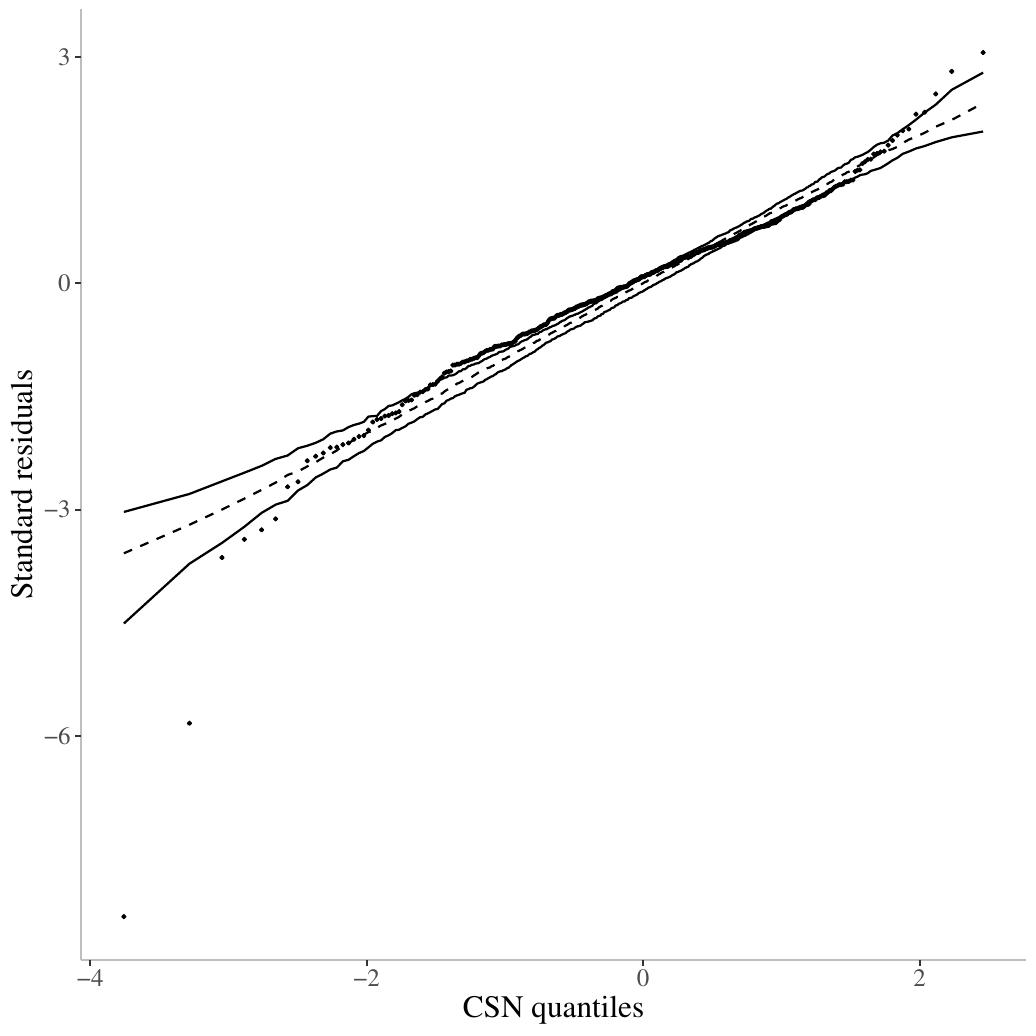}
			\caption{}
			\label{fig:residuos_cst_csn}
		\end{subfigure}
		\begin{subfigure}{.37\linewidth}
			\includegraphics[width=\linewidth]{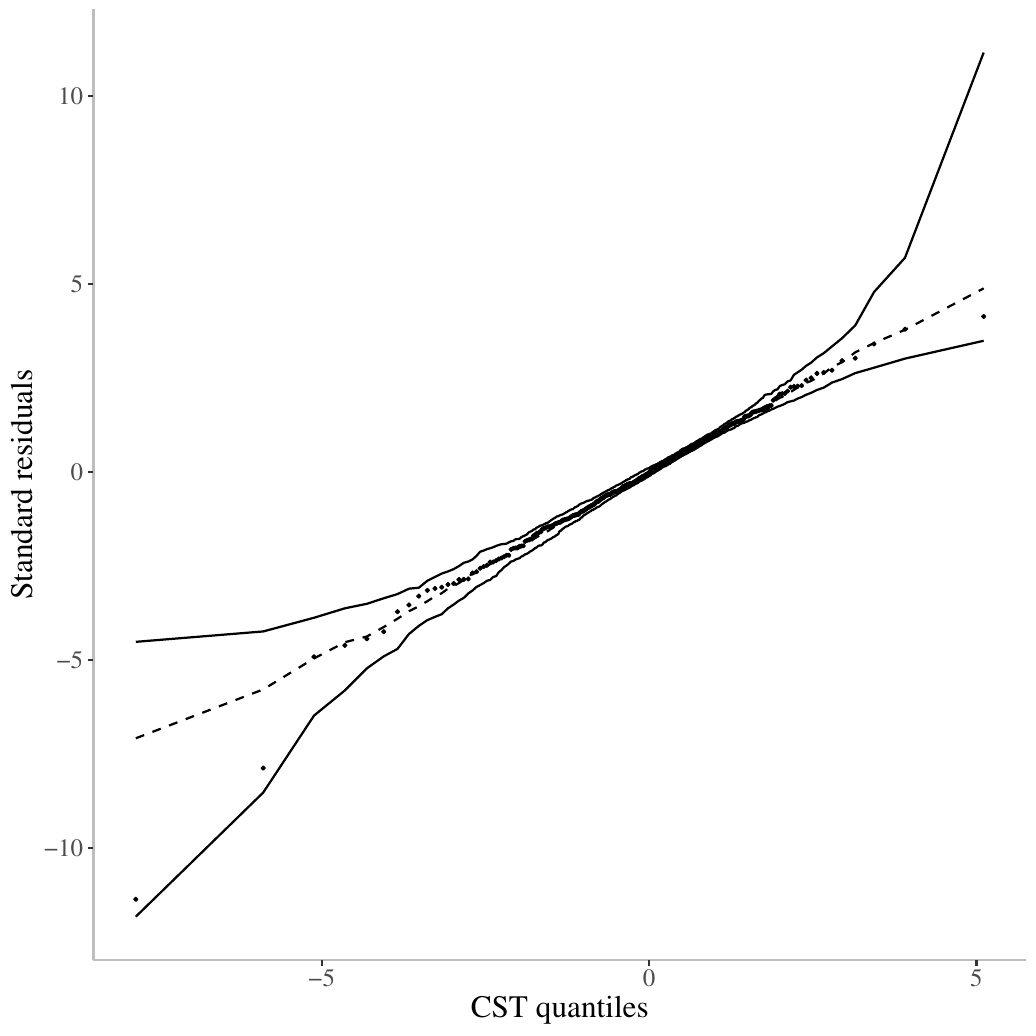}
			\caption{}
			\label{fig:residuos_cst_cst}
		\end{subfigure}\\
		\begin{subfigure}{.37\linewidth}
			\includegraphics[width=\linewidth]{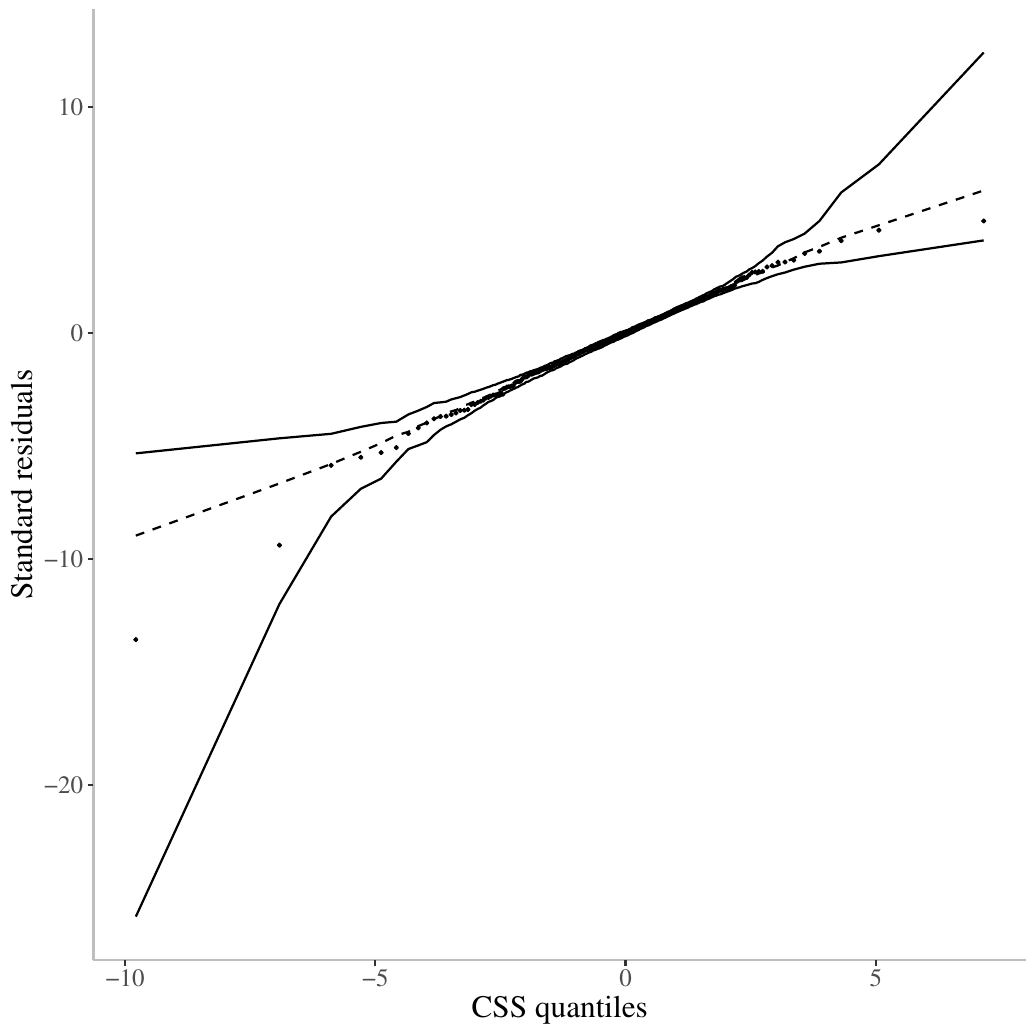}
			\caption{}
			\label{fig:residuos_cst_css}
		\end{subfigure}
		\begin{subfigure}{.37\linewidth}
			\includegraphics[width=\linewidth]{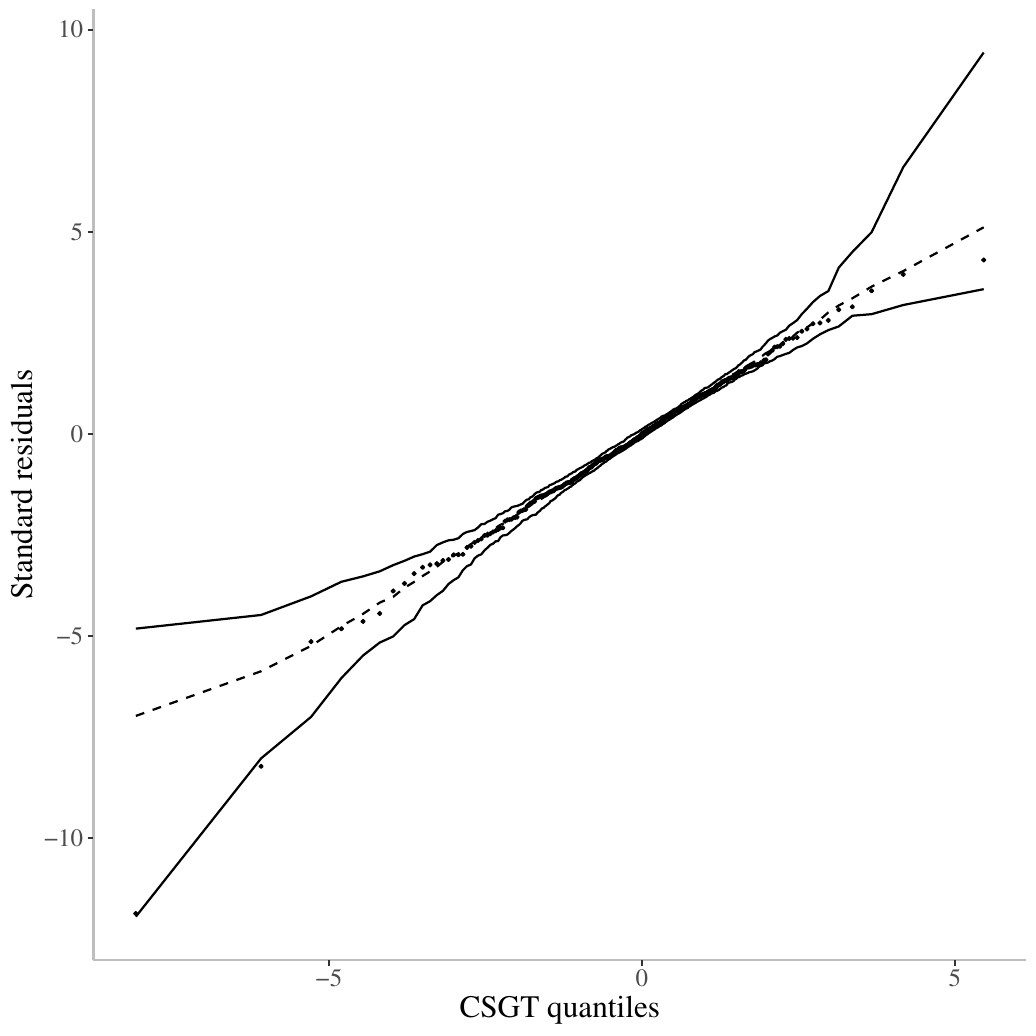}
			\caption{}
			\label{fig:residuos_cst_csgt}
		\end{subfigure}\\
		\begin{subfigure}{.37\linewidth}
			\includegraphics[width=\linewidth]{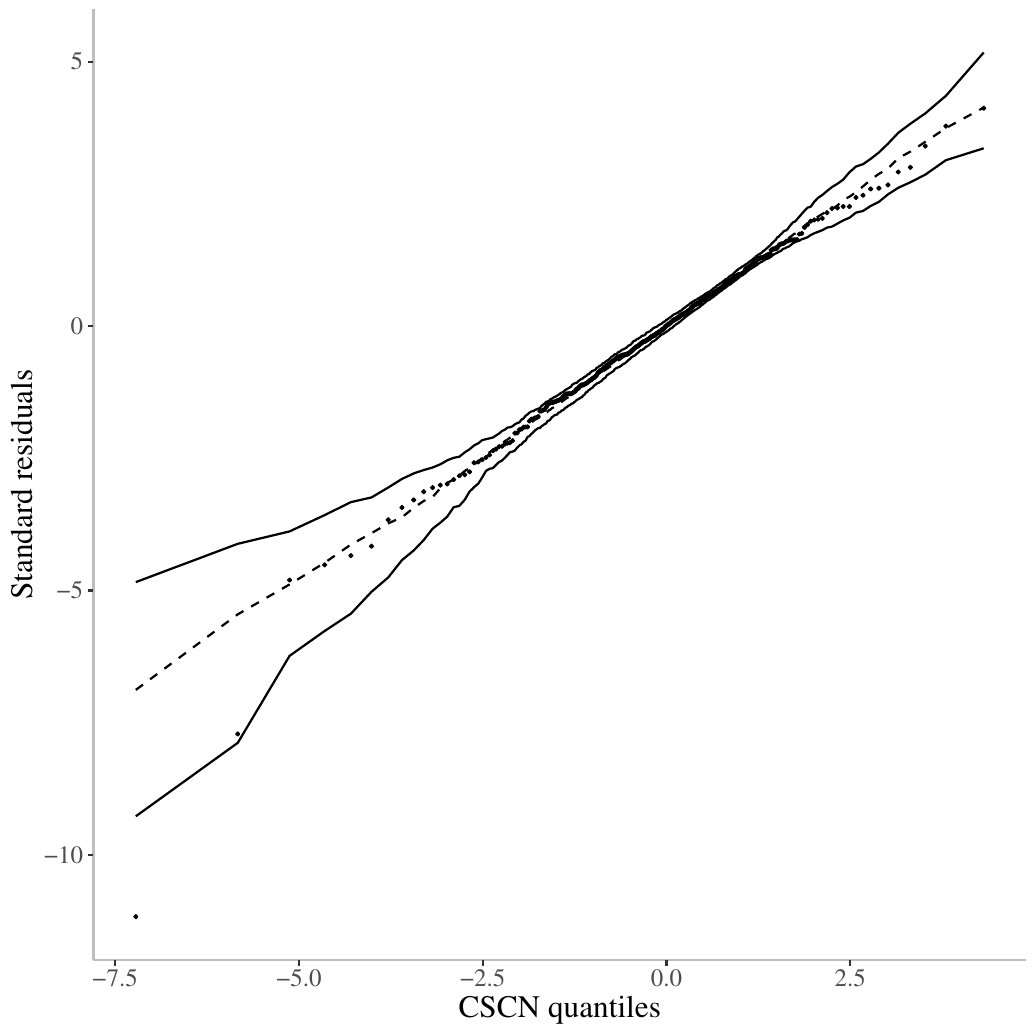}
			\caption{}
			\label{fig:residuos_cst_cscn}
		\end{subfigure}
		\end{center}
	\caption{QQ plots using the data set generated by the skew t distribution and adjusting by: skew normal (\subref{fig:residuos_cst_csn}), skew t (\subref{fig:residuos_cst_cst}), skew slash (\subref{fig:residuos_cst_css}), skew generalized t (\subref{fig:residuos_cst_csgt}) and skew contaminated normal (\subref{fig:residuos_cst_cscn})}
	\label{fig:residuos_cst}
\end{figure}

\begin{figure}[h!]
\begin{center}
		\begin{subfigure}{0.37\linewidth}
			\includegraphics[width=\linewidth]{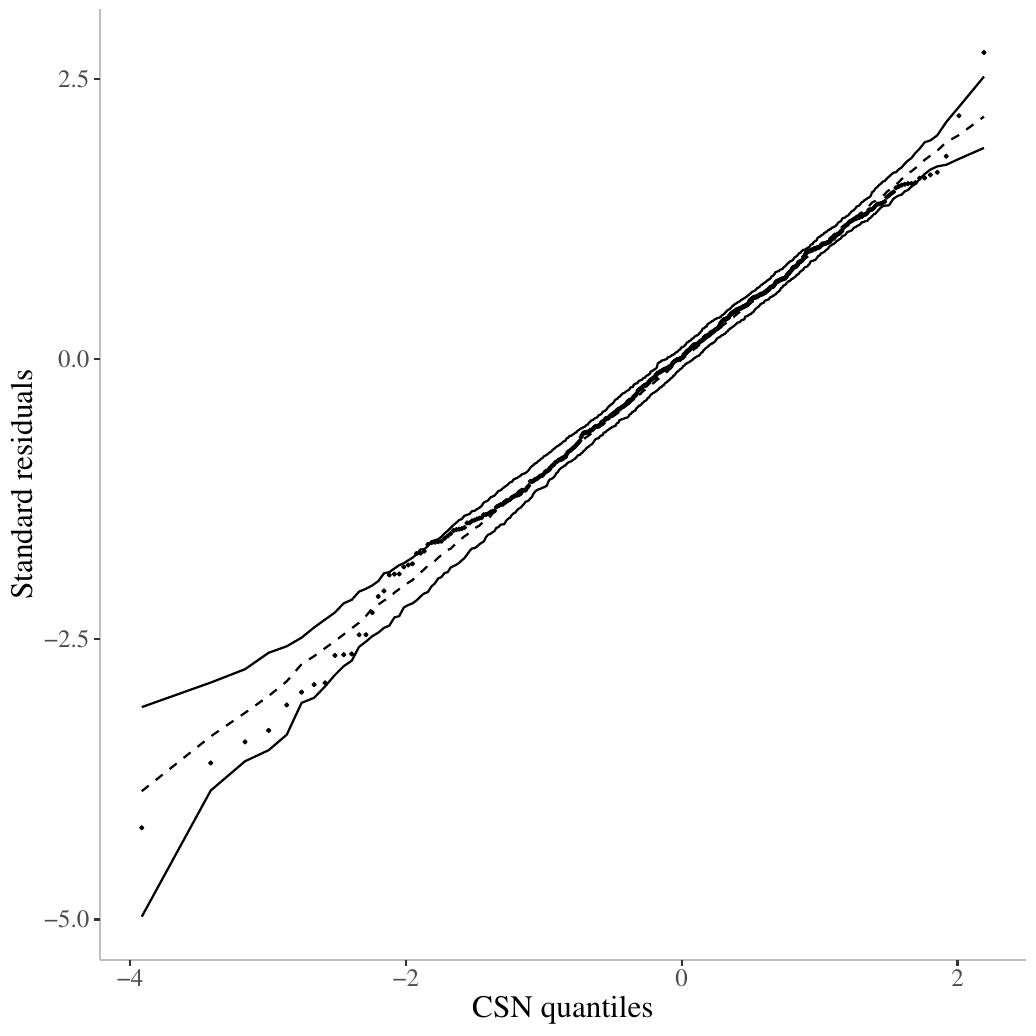}
			\caption{}
			\label{fig:residuos_css_csn}
		\end{subfigure}
		\begin{subfigure}{.37\linewidth}
			\includegraphics[width=\linewidth]{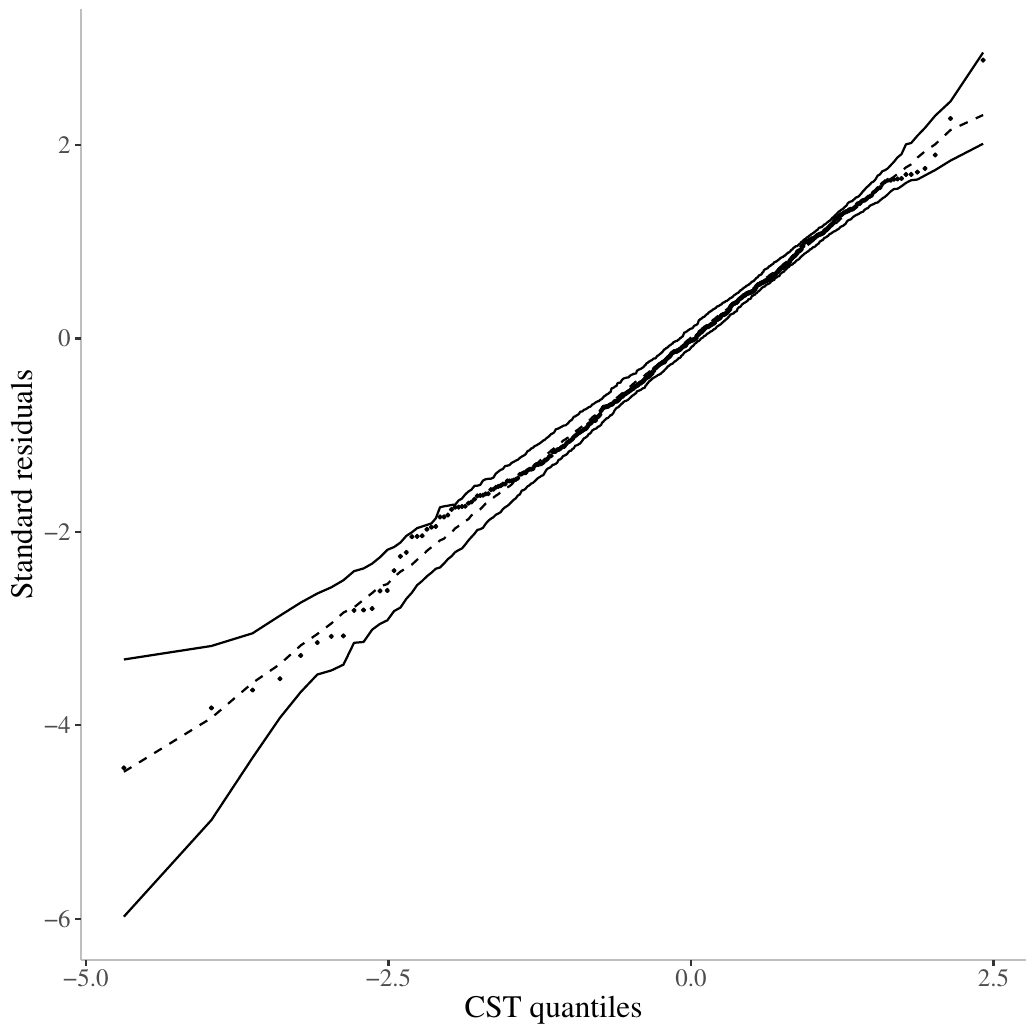}
			\caption{}
			\label{fig:residuos_css_cst}
		\end{subfigure}\\
		\begin{subfigure}{.37\linewidth}
			\includegraphics[width=\linewidth]{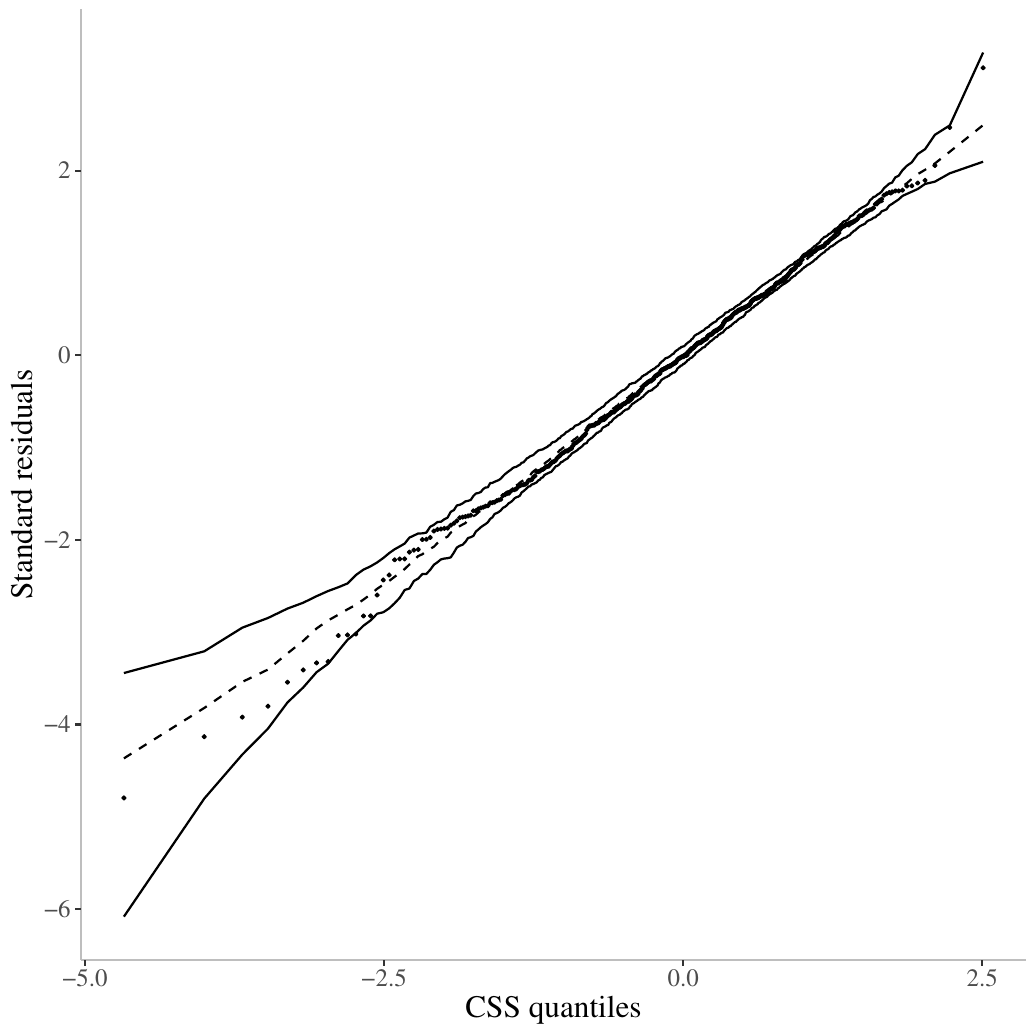}
			\caption{}
			\label{fig:residuos_css_css}
		\end{subfigure}
		\begin{subfigure}{.37\linewidth}
			\includegraphics[width=\linewidth]{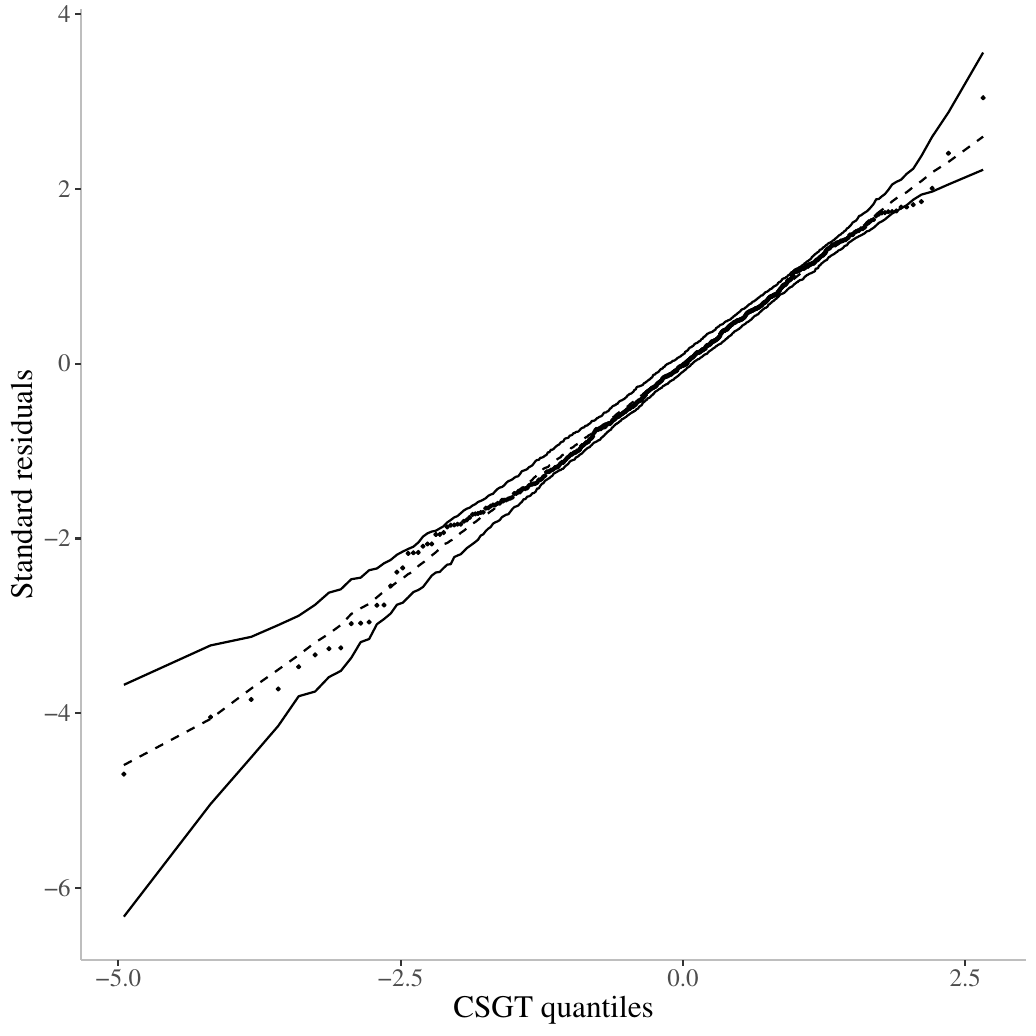}
			\caption{}
			\label{fig:residuos_css_csgt}
		\end{subfigure}\\
		\begin{subfigure}{.37\linewidth}
			\includegraphics[width=\linewidth]{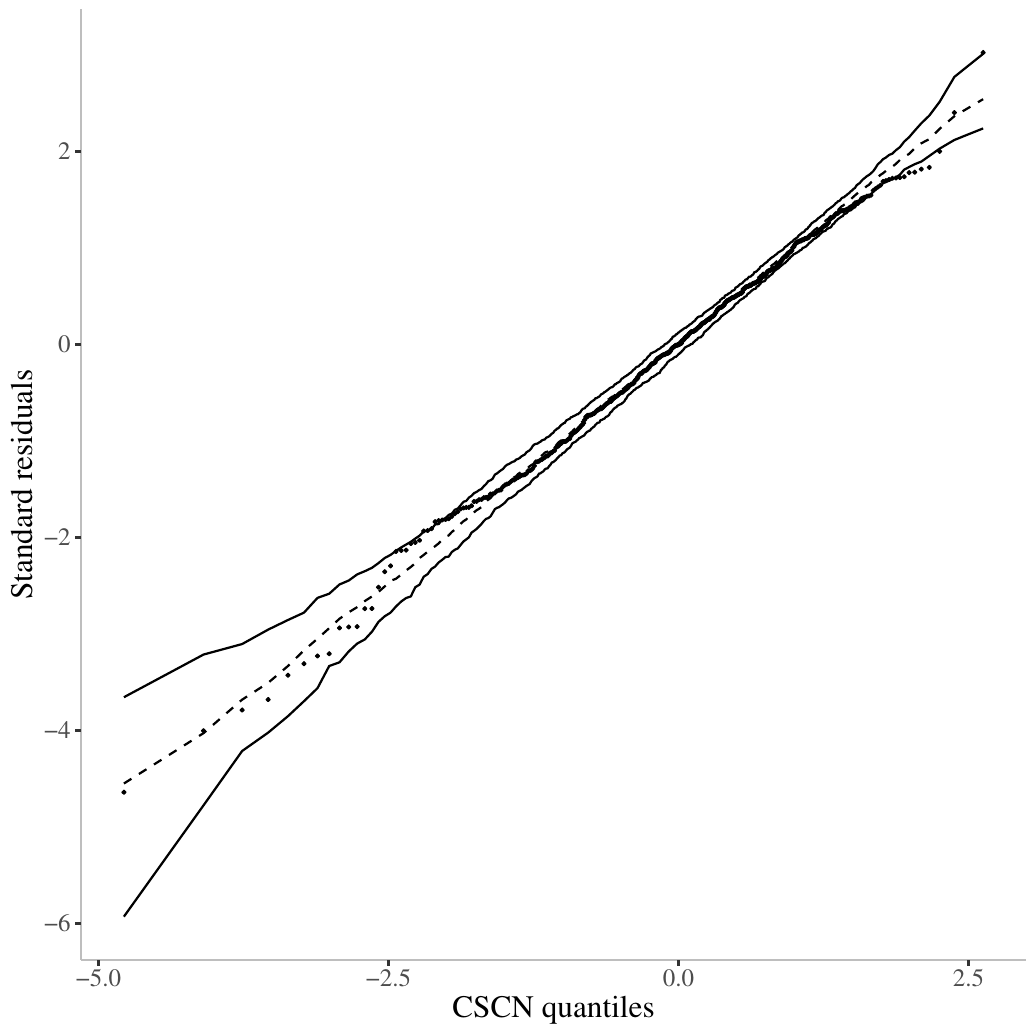}
			\caption{}
			\label{fig:residuos_css_cscn}
		\end{subfigure}
		\end{center}
	\caption{QQ plots using the data set generated by the skew slash distribution and adjusting by: skew normal (\subref{fig:residuos_css_csn}), skew t (\subref{fig:residuos_css_cst}), skew slash (\subref{fig:residuos_css_css}), skew generalized t (\subref{fig:residuos_css_csgt}) and skew contaminated normal (\subref{fig:residuos_css_cscn})}
	\label{fig:residuos_css}
\end{figure}

\begin{figure}[h!]
		\begin{center}
		\begin{subfigure}{.37\linewidth}
			\includegraphics[width=\linewidth]{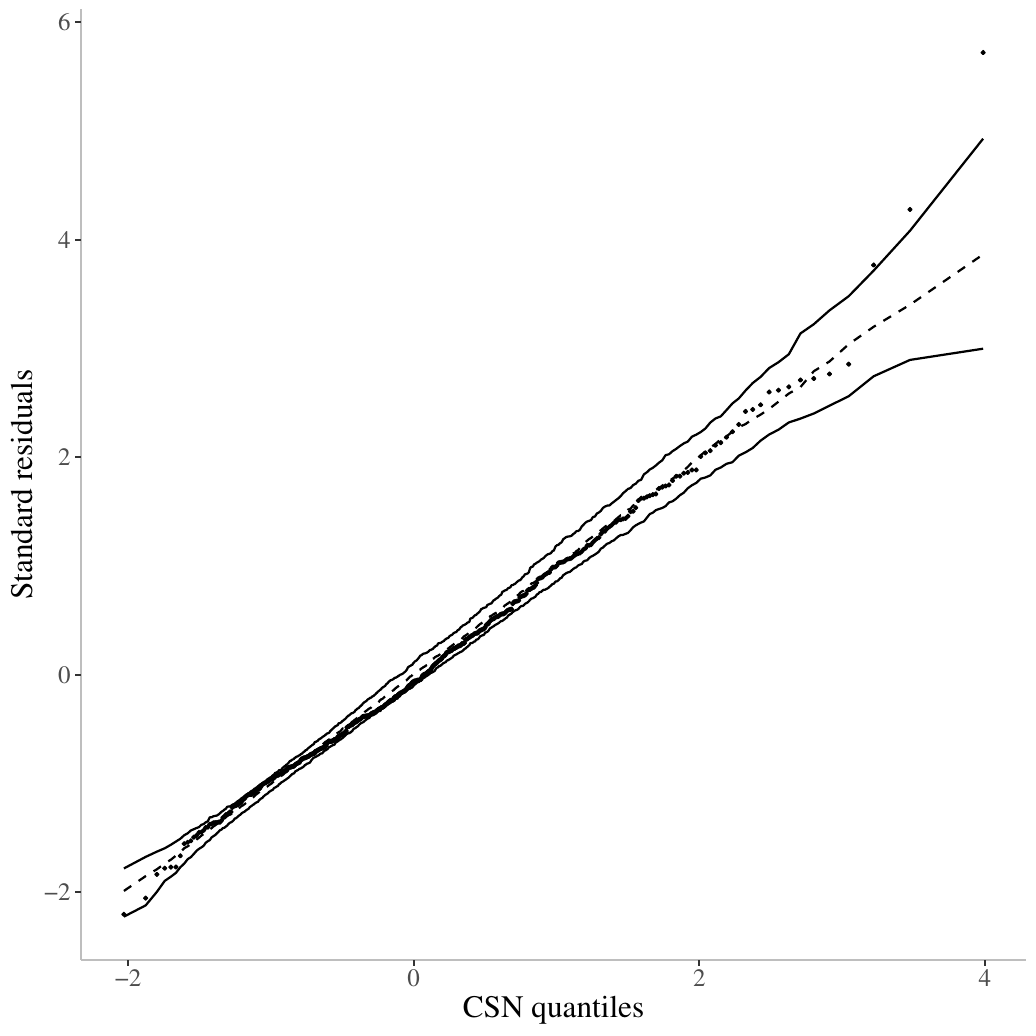}
			\caption{}
			\label{fig:residuos_csgt_csn}
		\end{subfigure}
		\begin{subfigure}{.37\linewidth}
			\includegraphics[width=\linewidth]{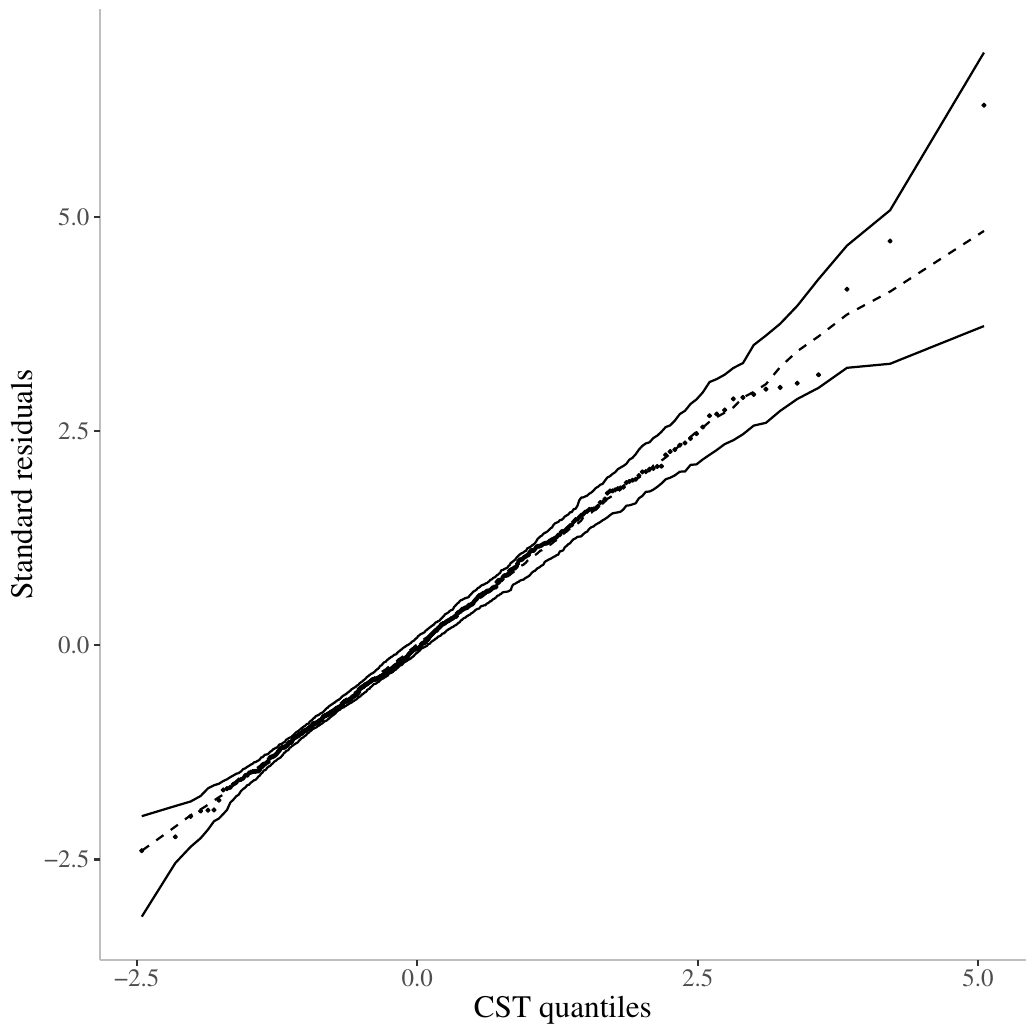}
			\caption{}
			\label{fig:residuos_csgt_cst}
		\end{subfigure}\\
		\begin{subfigure}{.37\linewidth}
			\includegraphics[width=\linewidth]{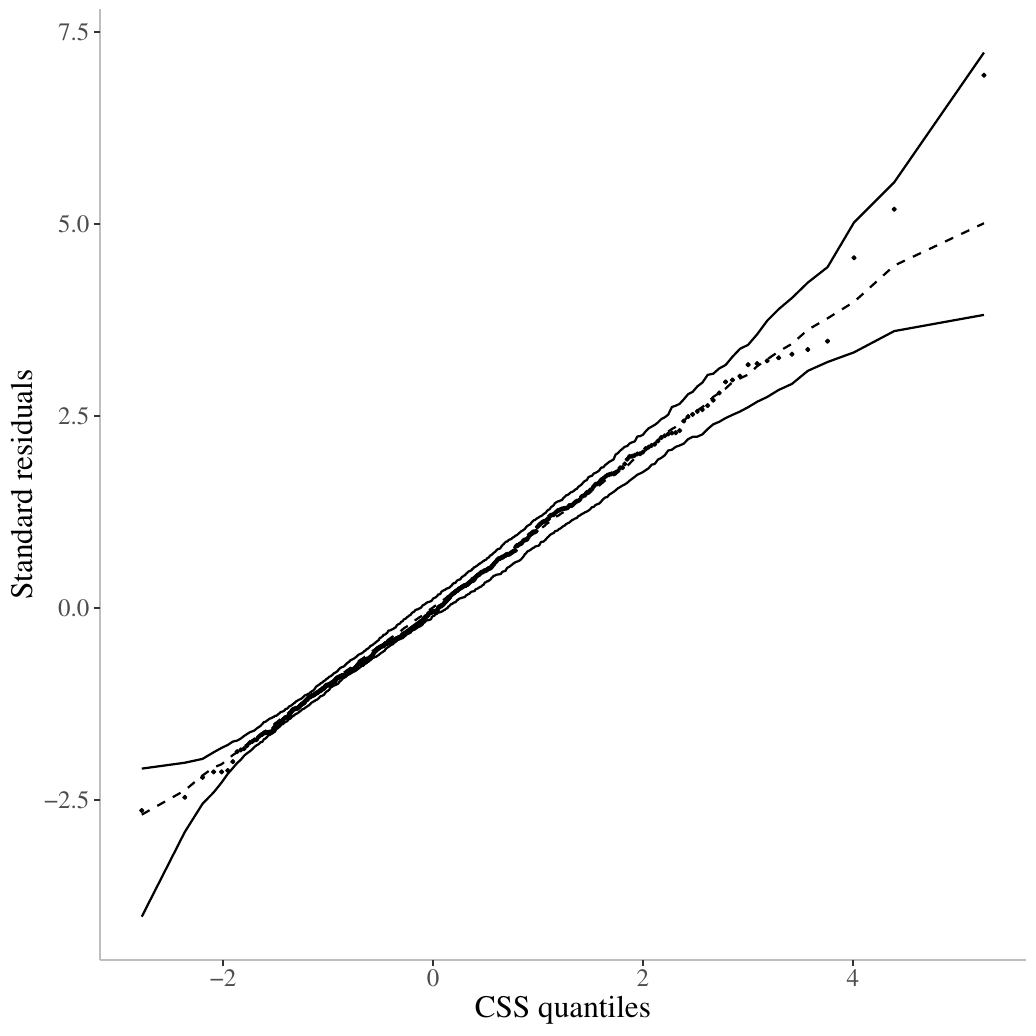}
			\caption{}
			\label{fig:residuos_csgt_css}
		\end{subfigure}
		\begin{subfigure}{.37\linewidth}
			\includegraphics[width=\linewidth]{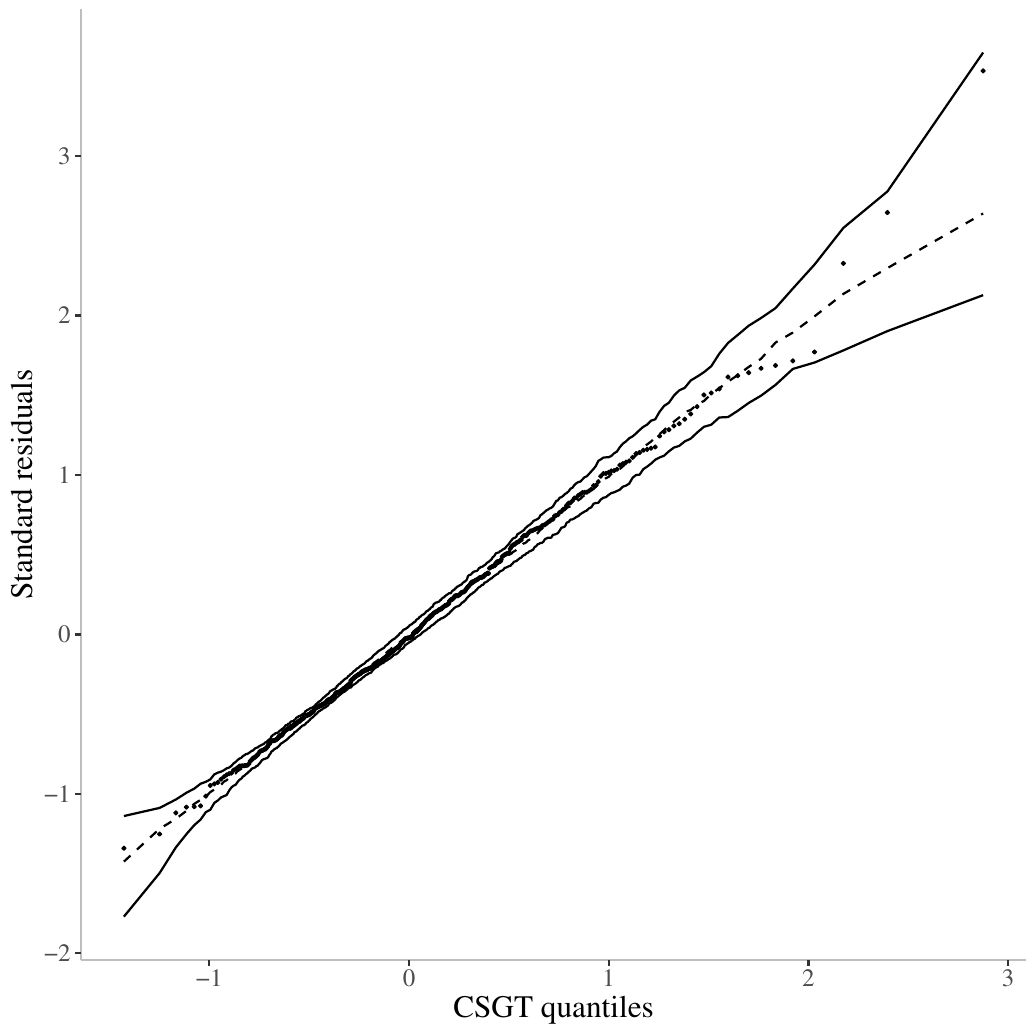}
			\caption{}
			\label{fig:residuos_csgt_csgt}
		\end{subfigure}\\
		\begin{subfigure}{.37\linewidth}
			\includegraphics[width=\linewidth]{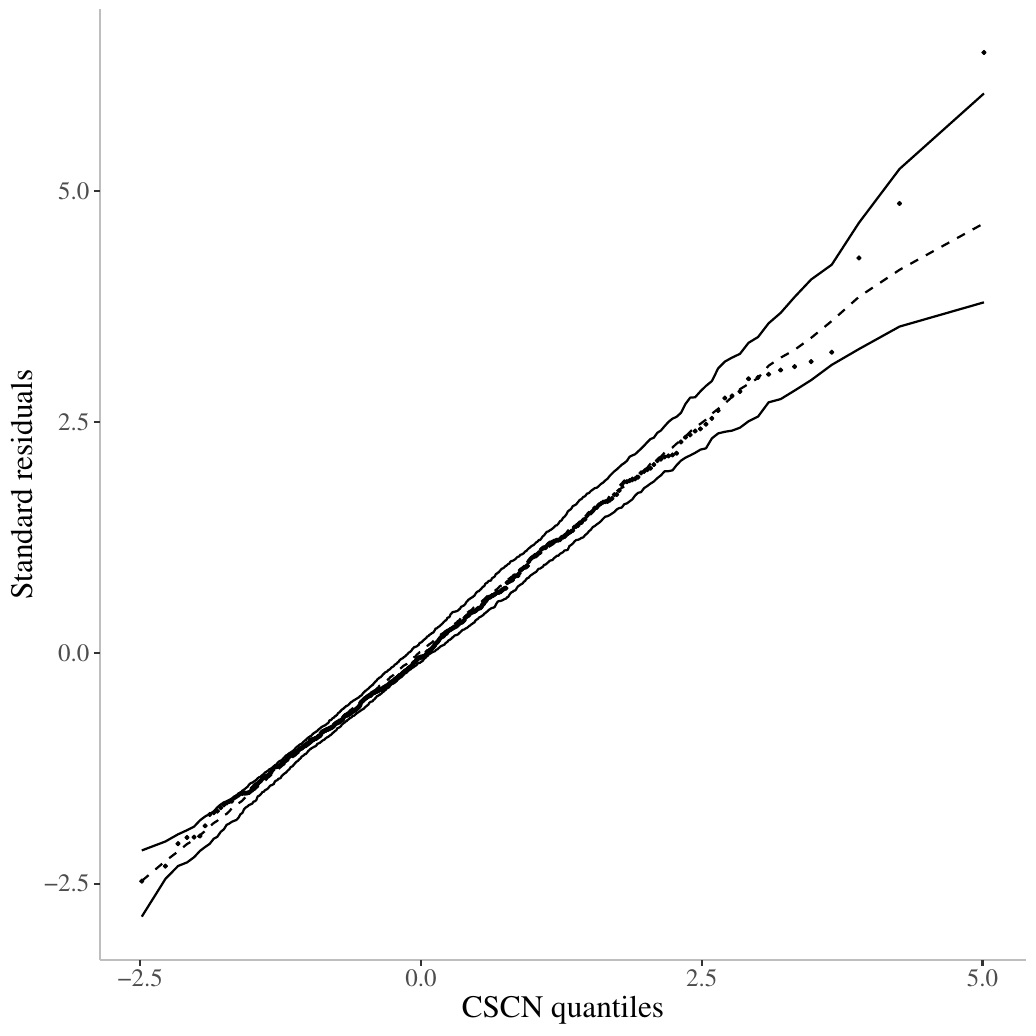}
			\caption{}
			\label{fig:residuos_csgt_cscn}
		\end{subfigure}
		\end{center}
	\caption{QQ plots using the data set generated by the skew generalized t distribution and adjusting by: skew normal (\subref{fig:residuos_csgt_csn}), skew t (\subref{fig:residuos_csgt_cst}), skew slash (\subref{fig:residuos_csgt_css}), skew generalized t (\subref{fig:residuos_csgt_csgt}) and skew contaminated normal (\subref{fig:residuos_csgt_cscn})}
	\label{fig:residuos_csgt}
\end{figure}

\begin{figure}[h!]
		\begin{center}
		\begin{subfigure}{.37\linewidth}
			\includegraphics[width=\linewidth]{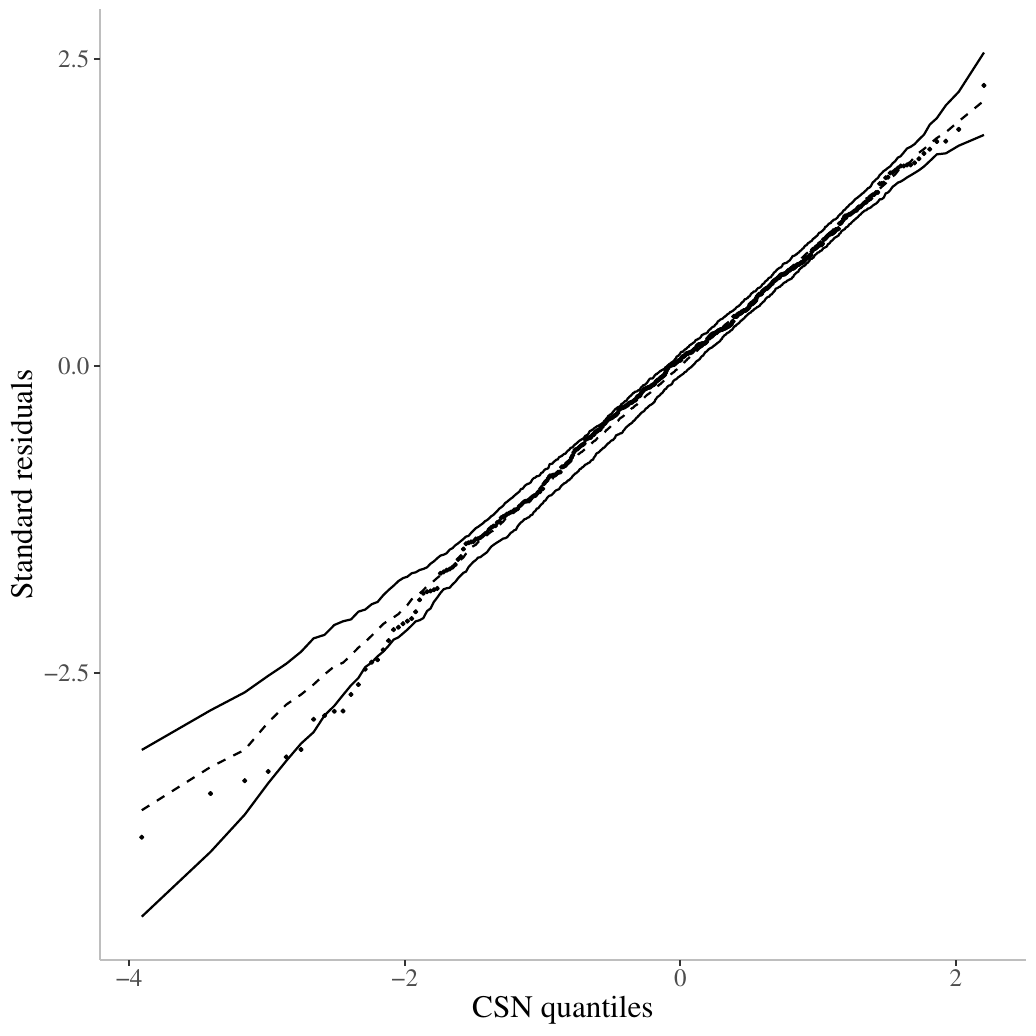}
			\caption{}
			\label{fig:residuos_cscn_csn}
		\end{subfigure}
		\begin{subfigure}{.37\linewidth}
			\includegraphics[width=\linewidth]{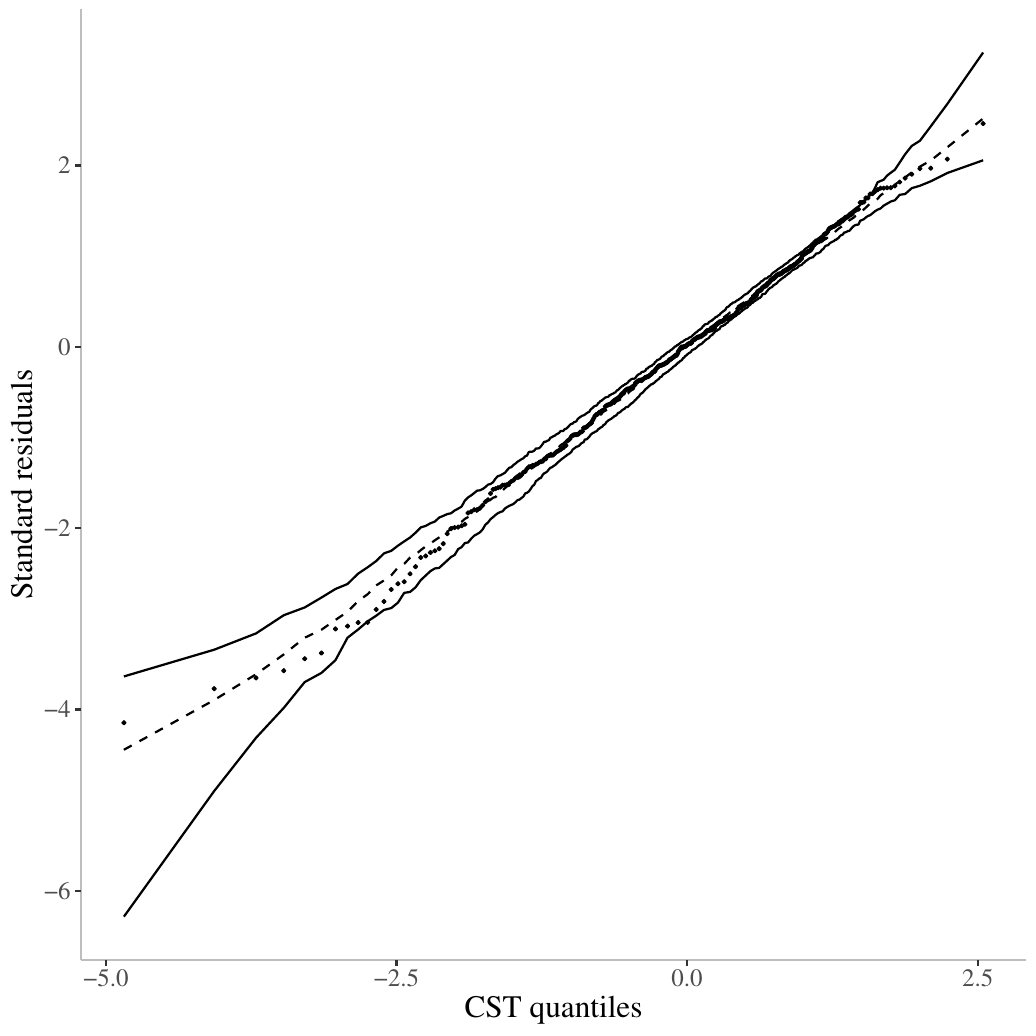}
			\caption{}
			\label{fig:residuos_cscn_cst}
		\end{subfigure}\\
		\begin{subfigure}{.37\linewidth}
			\includegraphics[width=\linewidth]{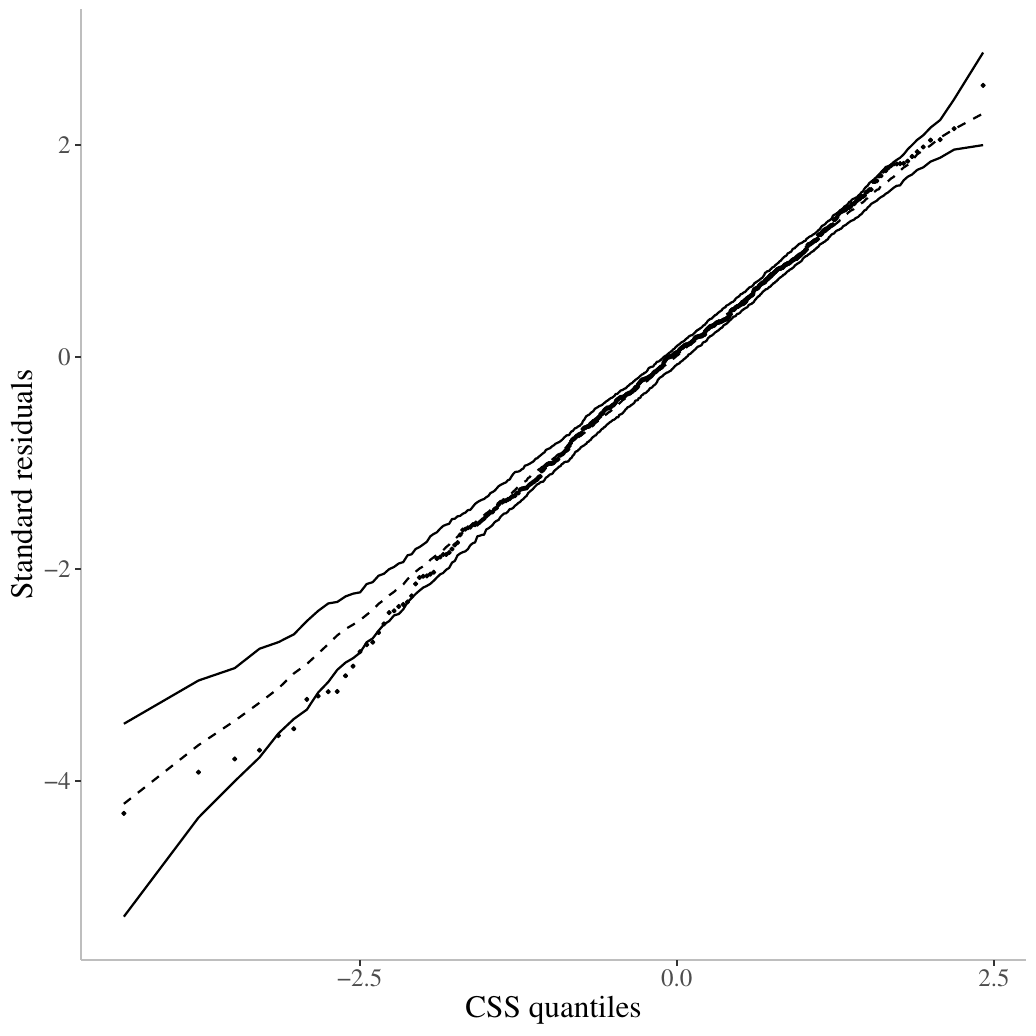}
			\caption{}
			\label{fig:residuos_cscn_css}
		\end{subfigure}
		\begin{subfigure}{.37\linewidth}
			\includegraphics[width=\linewidth]{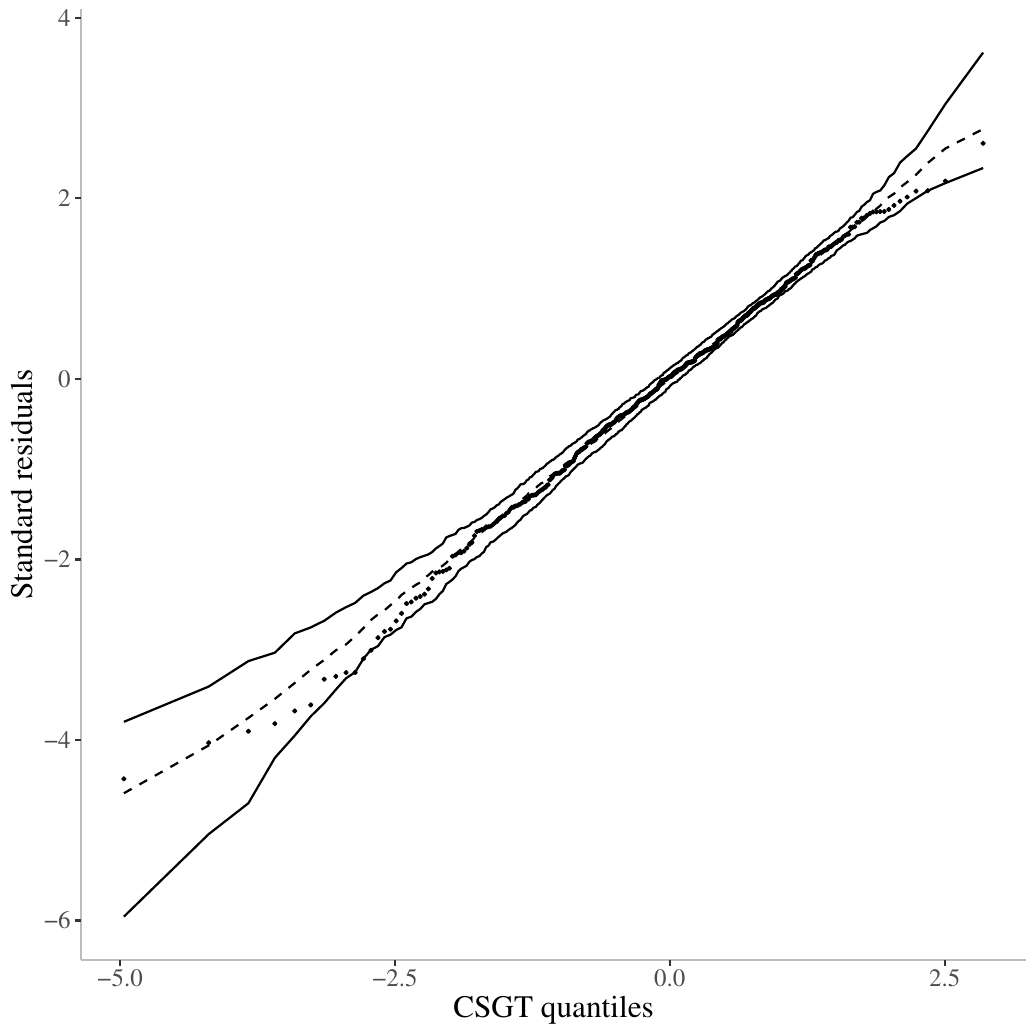}
			\caption{}
			\label{fig:residuos_cscn_csgt}
		\end{subfigure}\\

		\begin{subfigure}{.37\linewidth}
			\includegraphics[width=\linewidth]{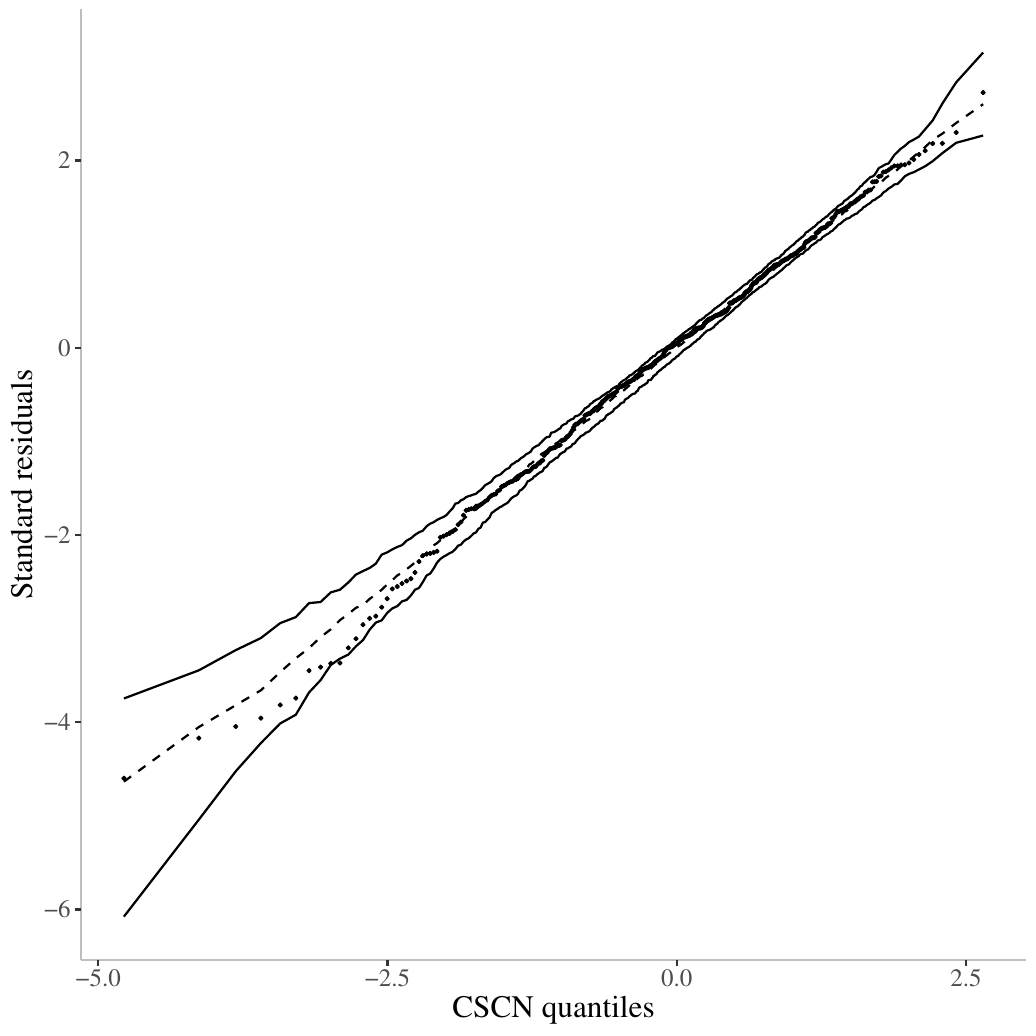}
			\caption{}
			\label{fig:residuos_cscn_cscn}
		\end{subfigure}
		\end{center}
	\caption{QQ plots using the data set generated by the skew contaminated normal distribution and adjusting by: skew normal (\subref{fig:residuos_cscn_csn}), skew t (\subref{fig:residuos_cscn_cst}), skew slash (\subref{fig:residuos_cscn_css}), skew generalized t (\subref{fig:residuos_cscn_csgt}) and skew contaminated normal (\subref{fig:residuos_cscn_cscn})}
	\label{fig:residuos_cscn}
\end{figure}


\bibliography{sn-bibliography.bib}